\documentclass{aa}

\usepackage{psfig}
\usepackage{amssymb}

\newcommand{\hei}{He~{\sc i} $\lambda$4471}
\newcommand{\heii}{He~{\sc ii}  $\lambda$4542}
\newcommand{\heib}{He~{\sc i} $\lambda$4388}
\newcommand{\heic}{He~{\sc i} $\lambda$4713}
\newcommand{\heid}{He~{\sc i} $\lambda$4920}
\newcommand{\heie}{He~{\sc i} $\lambda$4144}
\newcommand{\heif}{He~{\sc i} $\lambda$5016}
\newcommand{\heiib}{He~{\sc ii}  $\lambda$4200}
\newcommand{\heiivz}{He~{\sc ii}  $\lambda$4686}
\newcommand{\heired}{He~{\sc i} $\lambda$5876}
\newcommand{\heiired}{He~{\sc ii} $\lambda$5412}
\newcommand{\civ}{C~{\sc iv} $\lambda\lambda$1548,1551}
\newcommand{\nv}{N~{\sc v}  $\lambda$1240}
\newcommand{\niv}{N~{\sc iv}  $\lambda$1718}
\newcommand{\oiv}{O~{\sc iv} $\lambda\lambda$1339,1343}
\newcommand{\ov}{O~{\sc v}  $\lambda$1371}
\newcommand{\ovi}{O~{\sc vi} $\lambda\lambda$1032,1038}
\newcommand{\siiv}{Si~{\sc iv} $\lambda\lambda$1394,1403}

\newcommand{\heiiuv}{He~{\sc ii} $\lambda$ 1640}
\newcommand{\ha}{H$\alpha$}

\newcommand{\teff}{\ifmmode T_{\rm eff} \else $T_{\mathrm{eff}}$\fi}
\newcommand{\logg}{\ifmmode \log g \else $\log g$\fi}
\newcommand{\msun}{\ifmmode M_{\odot} \else M$_{\odot}$\fi}
\newcommand{\zsun}{\ifmmode Z_{\odot} \else Z$_{\odot}$\fi}
\newcommand{\lsun}{\ifmmode L_{\odot} \else L$_{\odot}$\fi}
\newcommand{\rsun}{\ifmmode R_{\odot} \else R$_{\odot}$\fi}

\newcommand{\mdot}{$\dot{M}$}
\newcommand{\myr}{M$_{\odot}$ yr$^{-1}$}
\newcommand{\vsini}{$V$ sin$i$}
\newcommand{\vinf}{$v_{\infty}$}
\newcommand{\vturb}{v$_{turb}$}
\newcommand{\vesc}{v$_{esc}$}
\newcommand{\kms}{km s$^{-1}$}

\def\aap{A\&A}
\def\aaps{A\&AS}
\def\aas{A\&AS}
\def\aj{AJ}
\def\apj{ApJ}

\def\apjs{ApJS}
\def\mnras{MNRAS}

\begin{document}

\title{O stars with weak winds: the Galactic case
\thanks{Partly based on
    observations collected with ESO-NTT telescope (program 72.D-0038(A))}}

\author{Fabrice Martins 
           \inst{1,2,3}
        \and Daniel Schaerer 
           \inst{1,2}
        \and D. John Hillier
           \inst{4}
        \and Fr\'ed\'eric Meynadier 
	   \inst{5}
        \and Mohammad Heydari-Malayeri 
	   \inst{5}
        \and Nolan R. Walborn 
	   \inst{6}}

\offprints{F. Martins, martins@mpe.mpg.de }

 \institute{
 Observatoire de Gen\`eve, 51 Chemin des Maillettes, CH-1290 Sauverny, 
 Switzerland
 \and
Laboratoire d'Astrophysique, Observatoire Midi-Pyr\'en\'ees,
 14 Av.  E. Belin, F-31400 Toulouse, France 
 \and 
Max Planck Instit$\ddot{\rm u}$t f$\ddot{\rm u}$r Extraterrestrische Physik, 
Postfach 1312, D-85741 Garching, Germany
 \and
 Department of Physics and Astronomy, University of Pittsburgh, 3941
 O'Hara Street, Pittsburgh, PA 15260, USA
  \and
 LERMA, Observatoire de Paris, 61 Rue de l'Observatoire, F-75012
 Paris, France
  \and
 Space Telescope Science Institute, 3700 San Martin Drive, Baltimore, MD
 21218, USA
}

\date{submitted 23 Feb. 2005 / accepted 5 Jul. 2005}

\titlerunning{Galactic O stars with weak winds}

\abstract{

We study the stellar and wind properties of a sample of Galactic O dwarfs 
to track the conditions under which weak winds (i.e mass loss rates lower than 
$\sim 10^{-8}$ \myr) appear. The sample is 
composed of low and high luminosity dwarfs including Vz stars and stars 
known to display qualitatively weak winds. Atmosphere models including 
non-LTE treatment, spherical expansion and line blanketing are computed 
with the code CMFGEN (Hillier \& Miller \cite{hm98}). Both UV 
and \ha\ lines are used to derive wind properties while optical H and He lines 
give the stellar parameters. We find that the stars of our sample are 
usually 1 to 4 Myr old. Mass loss rates of all stars are found to be 
lower than expected from the hydrodynamical predictions of Vink et al.\ 
(\cite{vink01}). For stars with $\log \frac{L}{L_{\odot}} \ga 5.2$, the 
reduction is by less than a factor 5 and is mainly due to the inclusion 
of clumping in the models. For stars with $\log \frac{L}{L_{\odot}} \la 5.2$ 
the reduction can be as high as a factor 100. The inclusion of X-ray 
emission (possibly due to magnetic mechanisms)
in models with low density is crucial to derive accurate mass loss
rates from UV lines, while it is found to be unimportant for 
high density winds. 
The modified wind momentum 
- luminosity relation shows a significant change of slope around this 
transition luminosity. Terminal velocities of low luminosity stars are also 
found to be low. Both mass loss rates and terminal velocities of 
low $L$ stars are consistent with a reduced line force parameter $\alpha$. 
However, the 
physical reason for such a reduction is still not clear although the 
finding of weak winds in Galactic stars excludes the role of a reduced 
metallicity. There may be a link between an early evolutionary state and a 
weak wind, but this has to be confirmed by further studies of Vz stars.
X-rays, through the change in the ionisation structure they imply, may 
be at the origin of a reduction of the radiative acceleration, leading 
to lower mass loss rates. A better understanding of the origin of X-rays 
is of crucial importance for the study of the physics of weak winds

}

\maketitle

\section{Introduction}
\label{s_intro}

Massive stars are known to develop winds so intense that mass loss 
rate turns out to be the main factor governing their evolution (e.g. 
Chiosi \& Maeder \cite{cm86}). The mechanism responsible for such 
strong outflows was first pointed out by Milne (\cite{milne}) when 
observations of winds were not yet available: the radiative 
acceleration in these bright objects was suspected to be large enough 
to overtake gravitational acceleration, creating expanding atmospheres. 
The first quantitative description of this process was given by Lucy 
\& Solomon (\cite{ls71}) who computed mass loss rates due to radiative 
acceleration through strong UV resonance lines. Castor, Abbott \& 
Klein (\cite{cak}) made a significant improvement in the understanding 
of winds of massive stars in their detailed calculation of radiative 
acceleration including an ensemble of lines by means of their now famous 
formalism and found mass loss rates $\sim$ 100 times larger than Lucy 
\& Solomon (\cite{ls71}). The theory of radiation driven winds developed
 by Castor, Abbott \& Klein was further improved by Pauldrach et al.\ 
(\cite{pauldrach86}) and Kudritzki et al.\ 
(\cite{kud89}) who included the effect of the finite size of the star
in the radiative acceleration. 

In parallel to theoretical studies, observational constraints on the
wind properties of massive stars were obtained. Most methods relied on
either the measurement of infrared and radio excess emitted in the
wind of such stars (Howarth \& Prinja \cite{hp89}, Leitherer
\cite{leith88}, Lamers \& Leitherer \cite{ll93}), or on the analysis
of UV and optical emission or P-Cygni lines (e.g. Leitherer
\cite{leith88}, Haser \cite{haser}, Puls et al.\ \cite{puls96}). The
results confirmed the prediction of the theory that the mass loss rate
should scale mainly as a power law of luminosity (e.g. Howarth \&
Prinja \cite{hp89}) and that the terminal velocities are directly
proportional to escape velocities (e.g. Lamers et al.\
\cite{lamers95}). Another success of the radiation driven wind theory
came from the so called modified wind momentum - luminosity relation
(hereafter WLR). Kudritzki et al.\ (\cite{kud95}) showed that the
quantity $\dot{M} v_{\infty} \sqrt{R}$ (with \mdot\ the mass loss
rate, \vinf\ the terminal velocity and R the stellar radius) should
depend only on luminosity (contrary to \mdot\ which also depends
slightly on the star mass) which was soon confirmed by the
spectroscopic analysis of O and B stars (Puls et al.\ \cite{puls96},
Kudritzki et al.\ \cite{kud99}). This finding was quite exciting since
once calibrated, the WLR could be used as a distance indicator up to
several Mpc (Kudritzki \cite{kud98}). Recent determinations of wind
parameters with sophisticated atmosphere codes confirm the good
agreement between observational constraints and theoretical
predictions for bright O stars, both in term of mass loss rate (for
which the most recent predictions are those of Vink et al.\
\cite{vink00}, \cite{vink01}) and WLR (see Herrero, Puls \& Najarro
\cite{hpn02}, Crowther et al.\ \cite{paul02}, Repolust et al.\
\cite{repolust}).

In spite of these encouraging results, the behaviour of the wind
properties of O stars with relatively low luminosity seems to be a
little more complicated. Martins et al.\ (\cite{iau212}, \cite{n81},
hereafter paper I) have shown that the stellar components of the star
forming region N81 of the SMC are O dwarfs with low luminosities and
surprisingly weak winds: the mass loss rates are lower than 10$^{-8}$
\myr\ and the modified wind momenta are nearly 2 orders of magnitude
lower than expected from the WLR obtained for bright stars. Bouret et
al.\ (\cite{jc03}) also found low mass loss rates for the faintest of
the NGC 346 dwarfs they analysed. Although all stars were in the SMC,
we showed in paper I that metallicity may not be the only factor
responsible for such a strong reduction of the wind strength. In
particular, we showed that a Galactic star - 10 Lac - displayed a
similar weak wind. One of the explanations we highlighted was a
possible link with the youth of the stars since most of them were (or
were suspected to be) Vz stars, i.e. young stars lying close to the
ZAMS (Walborn \& Parker \cite{wp92}). Another possibility was a break
down of the scaling relations (especially the WLR) at low
luminosity. This reduction of the wind strength at low luminosities
was in fact already mentioned by Chlebowski \& Garmany (\cite{chleb})
more than a decade ago.

In this paper, we try to investigate more deeply the wind properties of low 
luminosity Galactic stars. The aim is 1) to see if one can exclude the effect 
of metallicity to explain the weakness of the winds, 2) to test the hypothesis 
of the link between the weakness of the wind and the youth of the stars and 3) 
to quantify the wind properties of faint O stars and the luminosity below which 
such weak winds are observed. For this, we study a sample of O dwarfs with 
both low and high luminosities. Stars known to display qualitatively weak 
winds are included together with stars 
belonging to the Vz subclass. We selected stars showing weak
UV lines usually sensitive to winds (from the IUE atlas of Walborn et
al.\ \cite{walbornIUE})  
and/or with low mass loss rates from the study of Chlebowski \&
Garmany (\cite{chleb}). We also included Vz stars (N. Walborn,
private communication) and bright stars (two in common with the Repolust et al.\ 
\cite{repolust} sample) to examine the dependence of the wind
properties on luminosity. Finally, stars from the young star forming
region in the Rosette nebula were included. 

The remainder of the paper is organised as follows: In Sect.\ \ref{s_obs} we 
give information about the observational data we used; Sect.\ \ref{s_method} 
explains how we derived the stellar and wind parameters; Sect.\ 
\ref{s_results} gives the results for individual stars; Sect.\ \ref{X_rays} 
highlights the importance of X-rays and magnetic fields in weak wind stars, 
while Sect.\ 
\ref{s_uncertainty} discusses possible sources of uncertainty; the 
results are discussed in Sect.\ \ref{s_disc} and the conclusions are 
given in Sect.\ \ref{s_conc}.

\section{Observations}
\label{s_obs}

\subsection{Optical}
\label{s_optical}

Various sources have been used for the optical spectra of the
stars studied here. First, the VLT archive provided UVES spectra for
HD 152590, HD 38666 and HD 46202. The instrumental resolution varies
between 0.04 \AA\ and 0.1 \AA, due to different slit widths. The UVES
pipeline was used for the reduction of the data.
Second, optical data for HD 34078 and HD 15629 were retrieved from the
La Palma archive. Spectra 
obtained with the instrument ISIS on the WHT were reduced using
standard procedures under the ESO/MIDAS environment. The spectral
resolution is 0.9 \AA. Third, spectra of HD 93204, HD 93250 (EMMI) and
HD 15629 (La Palma) were provided
by Artemio Herrero and Danny Lennon and have a typical resolution of
0.95 \AA.
Finally for stars HD 93146, HD 93028, HD 46223 and HD 42088, we used EMMI
spectra obtained during the nights of 29, 30 and 
31 December 2003 on the ESO/NTT in La Silla, under the program
72.D-0038(A) (PI Martins). These spectra were
obtained in the red mode of the instrument and provided the
\ha\ profiles. The IRAF package was used for the data
reduction. 
For a few stars, we were left with several spectra of the same
wavelength range. In that case, we always chose the spectra with the
best resolution. The signal to noise ratio depends on the instrument
used but is usually larger than 100 in most lines of interest.

\subsection{UV}
\label{s_UV}

The IUE archive was used to retrieve the UV spectra of all the stars
of this study. Spectra in the range 1150-2000 \AA\ obtained with the
Short Wavelength Prime (SWP) camera were selected. The typical
instrumental resolution is 0.2 \AA\ and a S/N ratio of the order of
10. The normalisation was made ``by
eye'' and turned out to be somewhat uncertain below 1200 \AA.

We also retrieved FUSE spectra when available from the MAST
archive. The data are provided already reduced (without binning) and 
co-added by the CALFUSE 
pipeline (version 1.8.7), and we simply
normalised them by eye. Due to the strong Galactic interstellar
absorption, many broad absorption bands form H$_{2}$ render the
bluest part of the FUSE spectra useless for our purpose (e.g. Pellerin et al.\
\cite{pellerin}). We mostly used the 1100-1180 \AA\ range which has a
better signal to noise ratio than the IUE spectra for such wavelengths 
and extends to shorter wavelengths.

\begin{table*}
\caption{Adopted parameters for our program stars. The photomerty comes from Chlebowski \& Garmany (\cite{chleb}), Hiltner (\cite{hiltner}), Howarth \& Prinja (\cite{hp89}), Humphreys (\cite{humph}), Ma\'iz-Apell\'aniz et al.\ (\cite{MA}), Massey et al.\ (\cite{massey01}), Nicolet (\cite{nicolet}), Puls et al.\ (\cite{puls96}), Schild et al.\ (\cite{schild}), Walborn et al.\ (\cite{walborn02}). Distance modulus are taken from DeGioia-Eastwood et al.\ (\cite{degioia}), Humphreys (\cite{humph}), Ma\'iz-Apell\'aniz et al.\ (\cite{MA}), Markova et al.\ (\cite{mark04}), Massey et al.\ (\cite{massey01}), Mason et al.\ (\cite{mason}) and the Webda database. References to the observing data sets retrieved from archives are also given when available.}
\label{tab_prop_adopt}
\small
\center
\begin{tabular}{l|lllllllllllll}
HD & ST & V & E(B-V) & DM & M$_{V}$ & FUSE & IUE SWP & optical data &\\
\hline 
38666 & O9.5V & 5.15$^{+0.01}_{-0.01}$ & 0.05$^{+0.01}_{-0.01}$ & 8.63$^{+0.93}_{-0.63}$ & -3.64$^{+0.60}_{-0.93}$ & - & 6631 & ESO/UVES 65.H-0375 & \\
34078 & O9.5V & 5.99$^{+0.01}_{-0.04}$ & 0.54$^{+0.01}_{-0.02}$ & 8.25$^{+0.87}_{-0.62}$ & -3.92$^{+0.69}_{-0.96}$ & B063 & 54036 & WHT/ISIS & \\
46202 & O9V & 8.18$^{+0.02}_{-0.02}$ & 0.49$^{+0.01}_{-0.01}$ & 10.85$^{+0.05}_{-0.05}$ & -4.19$^{+0.10}_{-0.10}$ & - & 8845 & INT/IDS & \\
93028 & O9V & 8.36$^{+0.01}_{-0.01}$ & 0.26$^{+0.01}_{-0.01}$ & 12.09$^{+0.37}_{-0.38}$ & -4.54$^{+0.38}_{-0.37}$ & A118 & 5521 & ESO/EMMI 72.D-0038 &\\
152590 & O7.5Vz & 8.44$^{+0.02}_{-0.02}$ & 0.46$^{+0.10}_{-0.10}$ & 10.72$^{+0.69}_{-0.44}$ & -3.71$^{+0.51}_{-0.74}$ & - & 16098 & ESO/UVES 67.B-0504 & \\
93146 & O6.5V((f)) & 8.43$^{+0.02}_{-0.02}$ & 0.34$^{+0.01}_{-0.02}$ & 12.09$^{+0.37}_{-0.37}$ & -4.70$^{+0.45}_{-0.45}$ & - & 11136 & ESO/EMMI 72.D-0038 & \\
42088 & O6.5Vz & 7.55$^{+0.01}_{-0.01}$ & 0.46$^{+0.01}_{-0.01}$ & 11.20$^{+0.20}_{-0.23}$ & -4.66$^{+0.30}_{-0.33}$ & P102 & 7706 & ESO/EMMI 72.D-0038 & \\
93204 & O5V((f)) & 8.44$^{+0.02}_{-0.02}$ & 0.42$^{+0.01}_{-0.01}$ & 12.34$^{+0.45}_{-0.27}$ & -5.20$^{+0.34}_{-0.38}$ & - & 7023 & INT/IDS & \\
15629 & O5V((f)) & 8.42$^{+0.01}_{-0.01}$ & 0.74$^{+0.01}_{-0.01}$ & 11.38$^{+0.30}_{-0.30}$ & -5.25$^{+0.30}_{-0.30}$ & - & 10754 & INT/IDS & \\
46223 & O4V((f+)) & 7.27$^{+0.05}_{-0.02}$ & 0.54$^{+0.01}_{-0.01}$ & 10.85$^{+0.05}_{-0.05}$ & -5.25$^{+0.10}_{-0.07}$ & - & 8844 & ESO/EMMI 72.D-0038 & \\
93250 & O3.5V((f+)) & 7.38$^{+0.02}_{-0.02}$ & 0.48$^{+0.01}_{-0.01}$ & 12.34$^{+0.45}_{-0.27}$ & -6.45$^{+0.29}_{-0.48}$ & - & 22106 & INT/IDS & \\
\normalsize
\end{tabular}
\end{table*}

\section{Method}
\label{s_method}

Our main concern is to derive wind parameters (mass loss rates,
terminal velocities) and modified wind momenta. However, such
determinations require reliable stellar parameters, especially
effective temperatures. Indeed, any uncertainty in \teff\ can lead to
an error on \mdot. We thus first estimate the stellar parameters
using the optical spectra, and then we use the UV range +
\ha\ line to determine the wind properties.

\subsection{Stellar parameters}
\label{s_stellar_param}
The main stellar parameters have been determined from blue optical
spectra. As such spectra contain diagnostic lines which are formed
just above the photosphere and are not affected by winds,
plane-parallel models can be used for a preliminary analysis. Hence, we have
taken advantage of the recent grid of TLUSTY spectra (OSTAR2002,
Lanz \& Hubeny \cite{ostar2002}). This grid covers the log g - \teff\
plane for O stars and includes optical synthetic spectra computed with
a turbulent velocity of 10 \kms. The models include the main
ingredients of the modelling of O star atmospheres (especially non-LTE
treatment and 
line-blanketing) except that they do not take the wind into account
(see Hubeny \& Lanz \cite{hl95} for details). 

Our method has been the following:\\

\textit{- \vsini\ :}  we adopted the rotational velocities from the literature
(mostly Penny \cite{penny}) and refined them in the fitting process when possible.\\

\textit{- \teff\ :} the ratio of \hei\ to \heii\
equivalent widths gave the spectral type which was used to estimate
\teff\ from the \teff-scale of Martins et al.\
(\cite{teffscale}). Then, TLUSTY spectra with effective temperatures
bracketing this value were convolved to take into account the
rotational velocity and instrumental resolution, and the resulting
spectra were compared to the observed 
profiles of the \hei\ and \heii\ lines. The best fit led to the
constraint on \teff. As the OSTAR2002 grid has a relatively coarse sampling 
(2500 K steps), we have often interpolated line profiles of intermediate
temperatures. A simple linear interpolation was used. For the stars
for which the \hei\ and \heii\ lines were not available, we used
\heired\ and \heiired\ as the main indicators.  

Secondary \teff\ diagnostic lines such as \heib, \heic, \heid, \heie,
\heif\ and \heiib\ were also used to refine the determination (when
available). The uncertainty in \teff\ depends on the resolution of the
spectra and on the rotational broadening. Indeed, the broader the
profile, the lower the precision of the fit of the line. The typical
error on \teff\ is usually of $\pm$ 2000 K but can be reduced when
many optical He lines are available. Note that the errors we give are
$2 \sigma$ errors (we have $\teff\ - error < \teff\ < \teff\ +
error$).

We also checked that our final models including winds computed with
CMFGEN fitted correctly the optical lines. It turns out that the
agreement between TLUSTY and CMFGEN is very good as already noticed in
previous studies (e.g. Bouret et al.\ \cite{jc03}). The problem
recently highlighted by Puls et al.\ (\cite{puls05}) concerning the
weakness of the He~{\sc I} singlet lines between 35000 and 40000 K is
in fact related to subtle line blanketing effects and is solved when
both the turbulent velocity is reduced (in the computation of the
atmospheric structure) and other species (Neon, Argon, Calcium and
Nickel) are added into the models (see Sect.\ \ref{s_hd152590}).\\

\textit{- $\log g$ :} Fits of the wings of H$\gamma$ led to
constraints on \logg. Once again, interpolations between the OSTAR2002
spectra were made to improve the determination as the step size of the
OSTAR2002 grid is 0.25 in log g. H$\beta$, which behaves similarly to
H$\gamma$, was used as a secondary indicator. The typical uncertainty
in log g is 0.1 dex.\\

Once obtained, these parameters have been used to
derive $L$, $R$ and $M$:\\

\textit{- Luminosity :} with \teff\ known, we have estimated a bolometric 
correction according to 

\begin{equation}
BC(T_{\rm eff})=27.66-6.84 \times \log T_{\rm eff}
\end{equation}

\noindent which has been established by Vacca et al.\ (\cite{vacca}). Visual
magnitudes together with estimates of the reddening and the distance
modulus of the star have then lead to M$_{V}$ and $L$ from:

\begin{equation}
\log \frac{L}{L_\odot}=-0.4 (M_{\rm{V}}+BC-M_{\odot})
\end{equation}
\noindent the error on \teff\ leads to a typical error of 0.2 dex on BC.
Note that we have recently revised the calibration of bolometric corrections 
as a function of \teff\ (see Martins et al.\ \cite{calib05}), but it turns out 
that due to line-blanketing effects, BCs are reduced by only 0.1 dex, which 
translates to a reduction of $\log L$ by 0.04 dex, which is negligible here given 
the uncertainty in the distance. 

The solar bolometric magnitude was taken as equal to 4.75 (Allen
\cite{allen}). We want to caution here that for most of the stars of
this study, the distance is poorly known (with sometimes a difference
of 1 magnitude on the distance modulus between existing
determinations). This leads to an important error on the
luminosity. As this last parameter is crucial for the calibration of
the modified wind momentum - relation, we decided to take the maximum
error on $L$ by adopting the lowest (resp. highest) luminosity
(derived from the lowest -resp.  highest- extinction, distance modulus
and bolometric correction) as the boundaries to the range of possible
luminosities. The typical error on $L$ is $\sim$ $\pm$ 0.25 dex, and
the main source of uncertainty is the distance. \\

\textit{- Radius :} Once \teff\ and $L$ are known, $R$ is simply derived from

\begin{equation}
R = \sqrt{\frac{L}{4 \pi \sigma \teff^{4}}}
\end{equation}

\noindent where $\sigma$ is the Stefan Boltzmann constant. Standard errors
have been derived according to 

\begin{equation}
\Delta \log R = 0.5 \sqrt{(\Delta \log L)^{2} + (4 \Delta \log \teff)^{2}}
\end{equation}

\textit{- M :} The (spectroscopic) mass is derived from $g$ and $R$
according to

\begin{equation}
M = \frac{gR^{2}}{G} 
\end{equation}
\noindent and the standard error is given by

\begin{equation}
\Delta \log M = \sqrt{(\Delta \log g)^{2} + (2 \Delta \log R)^{2}}
\end{equation}

With this set of stellar parameters, we have run models including
winds to derive the mass loss rate and the terminal velocity (see
next section). The stellar parameters
giving the best agreement between observations and models with winds
were adopted as the final stellar parameters.

\subsection{Wind parameters}
\label{s_wind_param}
UV (and FUV when available) spectra and \ha\ profiles
were used to constrain the wind parameters. In the case where mass loss 
rates were low, priority was given to UV indicators since
\ha\ becomes insensitive to \mdot: for such situations, we checked
that the \ha\ line given by our models with \mdot\ estimated from UV
was consistent with the observed line. We want to stress here that it is 
only because metals are now included in a reliable way in new generation 
atmosphere models that such a study is possible. Indeed, UV metallic lines 
now correctly reproduced allow to push the limits of mass loss determination 
below $\sim 10^{-8}$ \myr.
 
Models including stellar winds were computed with the code CMFGEN
(Hillier \& Miller \cite{hm98}). This code allows for a non-LTE
treatment of the radiative transfer and statistical equilibrium
equations in spherical geometry and includes line blanketing effects
through a super-level approach. The temperature structure is computed
under the assumption of radiative equilibrium \footnote{Note that in
some models adiabatic cooling was also included, see Sect.\
\ref{s_results}}. At present, CMFGEN does not include
self-consistently the hydrodynamics of the wind so that the velocity
and density structures must be given as input (but hydrodynamical
quantities computed from the final atmosphere model are given as
output). In order to be as consistent as possible with the optical
analysis, we have used TLUSTY structures for the photosphere part and
we have connected them to a classical $\beta$ law ($v =
v_{\infty}(1-\frac{R_{\star}}{r})^{\beta}$) representing the wind
part. We chose $\beta = 1.0$ as the default value for our calculation
since it turns out to be representative of O dwarfs (e.g. Massa et
al.\ \cite{massa03}). The TLUSTY structures have been taken from the
OSTAR2002 grid or have been linearly interpolated from this grid for
\teff\ not included in OSTAR2002. This method has also been used by
Bouret et al.\ (\cite{jc03}) and has shown good consistency between
CMFGEN and TLUSTY photospheric spectra.

Clumping can be included in the wind models by means of a volume filling
factor $f$ parameterised as follows: 
$f = f_{\infty} + (1-f_{\infty})e^{-\frac{v}{v_{\rm init}}}$ where
$f_{\infty}$ is the value of $f$ at the top of the atmosphere and
$v_{\rm init}$ is the velocity at which clumping appears. As in Bouret et al.\ 
(\cite{jc03}), we chose $v_{\rm init} = 30$ \kms.

A depth independent microturbulent velocity can be included in the
computation of the atmospheric structure (i.e. temperature structure + 
population of individual levels). We chose a value of 20 \kms\ as 
the default value in our computations. Several tests (Martins et al.\
\cite{teffscale}, Bouret et al.\ \cite{jc03}) indicate that a
reasonable change of this parameter has little effect on the emergent
spectrum, except for some specific lines (see Sect.\ \ref{s_hd152590}). 
For the computation of the detailed spectrum resulting from 
a formal solution of the radiative transfer equation (i.e. with the
populations kept fixed), a depth dependent microturbulent velocity can 
be adopted. In that case, the microturbulent velocity follows the
relation
$v_{turb}(r) = v_{min} + (v_{max} - v_{min}) \frac{v(r)}{v_{\infty}}$
where $v_{min}$ and $v_{max}$ are the minimum and maximum
microturbulent velocities. 
By default, we chose $v_{min}$ = 5 \kms\ in the
photosphere, and $v_{max}$ = min (0.1 \vinf, 200) \kms\ at the top
of the atmosphere. For some stars, we had to increase $v_{min}$ from 5
to 10 \kms\ to be able to fit correctly the observed spectra. 

CMFGEN allow the possibility to include X-ray emission in the models.
In some cases (see next Section), we had to include such high energy
photons. Practically, as X-rays are thought to be emitted by shocks
distributed in the wind, two parameters are adopted to take them into
account: one is a shock temperature (chosen to be $3 \times 10^{6}$ K
since it is typical of high energy photons in O type stars,
e.g. Schulz et al.\ \cite{schulz03}, Cohen et al.\ \cite{cohen03}) to
set the wavelength of maximum emission, and the other is a volume
filling factor which is used to set the level of emission. With this
formalism, X-ray sources are distributed throughout the atmosphere and
the emissivities are taken from tables computed by a Raymond \& Smith
code (Raymond \& Smith \cite{rs77}).  We include X-rays in the models
for the four faintest stars and in test models for the strong lined
star HD 93250 as explained in Sect.\ \ref{X_rays}, using measured
X-ray fluxes or a canonical value of $L_{\rm X}/L_{\rm bol} = -7.0$.

The main wind parameters we have determined are the mass loss rate
($\dot{M}$) and the terminal velocity (v$_{\infty}$). Constraints on
the amount of clumping were also derived when possible.
The terminal velocities have been estimated from the blueward
extension of the absorption part of UV P-Cygni profiles 
which occurs up to \vinf\ + $v_{max}$ where $v_{max}$
is the maximum microturbulent velocity described above: fits of the UV
P-Cygni profiles using the above relation for microturbulent velocity
allows a direct determination of \vinf. Note that other definitions of 
the terminal velocity exist (see Prinja, 
Barlow \& Howarth (\cite{Prinja90})).. 
The typical uncertainty in our determination of \vinf\ is 200 \kms\
(depending on the maximum microturbulent velocity we adopt). 

Fits of strong UV lines such as \nv\, \civ\, \siiv\, \ov\ and \niv\
provide constraints on $\dot{M}$. \ha\ is also sensitive to
$\dot{M}$: in dwarfs with weak
winds, a quasi photospheric profile is expected but the line
can be used to estimate upper limits on the mass loss rate as it is
filled by emission when $\dot{M} \gtrsim 10^{-8}$ \myr. Also, 
in the case of weak winds \civ\ is actually the only line showing 
some sensitivity to wind and was in several cases our best \mdot\ 
estimator. Given this, we
tried to adjust the mass loss rate (and clumping
parameters) to get the best fit of both the UV wind sensitive lines
and \ha.

As regards the abundances, we have taken as default values the solar
determinations of Grevesse \& Sauval (\cite{gs98}) since the stars of
this study are are all Galactic stars. CNO solar abundances have been 
recently revised downward by Asplund (\cite{asplund04}). However, we 
preferred to rely on the Grevesse \& Sauval abundances since they have 
been widely used in previous studies of massive stars and are therefore 
more suited for comparisons. When necessary, 
we indicate if these abundances have been changed to get a better fit.

\section{Results}
\label{s_results}

In this Section, we present the results of the analysis for each star
(from the latest to the earliest type ones) and highlight the main
difficulty encountered in the fitting process.  The observed
properties and adopted parameters are given in Table
\ref{tab_prop_adopt}.  The derived stellar and wind parameters are
gathered in Table \ref{tab_prop}, while results from previous studies
of wind properties are given in Table \ref{tab_comp}.

Spectra from atmosphere models are convolved to include the
instrumental resolution of the observational data and the projected
rotational velocity of the star. The wavelength range between $\sim$ 1200 and 
$\sim$ 1225 \AA\ is not used in the spectral analysis since it suffers 
from a strong interstellar Lyman absorption.

A general comment concerning effective temperatures is that we find values 
lower than previous determinations (see table \ref{tab_comp}) since line-blanketing 
is included in our models. This effect is now well known and has been highlighted 
by several studies (Martins et al. \cite{teffscale}, \cite{calib05}, Crowther et al.\ 
\cite{paul02}, Markova et al.\ \cite{mark04}). Our mass loss rates are also 
generally lower than previous determination for reasons discussed in Sect.\ 
\ref{disc_mdot}.

\begin{table*}
\caption{Derived stellar and wind parameters of the Galactic stars studied here. 
The escape
  velocities were computed from the spectroscopic derived mass, radius 
  and luminosity. The evolutionary masses have been estimated from the 
  isochrones of Lejeune \& Schaerer (\cite{lejeune}). Projected rotational 
velocities were adopted from Penny (\cite{penny}), 
Howarth \& Prinja (\cite{hp89}), Howarth et al.\ (\cite{howarth97}) and Villamariz et 
al.\ (\cite{villamariz02}) and refined in the fitting process when possible.}
\label{tab_prop}
\small
\center
\begin{minipage}{\textwidth}
\begin{tabular}{l|lllllllll}
HD & 38666 & 34078 & 46202 & 93028 & 152590 & 93146 & 42088 &\\
\hline 
 & & \\
\teff (kK) & 33$\pm 1$ & 33$\pm 2$ & 33$\pm 2$ & 34$\pm 2$ & 36$\pm 1$ &
37$\pm 2$ & 38$\pm 2$ &\\
BC & -3.25$^{+0.10}_{-0.08}$ & -3.25$^{+0.10}_{-0.08}$ & -3.25$^{+0.10}_{-0.08}$ & -3.34$^{+0.18}_{-0.17}$ & -3.52$^{+0.08}_{-0.08}$ & -3.59$^{+0.13}_{-0.15}$ & -3.67$^{+0.16}_{-0.15}$ &\\
log g & 4.0$\pm 0.1$ & 4.05$\pm 0.1$ & 4.0$\pm 0.1$ & 4.0$\pm 0.1$ & 4.10$\pm 0.1$ & 4.0$\pm 0.1$ & 4.0$\pm 0.1$ &\\
\vsini (\kms) & 111 & 40 & 30 & 50 & 66 & 80 & 60 &\\ 
log $\frac{L}{L_{\odot}}$ & 4.66$^{+0.40}_{-0.30}$ & 4.77$^{+0.41}_{-0.32}$ & 4.87$^{+0.07}_{-0.07}$ & 5.05$^{+0.22}_{-0.22}$ & 4.79$^{+0.33}_{-0.24}$ & 5.22$^{+0.23}_{-0.25}$
& 5.23$^{+0.19}_{-0.19}$ &\\ 
R (\rsun) & 6.58$^{+3.89}_{-2.45}$ & 7.47$^{+4.55}_{-2.83}$ & 8.38$^{+0.88}_{-0.81}$ & 9.71$^{+3.11}_{-2.38}$ & 6.42$^{+3.00}_{-2.05}$ & 9.97$^{+3.29}_{-2.49}$ & 9.56$^{+2.61}_{-2.07}$ &\\
M$_{spectro}$ (M$_{\odot}$) & 16$^{+0.25}_{-0.10}$ & 23$^{+38}_{-11}$ & 26$^{+9}_{-7}$ & 34$^{+28}_{-16}$ & 19$^{+23}_{-10}$
& 36$^{+31}_{-17}$ & 33$^{+24}_{-14}$ &\\
M$_{evol}$ (M$_{\odot}$) & 19 & 20 & 21 & 25 & 22
& 30 & 31 &\\
v$_{esc}$ (\kms) & 920 & 1043 & 1046 & 1112 & 1015 & 1106 & 1100 &\\
 & & \\
\vinf (\kms) & 1200 & 800 & 1200 & 1300 & 1750 & 2800 & 1900 &\\
$\log$ \mdot (\myr)  & $-$9.5$\pm 0.7$ & $-$9.5$\pm 0.7$ & $-$8.9$\pm 0.7$ & $-$9.0$\pm 0.7$ & $-$7.78$\pm 0.7$ & $-$7.25$\pm 0.7$ & $-$8.0$\pm 0.7$ &\\
\vturb \footnote{the first value is the minimum and the second one the maximum microturbulent velocity (See Sect.\ \ref{s_wind_param})} (\kms) & 5-120 & 5-80 & 5-120 & 5-130 & 10-175 & 5-200 & 5-190 &\\
$f_{\infty}$ & 1.0 & 1.0 & 1.0 & 1.0 & 1.0 & 1.0 & 1.0 &\\
$\log \dot{M}_{Vink}$ (\myr) & $-$7.41 & $-$7.38 & $-$7.23 & $-$6.97 & $-$7.15 &
$-$6.58 & $-$6.17 &\\
$\log \dot{M} v_{\infty} \sqrt{R}$ & 24.79 & 24.64 & 25.44 & 25.41 &
26.67 & 27.50 & 26.57 &\\

\normalsize
\end{tabular}
\end{minipage}
\end{table*}

\begin{table*}
\setcounter{table}{1}
\caption{Continued.}
\label{tab_prop}
\small
\center
\begin{minipage}{\textwidth}
\begin{tabular}{l|llllll}
HD & 93204 & 15629 & 46223 & 93250 &\\
\hline 
 & & \\
\teff (kK) & 40$\pm 2$ & 41$\pm 2$ & 41.5$\pm 2$ & 44$\pm 2$ &\\
BC & -3.82$^{+0.14}_{-0.16}$ & -3.89$^{+0.15}_{-0.15}$ & -3.93$^{+0.15}_{-0.15}$ & -4.10$^{+0.12}_{-0.12}$ &\\
log g & 4.0$\pm 0.1$ & 3.75$\pm 0.1$ & 4.0$\pm 0.1$ & 4.0$\pm 0.1$ &\\
\vsini (\kms) & 130 & 90 & 130 & 110 &\\ 
log $\frac{L}{L_{\odot}}$ & 5.51$^{+0.25}_{-0.20}$ & 5.56$^{+0.18}_{-0.18}$ & 5.57$^{+0.09}_{-0.10}$ & 6.12$^{+0.25}_{-0.17}$ &\\ 
R (\rsun) & 11.91$^{+4.23}_{-3.14}$ &
12.01$^{+3.08}_{-2.47}$ & 11.86$^{+1.78}_{-1.58}$ & 19.87$^{+6.98}_{-3.87}$ &\\
M$_{spectro}$ (M$_{\odot}$) & 52$^{+47}_{-25}$ & 30$^{+20}_{-12}$ & 51$^{+22}_{-16}$ & 144$^{+130}_{-56}$ &\\
M$_{evol}$ (M$_{\odot}$) & 41 & 44 & 45 & 105 &\\
v$_{esc}$ (\kms) & 1178 & 799 & 1157 & 1461 &\\
 & & \\
\vinf (\kms) & 2900 & 2800 & 2800 & 3000 &\\
$\log$ \mdot (\myr) & $-$6.75$\pm 0.7$ & $-$6.5$\pm 0.7$ & $-$6.5$\pm 0.7$ & $-$6.25$\pm 0.7$ &\\
\vturb (\kms) & 5-200 & 10-200 & 10-200 & 10-200 &\\
$f_{\infty}$ & 0.1 & 0.1 & 0.1 & 0.01&\\
$\log \dot{M}_{Vink}$ (\myr) & $-$6.11 & $-$5.74 & $-$5.97 & $-$5.25 &\\
$\log \dot{M} v_{\infty} \sqrt{R}$ & 28.05 & 28.29 & 28.28 & 28.68 &\\

\normalsize
\end{tabular}
\end{minipage}
\end{table*}

\begin{table}[!h]
\caption{Comparison between our derived wind parameters (\mdot,
  \vinf, shown in bold text) and previous determinations. (1) Leitherer (\cite{leith88}),
  (2) Bernabeu et al.\ (\cite{bernabeu}), (3) Howarth \& Prinja
  (\cite{hp89}), (4) Prinja et al.\ (\cite{Prinja90}), (5) Chlebowski
  \& Garmany (\cite{chleb}), (6) Lamers \& Leitherer (\cite{ll93}),
  (7) Lamers et al.\ (\cite{lamers95}), (8) Puls et al.\
  (\cite{puls96}), (9) Howarth et 
  al.\ (\cite{howarth97}), (10) Lamers et al.\ (\cite{lamers99}),
 (11) Repolust et al.\ (\cite{repolust}), (12)
  Markova et al.\ (\cite{mark04}).}
\label{tab_comp}
\small
\center
\begin{tabular}{l|lllll}
HD & $\log$ \mdot\ & & \vinf\ \\
\hline
38666 & $<$ $-$7.22 (1), $-$7.8 (3)  & & 2000 (1), 1000 (3) \\
      & $-$8.31 (5)     & &          \\
      & \textbf{$-$9.5} & & \textbf{1200} \\
& & \\
34078 & $-$6.6 (3)      & & 750 (3)  \\
      & \textbf{$-$9.5} & & \textbf{800} \\
& & \\
46202 & $<$ $-$6.87 (1), $-$7.2 (3) & & 2100 (1), 750 (3) \\
      & $-$8.10 (5)     & & 1150 (5), 1590 (2) \\
      & \textbf{$-$8.9} & & \textbf{1200} \\
& & \\
93028 & $-$7.0 (3)      & & 1500 (3), 1780 (2) \\
      & \textbf{$-$9.0} & & \textbf{1300} \\
& & \\
152590 & $-$6.9 (3), $-$7.36 (5)     & & 2150 (3), 2300 (5) \\
       &              & & 1785 (9) \\
       & \textbf{$-$7.78} & & \textbf{1750}  \\
& & \\
93146 & $-$6.9 (3)      & & 2975 (3), 3200 (2)  \\
      &               & & 2565 (4), 2640 (9)  \\
      & \textbf{$-$7.25} & & \textbf{2800} \\
& & \\
42088 & $-$6.35 (1), $-$7.0 (3)     & & 2550 (1), 2030 (3)  \\
      & $-$6.82 (5), $-$6.42 (12)     & & 2300 (5), 2420 (2)  \\
      &               & & 2155 (4), 2215 (9) \\
      &               & & 2200 (12) \\
      & \textbf{$-$8.0} & & \textbf{1900} \\
& & \\
93204 & $-$6.1 (3)      & & 3250 (3), 3180 (2) \\
      &               & & 2890 (4), 2900 (7) \\
      & \textbf{$-$6.75} & & \textbf{2900} \\
& & \\
15629 & $-$5.89 (11)    & & 3200 (11), 3220 (2) \\
      &               & & 2810 (9) \\
      & \textbf{$-$6.5} & & \textbf{2800} \\
& & \\
46223 & $-$5.75 (1), $-$5.8 (3)     & & 3100 (1), 3100 (3) \\
      & $-$5.62 (5), $-$5.85 (6)    & & 3100 (5), 2800 (6) \\
      & $-$5.68 (10)    & & 2800 (10), 3140 (2) \\
      &               & & 2910 (4), 2900 (7) \\
      & \textbf{$-$6.5} & & \textbf{2800} \\
& & \\
93250 & $-$4.9 (8), $-$4.6 (3)      & & 3250 (8), 3350 (3) \\
      & $-$5.46 (11)    & & 3250 (11), 3470 (2) \\
      &               & & 3230 (9) \\
      & \textbf{$-$6.25} & & \textbf{3000} \\

\normalsize
\end{tabular}
\end{table}

\subsection{HD38666}
\label{s_hd38666}

\begin{figure}[ht]
\centerline{\psfig{file=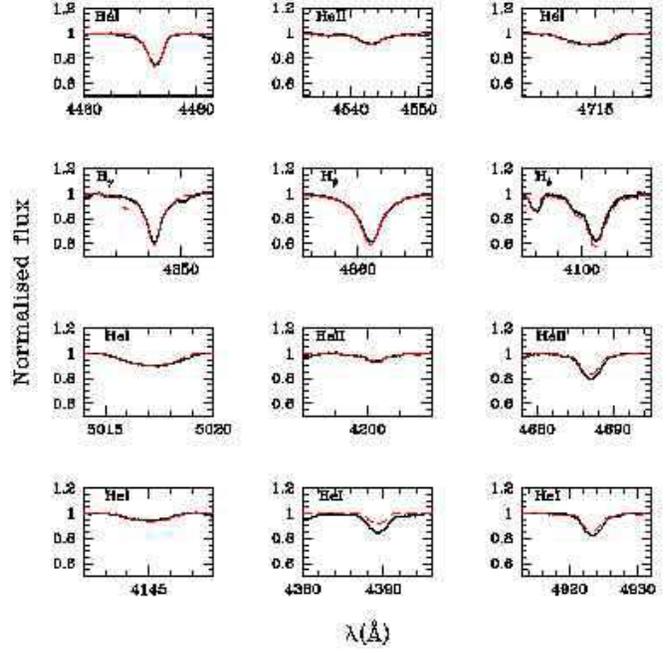,width=9cm}}
\caption{Best fit (red dashed line) of the optical spectrum (black
  solid line) of HD 38666. The effective temperature is 33000 K,
  $\log g$ = 4.0 and \vsini\ = 110 \kms. 
}
\label{hd38666_opt}
\end{figure}

\begin{figure}
\centerline{\psfig{file=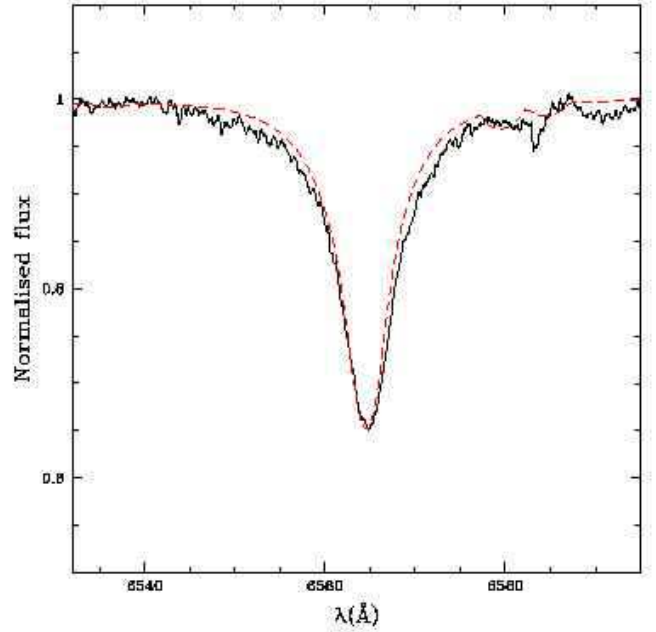,width=9cm}}
\caption{Best fit (red dashed line) of the observed \ha\ line 
  (black solid line) of HD 38666. We have derived \mdot\ = $10^{-9.5}$
  \myr\ and \vinf\ was 1200 \kms. 
}
\label{hd38666_ha}
\end{figure}

\begin{figure}
\centerline{\psfig{file=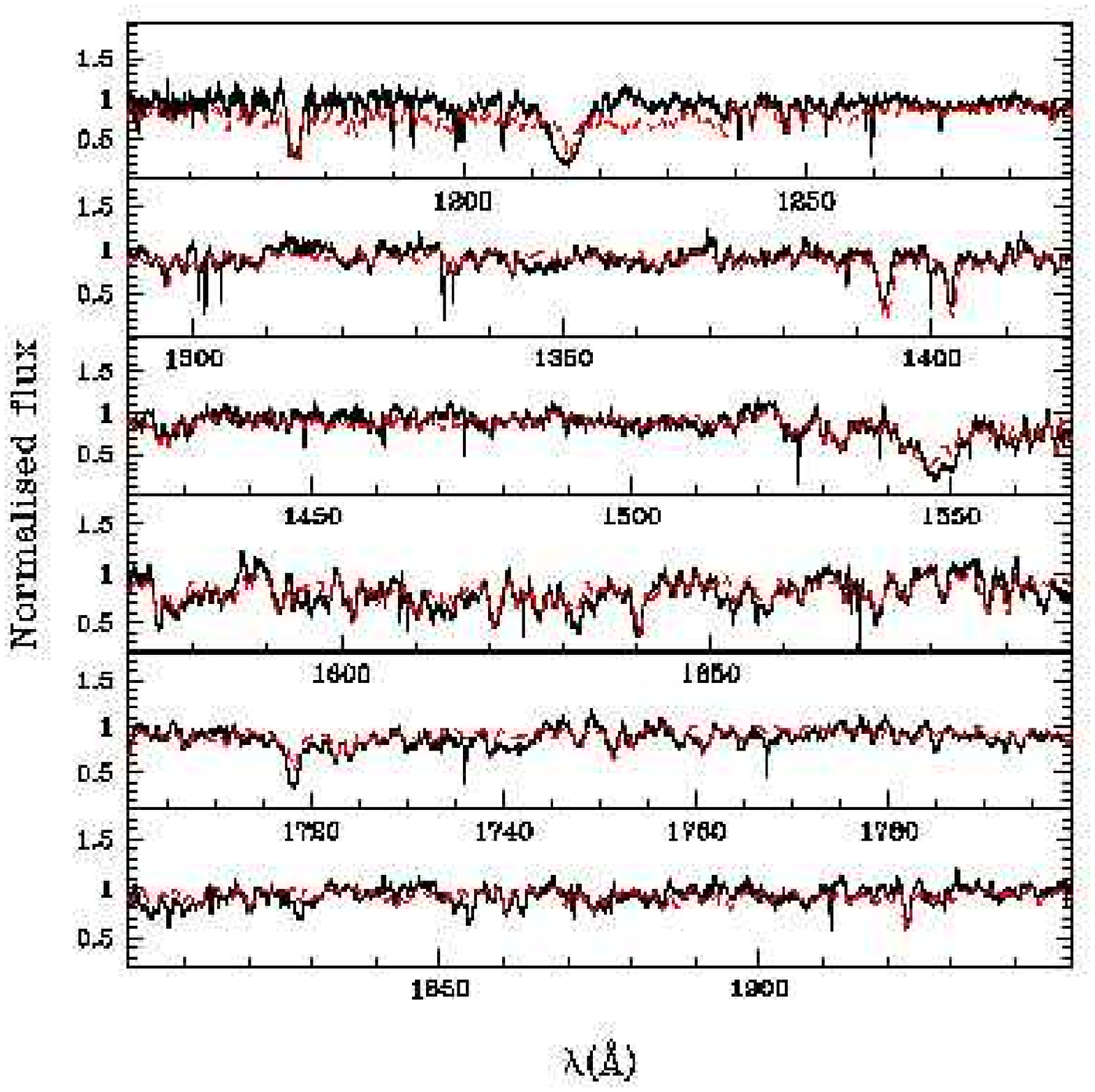,width=9cm}}
\caption{Best fit (red dashed line) of the UV spectrum (black solid
  line) of HD 38666. For this model, \mdot\ = $10^{-9.5}$
  \myr, \vinf\ = 1200 \kms\ and $\log L_{\rm X}/L_{\rm bol} = -6.87$  
}
\label{hd38666_UV}
\end{figure}

HD 38666 (also $\mu$ Col) is an O9.5V runaway star for which we derive 
an effective temperature of 33000 K from the fit of the
optical He lines, as shown in Fig.\ \ref{hd38666_opt}.
A value of $\log g = 4.0$
is derived from the Balmer lines.

The \ha\ and UV fits are given in Fig.\
\ref{hd38666_ha} and Fig.\ \ref{hd38666_UV}. The
best fits are obtained for \mdot\ = $10^{-9.5}$ \myr\ and \vinf\ = 1200 
\kms. Importantly, X-rays have been included in the modelling 
with $\log \frac{L_{\rm X}}{L_{\rm bol}} = -6.87$ as indicated 
by the observed X-ray emission (see Table \ref{tab_X}). If this 
high energy component is not included, we need a mass loss rate 10 times 
lower to fit the \civ\ line. The reason for this is that the ionisation 
structure of the wind is increased when X-rays are present, leading to a 
lower C~{\rm IV} ionisation fraction, and thus requiring a higher mass loss 
rate to reproduce the observed line profile (see Sect.\ \ref{X_rays} for a 
more complete discussion). 
Note that the fit of the \civ\ profile
is not perfect. This is due to the presence of interstellar absorption
which adds to the photospheric component. However, the fit of the
blueshifted wind part of the line is good and is not affected by
interstellar absorption (see also paper I). 
Previous estimates of \mdot\ range between 
$10^{-8.31}$ and $10^{-7.22}$ \myr\ (see Table \ref{tab_comp}).
Our estimate 
is lower than all these determinations. The determination of Leitherer 
(\cite{leith88}) relies only on the \ha\ wind emission, which in the case of low 
mass rates is very small and difficult to disentangle from the
photospheric absorption. The studies of
Chlebowski \& Garmany (\cite{chleb}) and Howarth \& Prinja
(\cite{hp89}) are based on the fit of UV resonance lines with the following 
method: the optical depth as a function of the velocity (only for 
unsaturated profiles) is determined by profile fitting; from this, the 
determination of the mass loss rate requires the adoption of an ionisation 
structure which may or may not be representative of the real ionisation 
in the atmosphere. This assumption may affect the \mdot\ determination. 

As for \vinf, a higher terminal velocity leads to a 
too-much-extended blueward absorption in \civ. The value of \vinf\
we derive is just above the escape velocity. Leitherer
(\cite{leith88}) estimated \vinf\ = 2000 \kms\, while Howarth \&
Prinja (\cite{hp89}) found 1000 \kms\ (see Table \ref{tab_comp}), illustrating the
uncertainty in the exact value of the terminal velocity of
HD 38666.

\subsection{HD34078}
\label{s_hd34078}

\begin{figure}
\centerline{\psfig{file=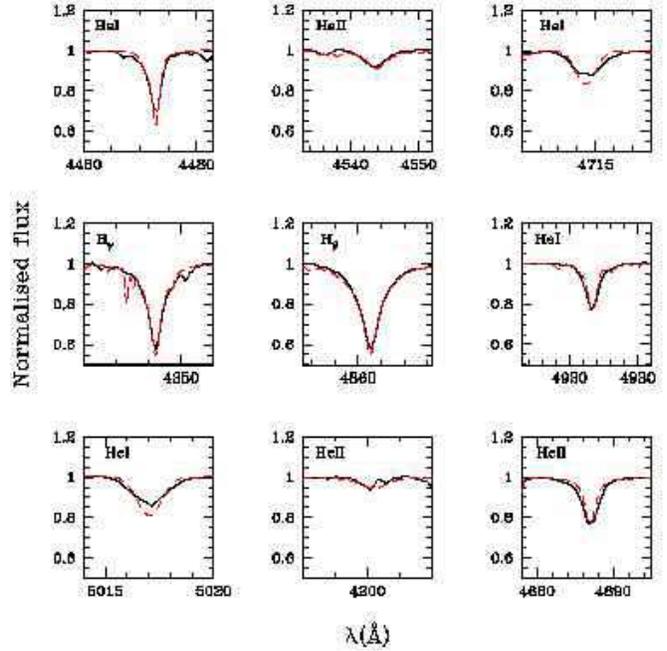,width=9cm}}
\caption{Best fit (red dashed line) of the optical spectrum (black
  solid line) of HD 34078. Here, \teff\ = 33000 K, $\log g$
  = 4.05 and \vsini\ = 40 \kms. 
}
\label{hd34078_opt}
\end{figure}

\begin{figure}
\centerline{\psfig{file=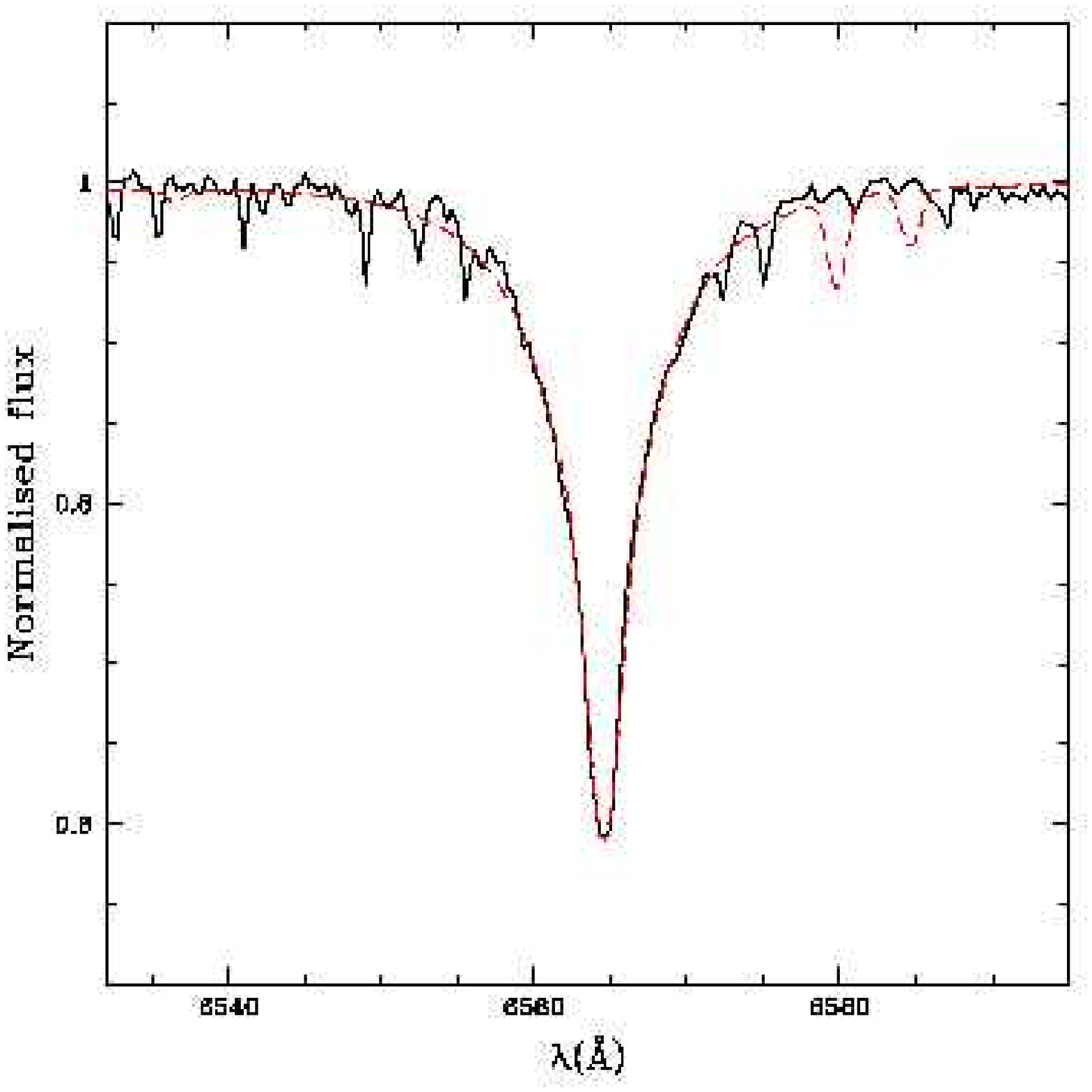,width=9cm}}
\caption{Best fit (red dashed line) of the observed \ha\ line 
  (black solid line) of HD 34078. Here, \mdot\ = $10^{-9.5}$ 
  \myr\ and \vinf\ = 800 \kms. 
}
\label{hd34078_ha}
\end{figure}

\begin{figure}
\centerline{\psfig{file=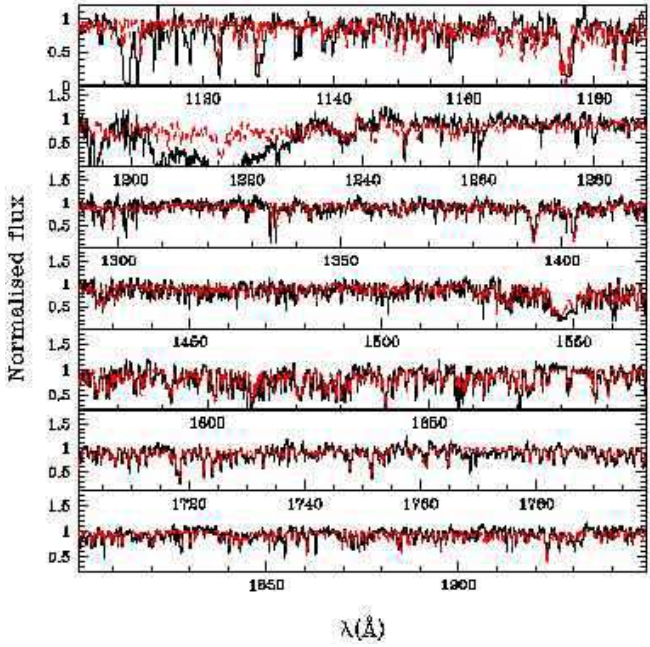,width=9cm}}
\caption{Best fit (red dashed line) of the UV spectrum (black solid
  line) of HD 34078. The wind parameters are: \mdot\ = $10^{-9.5}$ 
  \myr, \vinf\ = 800 \kms\ and $\log L_{\rm X}/L_{\rm bol} = -7.0$ 
}
\label{hd34078_UV}
\end{figure}

\begin{figure}
\centerline{\psfig{file=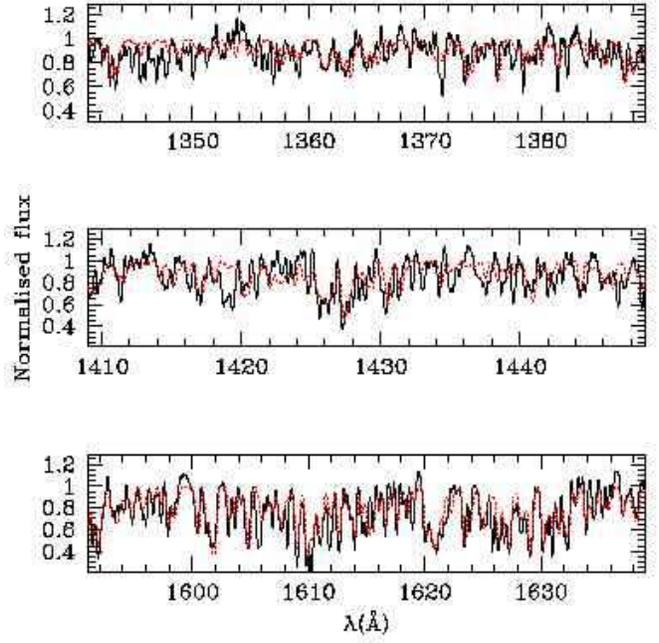,width=9cm}}
\caption{Zoom on the Iron line forests from Fig.\ \ref{hd34078_UV}
  showing the good agreement between the predicted spectrum (dotted
  line) and the observed spectrum (solid line) and confirming the
  \teff\ estimate. 
}
\label{hd34078_Fe}
\end{figure}

HD 34078 (also AE Aur) is a runaway O9.5V star possibly formed as a
binary (with $\mu$ Col, see Hoogerwerf et al.\ \cite{hbz01}) and
ejected after a binary - binary interaction with $\iota$ Ori (see
Sect.\ \ref{s_hd38666}).  Fig.\ \ref{hd34078_opt} shows the best fit
of the optical spectrum. From this best fit model, we derive an
effective temperature of 33000 K. This is confirmed by the good fit of
the iron lines shown in Fig.\ \ref{hd34078_Fe}. Note that the presence
of C~{\sc ii} $\lambda\lambda$6578,6582 in the model (see Fig.\
\ref{hd34078_ha}) may indicate a slightly too low effective
temperature. Test models reveals that increasing \teff\ to 34000 K
weakens this doublet. However, since these lines also depends on the C
abundance, we prefer to rely on the He lines and UV iron forests
estimate.  Also, improving the model atom for C~{\sc ii} produces a
weaker line since over recombination routes are available, reducing
the populations of the C~{\sc ii} $\lambda\lambda$6578,6582 transition
levels.  This shows that the value for \teff\ reported in Table
\ref{tab_prop} should be considered with its uncertainty of $\pm$ 2000
K.  Our modelling indicates that \vsini\ = 40 \kms\ seems to better
reproduce the observation, especially the optical spectra. The gravity
determined by Villamariz et al.\ (\cite{villamariz02}) gives a good
fit of the Balmer lines, so that we adopt $\log g = 4.05$ The other
stellar parameters are gathered in Table \ref{tab_prop}.

Figs.\ \ref{hd34078_ha} and \ref{hd34078_UV} show the fit of the
\ha\ line and UV spectrum of HD 34078. 
As we have shown that X-rays seem to be important for weak winds (see also 
next stars) and as 
HD 34078 shows no sign of a strong wind, we have adopted
$\log L_{\rm X}/L_{\rm bol} = -7.0$. Indeed, no X-ray measurement 
exists for HD 34078 and we have thus adopted the classical value for O 
stars (e.g. Chlebwoski \& Garmany \cite{chleb}).
A reasonable agreement
between the two types of mass loss indicators (\ha\ and UV lines) is found 
for \mdot =
$10^{-9.5}$ \myr and a terminal velocity of 800 \kms.   
Again, our value of \mdot\ is lower than previous determinations (see 
Table \ref{tab_comp}).

Surprisingly, the derived terminal velocity is similar to or even
lower than the escape velocity (1043 \kms). However, given the large
error on M and R, the escape velocity is also very uncertain. Also, a
value of \vinf\ lower than \vesc\ is possible since the escape
velocity quoted here is the photospheric escape velocity, and a
velocity in the wind of the order \vinf\ is obtained only in the outer
atmosphere where the local escape velocity is much lower. Moreover,
the weakness of the wind features may actually lead to underestimate
of the terminal velocity (see also Sect.\ \ref{disc_vinf}).

For HD 34078, we have used the He and CNO abundances of
Villamariz et al.\ (\cite{villamariz02}). They are nearly solar, except
for C which is found to have an abundance of $\sim$ 1/2 solar.

\subsection{HD46202}
\label{s_hd46202}

\begin{figure}
\centerline{\psfig{file=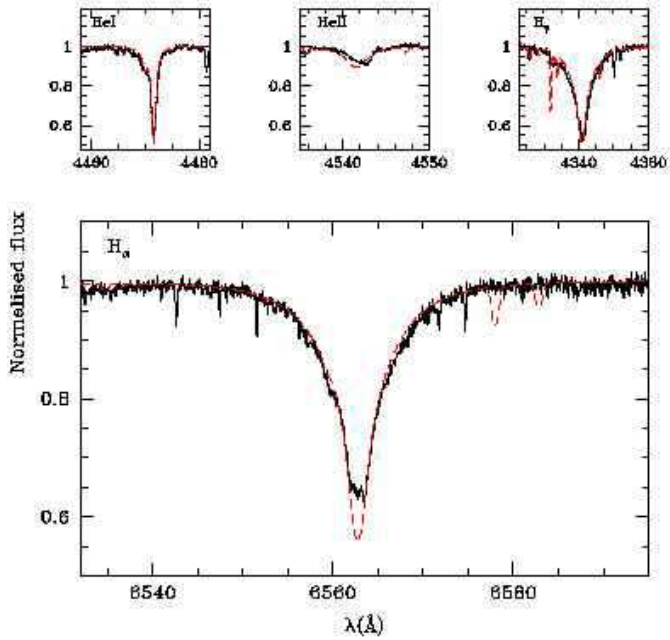,width=9cm}}
\caption{Best fit (red dashed line) of the optical spectrum (black
  solid line) of HD 46202. Here, \teff\ = 33000 K, $\log g$
  = 4.0, \vsini\ = 30 \kms, \mdot\ = $10^{-8.9}$ \myr\ and \vinf\ = 1200 \kms.
  The observed core of \ha\ is likely contaminated by small 
  interstellar emission.   
}
\label{hd46202_haopt}
\end{figure}

\begin{figure}
\centerline{\psfig{file=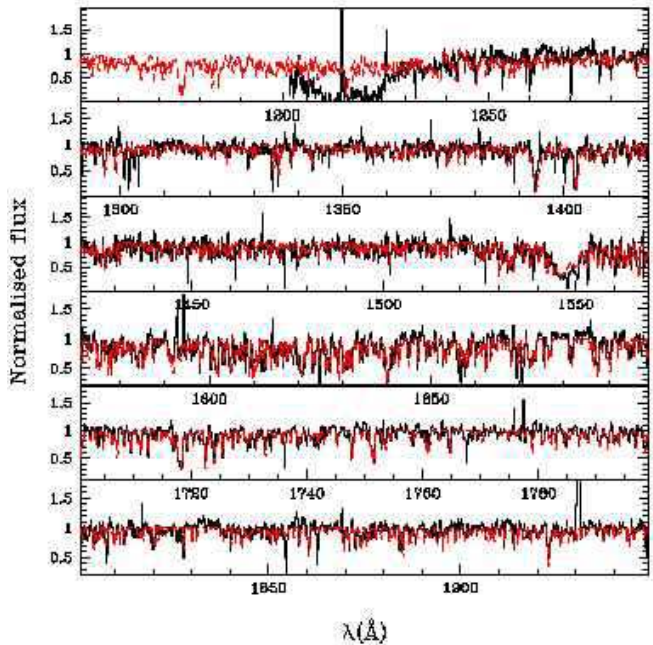,width=9cm}}
\caption{Best fit (red dashed line) of the UV spectrum (black solid
  line) of HD 46202. The wind parameters are: \mdot\ = $10^{-8.9}$ 
  \myr\ and \vinf\ = 1200 \kms. X-rays are included so that 
  $\log L_{\rm X}/L_{\rm bol} = -6.05$. The IUE spectrum below 1200 \AA\ is not 
  shown since the low S/N ratio does not allow any reliable comparison. 
}
\label{hd46202_UV}
\end{figure}

HD 46202 is an O9 V star situated in the Rosette nebula. 
An effective
temperature of 33000 K gives the best fit of the optical He lines
as shown in Fig.\ \ref{hd46202_haopt}. As for HD34078, the too 
deep C~{\sc ii} $\lambda\lambda$6578,6582 doublet may indicate a 
slightly too low \teff: once again, increasing \teff\ by 1000 K improves 
the fit but gives a worse fit of the He lines, so that we think 
the value of 33000 K is reasonable within its uncertainty of $\pm$ 2000 K.
A gravity of $\log g = 4.0$ gives a good fit of the H$\gamma$
line (see Fig.\ \ref{hd46202_haopt}).

Figures \ref{hd46202_haopt} and \ref{hd46202_UV} show the fit of the
wind sensitive lines from which we derive a mass loss rate of
10$^{-8.9}$ \myr\ and a terminal velocity of 1200 \kms.  According to
the X-ray detection, we have chosen $\log \frac{L_{\rm X}}{L_{\rm
bol}} = -6.10$ (see Table \ref{tab_X}).  If X-rays are not included, a
value of \mdot\ as low as 10$^{-10}$ \myr\ is required to fit the wind
part of \civ. Note that the core of the \ha\ line is stronger in the
model, but as the observed profile seems to be somewhat contaminated
(possibly by a small nebular contribution), we did not try to fit this
core.  As \civ\ is the main \mdot\ indicator and as $\log \frac{L_{\rm
X}}{L_{\rm bol}}$ is quite high for this star, we have run test models
including the high ionisation states C~{\sc V} and C~{\sc VI} to check
if the C ionisation was modified. They show that the C ionisation is
indeed slightly increased, which implies to increase \mdot\ by a
factor of $\sim$ 2 in order to fit \civ. Hence, given the uncertainty
in $\log \frac{L_{\rm X}}{L_{\rm bol}}$ (due to both uncertainties in
$L_{\rm X}$ and $L_{\rm bol}$), we think this effect is negligible
compared to other sources of errors for the \mdot\ determination (see
Sect.\ \ref{s_uncertainty}).  We have also run test models for which
the X-ray temperature was increased from 3 $10^{6}$ K to 7 $10^{6}$
K. Fitting \civ\ with this new X-ray temperature required a slight
increase ($\sim 0.3$ dex) of the mass loss rate.  All previous studies
give higher values of \mdot\ (see Table \ref{tab_comp}).  As for
\vinf, the range of values derived by other authors is quite large and
encompasses our estimate. This shows the difficulty of deriving \vinf\
from weak wind line profiles. The low terminal velocity will be
discussed in Sect.\ \ref{disc_vinf}.

\subsection{HD93028}
\label{s_hd93028}

\begin{figure}
\centerline{\psfig{file=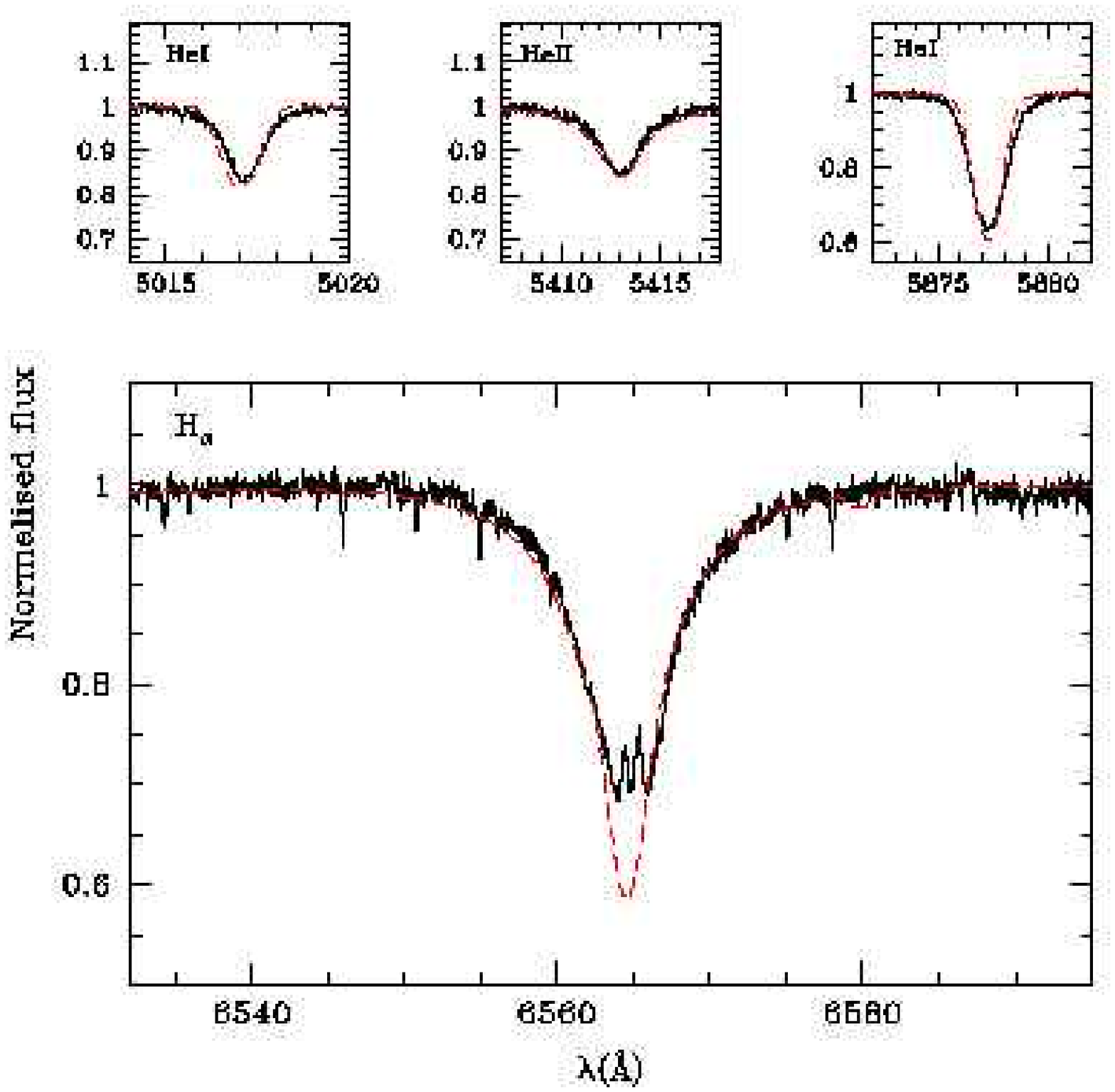,width=9cm}}
\caption{Best fit (red dashed line) of the observed He 
  and \ha\ lines 
  (black solid line) of HD 93028. The effective temperature is 34000 K, 
  $\log g$ = 4.0 and \vsini\ = 50 \kms. 
}
\label{hd93028_haopt}
\end{figure}

\begin{figure}
\centerline{\psfig{file=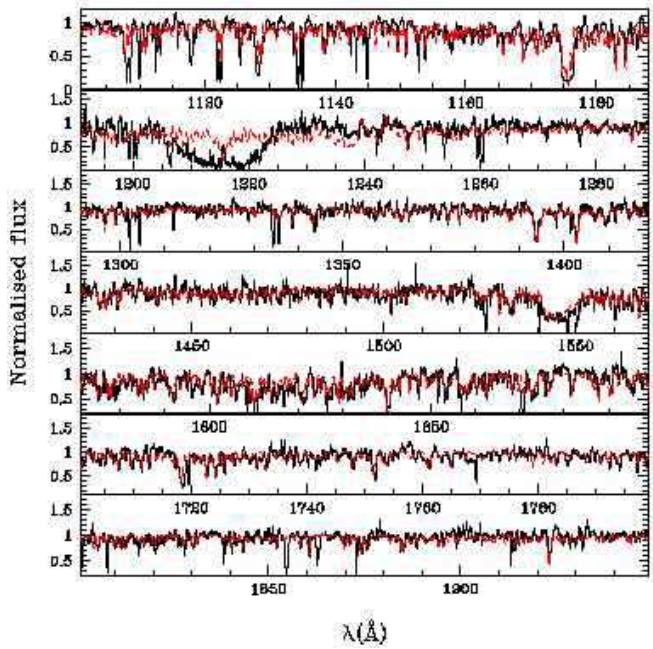,width=9cm}}
\caption{Best fit (red dashed line) of the UV spectrum (black solid
  line) of HD 93028. For this model, \mdot\ = $10^{-9.0}$ \myr,
  \vinf\ = 1300 \kms and we adopted $\log \frac{L_{X}}{L_{bol}} = -7.0$. 
}
\label{hd93028_UV}
\end{figure}

HD 93028 has a spectral type O9V and belongs to the
young cluster Collinder 228 in the Carina nebula. 
A value of \vsini\ of 50 \kms\ was deduced from optical lines fits 
(Fig.\ \ref{hd93028_haopt}) and 
from previous studies (Table \ref{tab_prop}).
The effective
temperature we derive from He optical lines is 34000 K.

From the \civ\ line, we derive a terminal velocity of 1300 km
s$^{-1}$, slightly lower than other estimates (see Table \ref{tab_comp}). 
We find that a mass loss rate of 10$^{-9.5}$ \myr\ gives a
good fit of the far UV, UV and \ha\ spectrum without X rays. However, 
as we have shown previously, X-rays influences strongly the determination 
of \mdot\ in stars with weak winds (see Sect.\ \ref{s_hd38666}, \ref{s_hd46202}). 
Hence, although there is no measurement of X-rays for HD 93028, we adopted the 
classical value $\log \frac{L_{X}}{L_{bol}} = -7.0$ (Chlebowski \& Garmany 
\cite{chleb}) and then derived \mdot\ = 
10$^{-9.0}$ \myr\ as shown in Figs.\ 
\ref{hd93028_haopt} and \ref{hd93028_UV}. 
The core of \ha\ is a little too strong in our best fit
model, but the observed line shows evidences of interstellar
contamination, which is natural in a star forming region (see also the 
\ha\ profile of HD 93146). The only previous determination of
mass loss rate for HD 93028
was made by Howarth \& Prinja (\cite{hp89}) who found \mdot\ $=
10^{-7}$ \myr, more than two orders of magnitude higher than our
value.

\subsection{HD152590}
\label{s_hd152590}

\begin{figure}
\centerline{\psfig{file=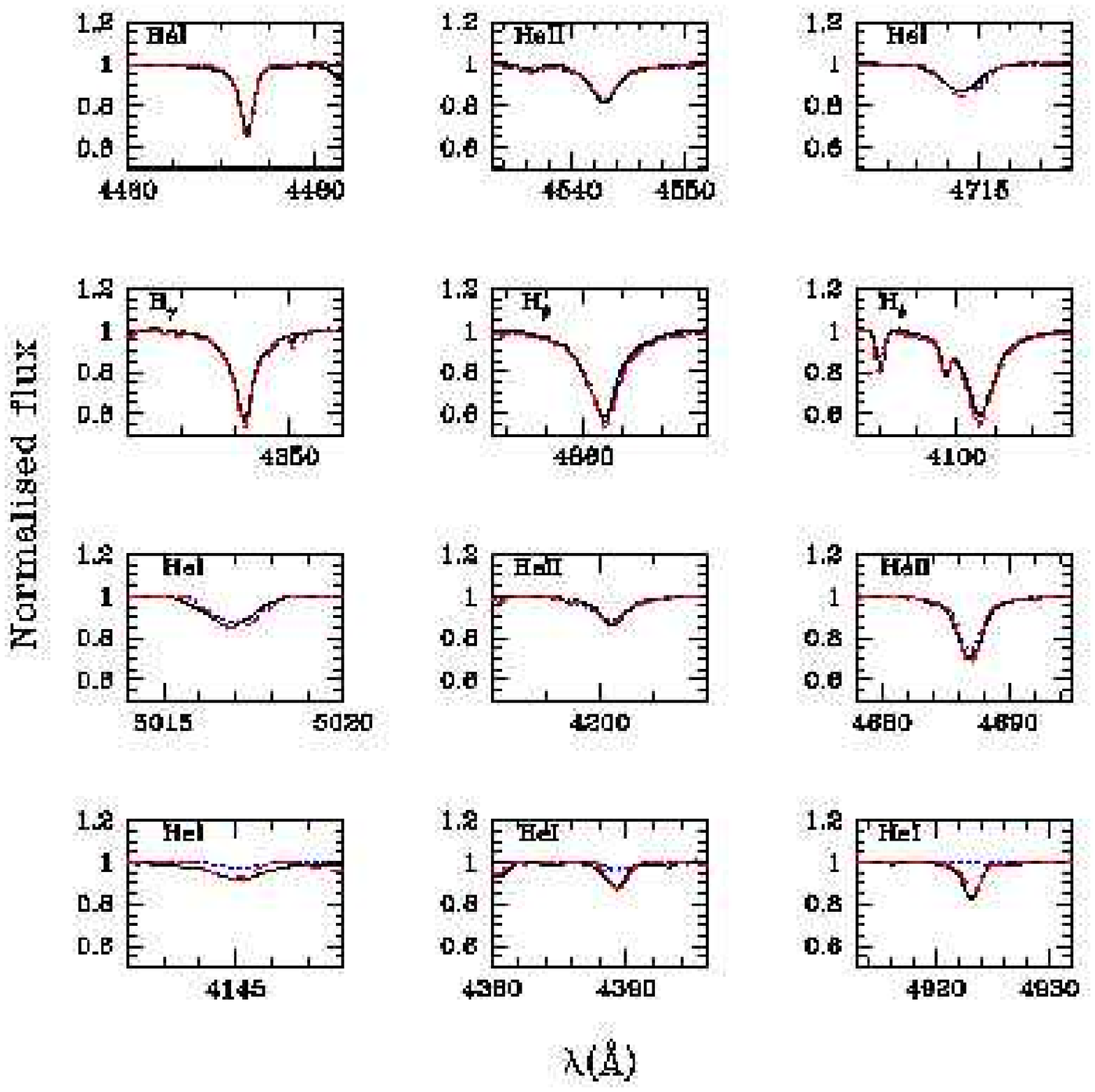,width=9cm}}
\caption{Best fit  of the optical spectrum (black
  solid line) of HD 152590. The effective temperature is 36000 K,
  $\log g$ = 4.1 and \vsini\ = 66 \kms. The blue dotted line is for 
a standard model (\vturb\ = 20 \kms\ in the computation of the atmospheric structure and \vturb\ from 5 to 175 \kms\ for the spectrum) while 
the red dashed line is for a model with \vturb\ = 10 \kms\ (atmospheric structure) \vturb\ = 10-175 \kms\ for the spectrum and 
additional metals (Ne, Ar, Ca and Ni). We see that this improved 
model leads to better fits of the He~{\sc I} singlet lines, leaving 
all other lines basically unchanged. 
}
\label{hd152590_opt}
\end{figure}

\begin{figure}
\centerline{\psfig{file=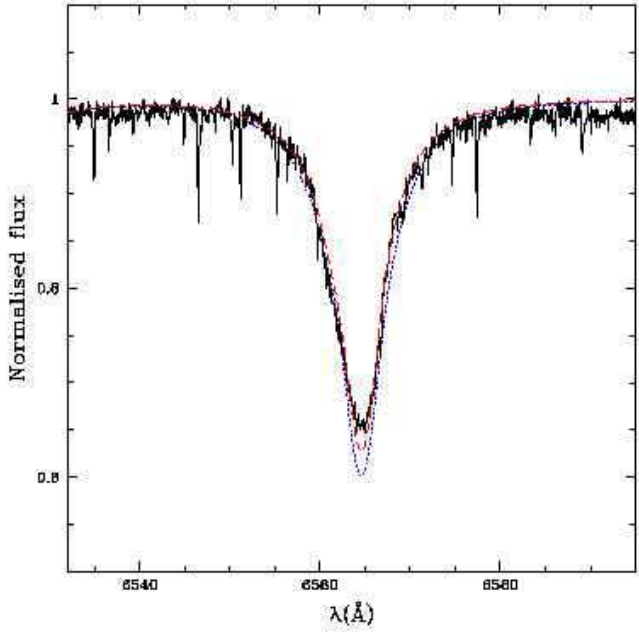,width=9cm}}
\caption{Fits of the observed \ha\ line 
  (black solid line) of HD 152590 for a model with \mdot\ = $10^{-7.78}$
  \myr\ (red dashed line) and for a model with \mdot\ = $10^{-8.75}$
  \myr\ (blue dotted line). 
  The terminal velocity in both models is 1750 \kms. Note the insensitivity
  of \ha\ line profile to the mass loss rate.
}
\label{hd152590_Ha}
\end{figure}

\begin{figure}
\centerline{\psfig{file=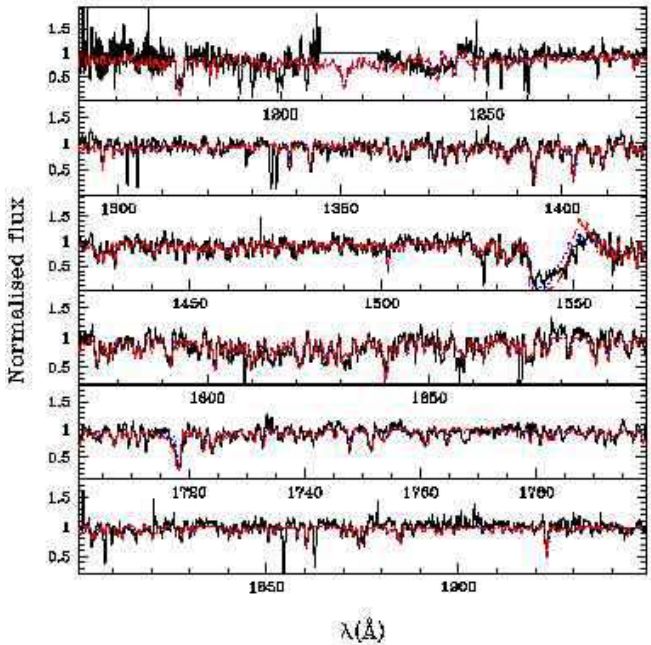,width=9cm}}
\caption{Same as Fig.\ \ref{hd152590_Ha} for the UV range.   
}
\label{hd152590_UV}
\end{figure}

HD 152590 is an O7.5Vz star. The 
distance estimate is difficult since it membership to 
Trumpler 24, Sco OB1 or NGC 6231 is not completely established. Given 
the uncertainty in the distance, we simply adopt the mean value (see 
Table \ref{tab_prop_adopt}).

We adopted \vsini\ = 66 \kms\ from Penny (\cite{penny}).  The optical
spectrum shown in Fig.\ \ref{hd152590_opt} is correctly reproduced
with an effective temperature of 36000 K.  Note that initially, we had
a problem to reproduce the He~{\sc I} singlet lines which were too
weak in our models wheras all other lines were very well
reproduced. This problem has been recently noted by Puls et al.\
(\cite{puls05}) when they put forward a discrepancy between CMFGEN and
FASTWIND for these lines between 36000 and 41000 K for
dwarfs. However, a more complete treatment of line blanketing appeared
to solve this problem. Indeed, if we reduce the microturbulent
velocity from 20 to 10 \kms\ in the computation of the atmospheric
structure AND if we add some more species (Neon, Argon, Calcium and
Nickel) we greatly improve the fit of the He~{\sc I} singlet lines
without modifying the strength of other H and He lines (Hillier et
al.\ \cite{hil03} already noted that the He~{\sc I} singlet lines were
much more sensitive to details of the modelling than the triplet
lines). This is shown in Fig.\ \ref{hd152590_opt}. Note that
increasing the microturbulent velocity from 5 to 10 \kms\ in the
computation of the spectrum changes only marginally the line
profiles. Hence, we attribute the origin of the discrepancy pinpointed
by Puls et al.\ (\cite{puls05}) to a subtle line-blanketing effect in
this particular temperature range, and concerning only the He~{\sc I}
singlet lines \footnote{The problem occurs only for the He~{\sc i}
singlet lines which have the 2p 1p$^{o}$ state as their lower
level. Thus the singlet problem is most likely related to the
treatment of blanketing in the neighbourhood of the He~{\sc i}
resonance transition at 584 \AA. Detailed testing by F. Najarro \&
J. Puls (private communication) supports these ideas.}. Note that
reducing \vturb\ without including additional metals strengthens the
singlet lines, but not enough to fit the observed spectrum. Hence the
additional line-blanketing effects of Ne, Ar, Ca and Ni, although
small (most lines are unchanged) is crucial to fit the He~{\sc I}
lines around \teff\ = 36000 K. Note that we usually restrict ourselves
to models with \vturb\ = 20 \kms\ and no Ne, Ca, Ar or Ni since the
computational time is much more reasonable. For HD 152590, we found
that a gravity $\log g =4.10 \pm 0.1$ gives the best fit of the Balmer
lines (in particular H$\gamma$)

The terminal velocity of HD 152590 estimated from \civ\ is 1750 \kms, in 
good agreement or lower than previous estimates (Table \ref{tab_comp}). 
The estimate of the 
mass loss rate is much more difficult for this star. In fact, we have 
not been able to fit simultaneously the UV lines and \ha. If the former 
are correctly reproduced (with \mdot\ = $10^{-8.75}$ \myr), then the 
later has a too strong absorption in its core, and if \ha\ is fitted 
(with \mdot\ = $10^{-7.78}$ \myr), \civ\ is too strong. This is shown 
in Fig.\ \ref{hd152590_Ha} and \ref{hd152590_UV}. We have tried without 
success to increase the $\beta$ parameter to improve the fit (an increase 
of $\beta$ leading to a weaker \ha\ absorption). The fits of Fig.\ 
\ref{hd152590_Ha} and \ref{hd152590_UV} are for $\beta$ = 1.2 and even for 
this quite high value for a dwarf star, the \ha\ core is not 
perfectly reproduced. A possible 
explanation is the presence of a companion for HD 152590 (Gieseking \cite{gk82}). 
In that 
case, \ha\ may be diluted by the continuum of this secondary whereas the 
UV spectrum may be unaffected provided the companion is a later type star 
than HD 152590 without strong UV lines. However, adopting a
conservative approach, we adopt the \ha\ 
mass loss rate (10$^{-7.78}$ \myr) as typical, keeping in mind that it 
may well be only an upper limit.

\subsection{HD93146}
\label{s_hd93146}

\begin{figure}
\centerline{\psfig{file=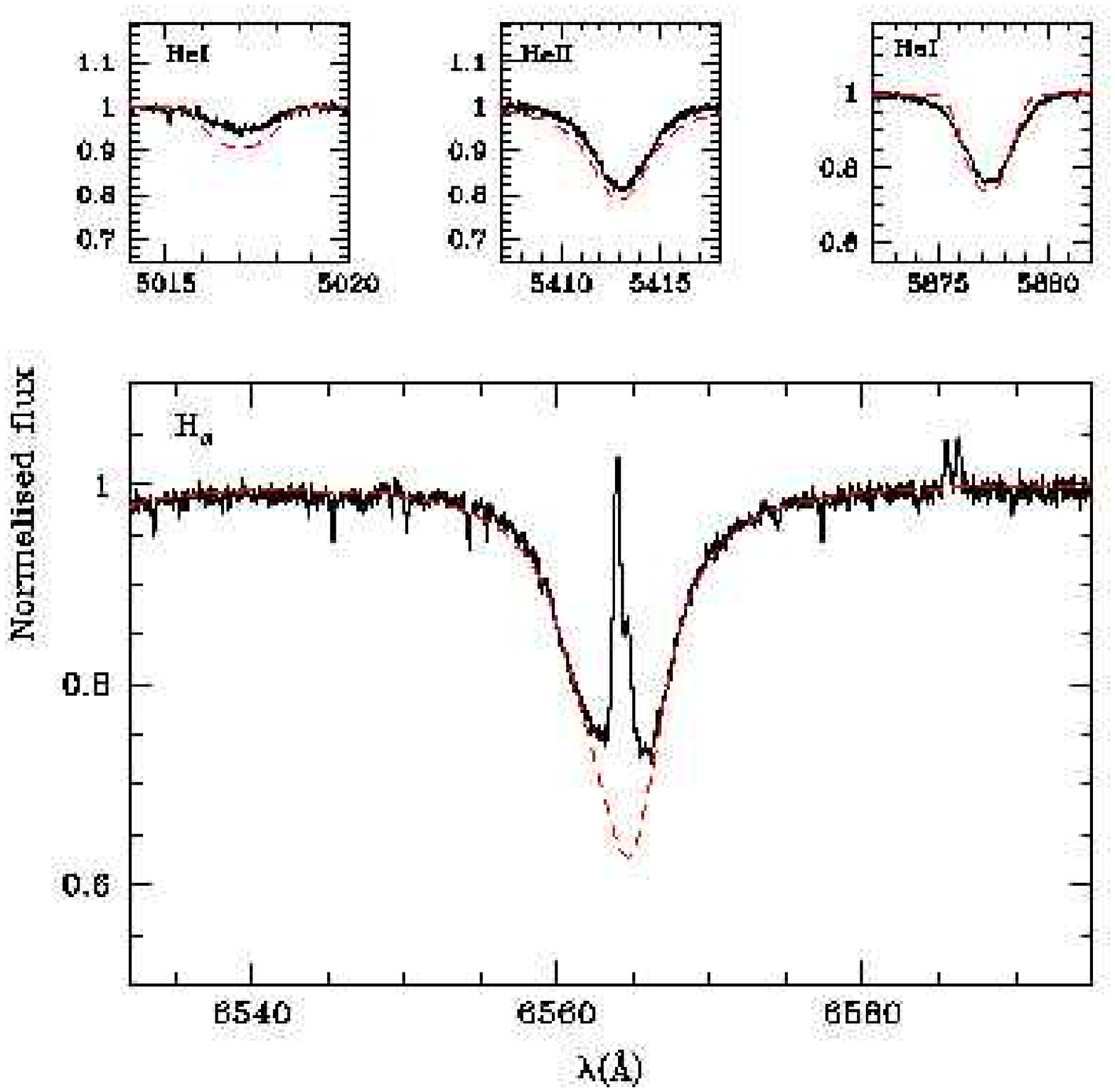,width=9cm}}
\caption{Best fit (red dashed line) of the observed He 
  and \ha\ lines 
  (black solid line) of HD 93146. The effective temperature is 37000 K, 
  $\log g$ = 4.0 and \vsini\ = 80 \kms. 
}
\label{hd93146_haopt}
\end{figure}

\begin{figure}
\centerline{\psfig{file=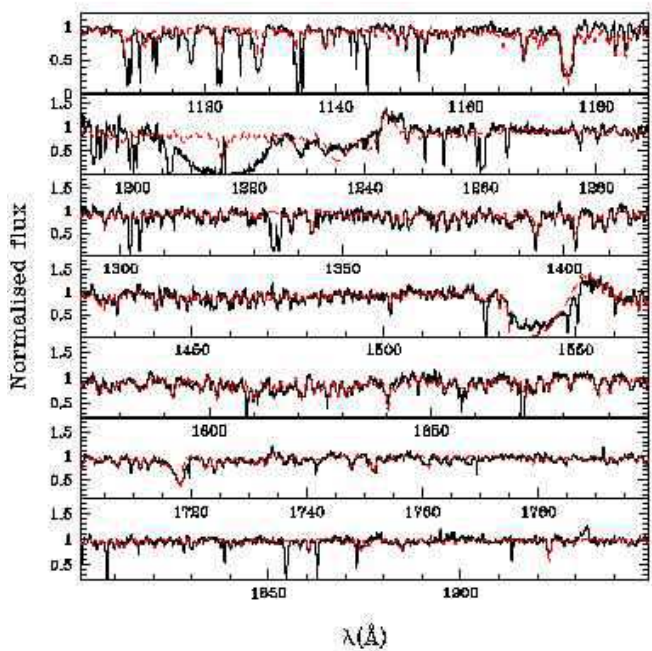,width=9cm}}
\caption{Best fit (red dashed line) of the UV spectrum (black solid
  line) of HD 93146. For this model, \mdot\ = $10^{-7.25}$ \myr and
  \vinf\ = 2800 \kms. 
}
\label{hd93146_UV}
\end{figure}

HD 93146 is an O6.5V((f)) star in the Carina nebula
and belongs to the cluster Cr 228. 

We adopt \vsini\ = 80 \kms\ from our fits and previous determinations 
(see Table \ref{tab_prop}). Fig.\ \ref{hd93146_haopt} shows our
best fit to the He optical spectrum between 5000 and 6000 \AA\ for which an
effective temperature of 37000 K is derived. 
Notice that this fit is not perfect, but it 
is actually the best we could get. Increasing \teff\ may help reduce 
the He~{\sc I} absorption, but it increases too much the He~{\sc II} 
strength. Moreover the UV photospheric lines are very well 
reproduced with this \teff\ (see Fig.\ \ref{hd93146_UV}). 
As we do not have reliable gravity estimators, we assume
$\log g = 4.0$ since this value is typical of dwarfs
(Vacca et al. \cite{vacca}, Martins et al.\ \cite{calib05}). 

Fig.\ \ref{hd93146_UV} shows our best fit of the (far) UV spectrum 
of HD 93146. The terminal velocity is 2800 \kms\ and the mass
loss rate is 10$^{-7.25}$ \myr. For higher values, \niv\ displays a 
too strong blueshifted absorption. The \ha\ profile of Fig.\
\ref{hd93146_haopt} confirms partly this value of \mdot\ since the line 
is correctly reproduced, under the uncertainty of the exact depth of
the core which is contaminated by nebular emission.
Previous estimates are in failry good agreement with the present one 
(Table \ref{tab_comp}).

\subsection{HD42088}
\label{s_hd42088}

\begin{figure}
\centerline{\psfig{file=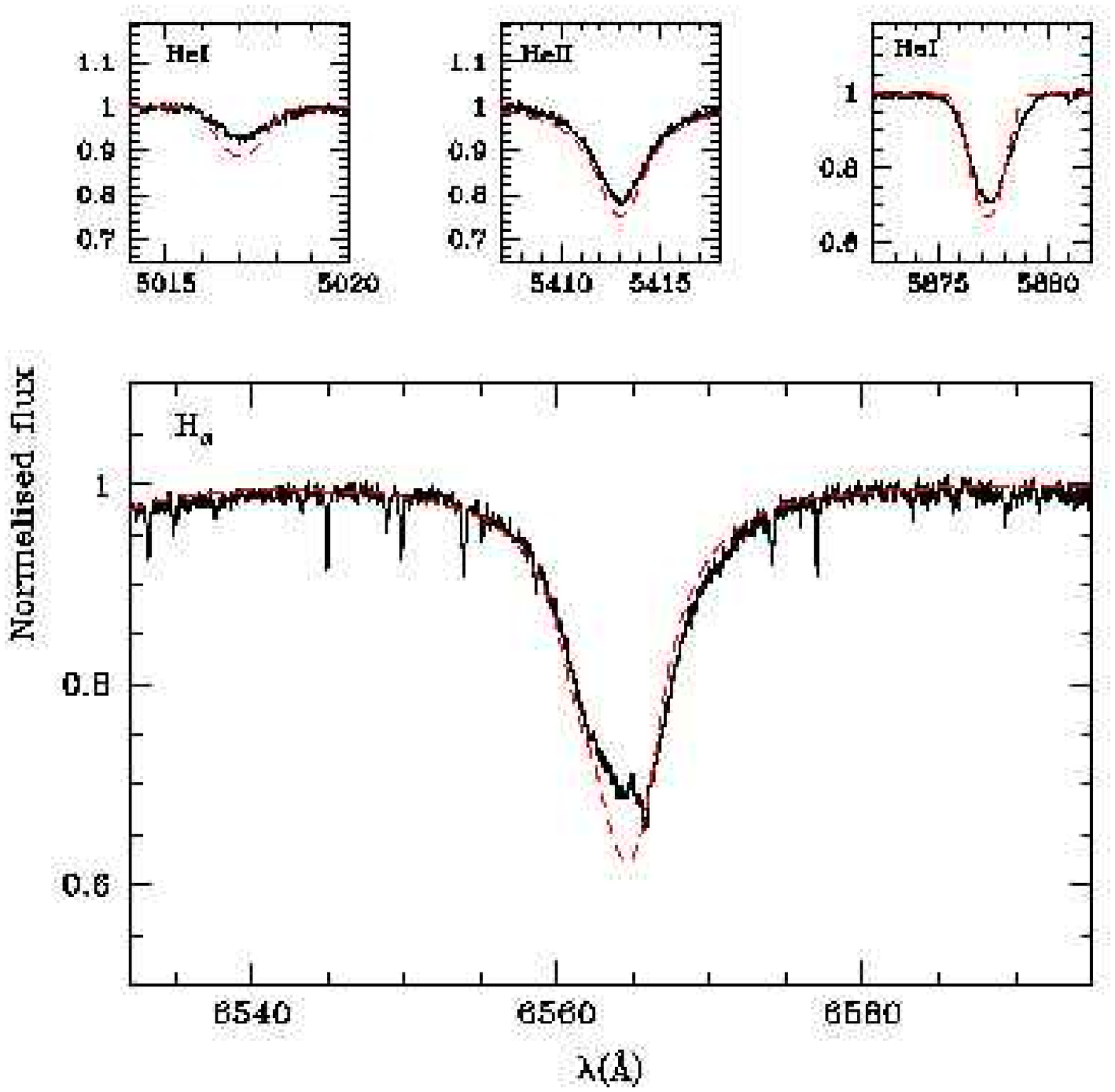,width=9cm}}
\caption{Best fit (red dashed line) of the observed He 
  and \ha\ lines 
  (black solid line) of HD 42088. The effective temperature is 38000 K, 
  $\log g$ = 4.0 and \vsini\ = 60 \kms. 
}
\label{hd42088_haopt}
\end{figure}

\begin{figure}
\centerline{\psfig{file=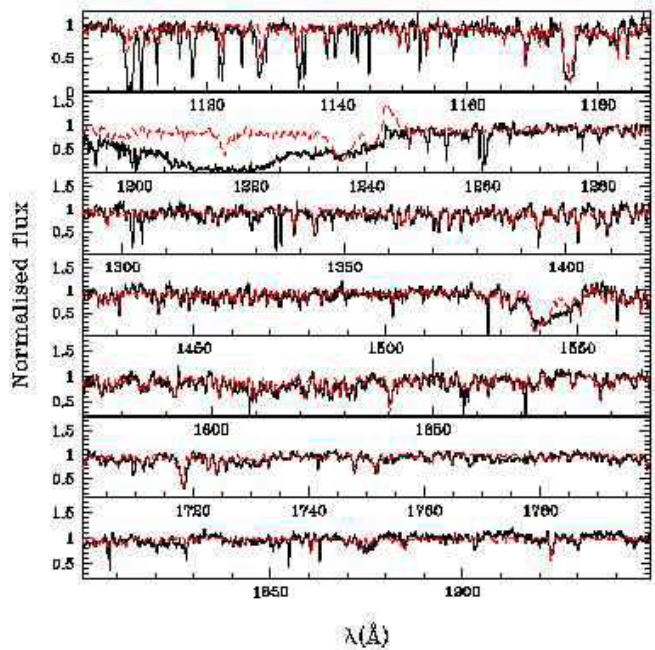,width=9cm}}
\caption{Best fit (red dashed line) of the UV spectrum (black solid
  line) of HD 42088. For this model, \mdot\ = $10^{-8}$ \myr and
  \vinf\ = 1900 \kms. 
}
\label{hd42088_UV}
\end{figure}

HD 42088 is a O6.5 V star associated with the H II region NGC 2175. It 
also belongs to the class of Vz stars. 
Note that the distance to this star is poorly constrained so that its 
luminosity is the least well known of all stars of our sample.
The rotational velocity is 
chosen to be 60 \kms\ in view of the determinations of Penny (\cite{penny}) 
- 62 \kms\ - and Howarth et al.\ (\cite{howarth97}) - 65 \kms. The fit 
of optical He lines above 5000 \AA\ leads to an estimate of the 
effective temperature which is found to be $\sim$ 38000 K as shown 
by Fig.\ \ref{hd42088_haopt}. This estimate also relies on the fit 
of UV lines since the number of optical indicators is small. We adopt 
$\log g = 4.0$ (from Vacca et al.\ \cite{vacca}) since we do not have 
strong gravity indicators.

The terminal velocity is derived from the blueward extension of the 
absorption in \civ\ and is 1900 \kms. Previous determinations go 
from 2030 \kms to 2550 \kms 
Given the fact that we adopted a microturbulent velocity 
of 190 \kms in the outer wind (10 \% of \vinf), the absorption actually 
extends up to 2100 \kms in the model, in good agreement with other 
determinations. Concerning the mass loss rate, it turns out that a 
value of 10$^{-8}$ \myr gives a reasonable fit of the main UV lines 
and \ha, although for the latter the very core is not correctly fitted 
but may suffer from nebular contamination (see Fig.\ 
\ref{hd42088_haopt}). The best fit model is shown in Fig.\ 
\ref{hd42088_UV}. 
Our mass loss rate determination
based on both \ha\ and UV lines gives 
a much lower value than ever found for this star (Table \ref{tab_comp}). 
But the UV lines 
produced by models with mass loss rates much higher than our adopted 
value are much too strong compared to the observed spectrum, forcing 
us to adopt such a low \mdot.

\subsection{HD93204}
\label{s_hd93204}

\begin{figure}
\centerline{\psfig{file=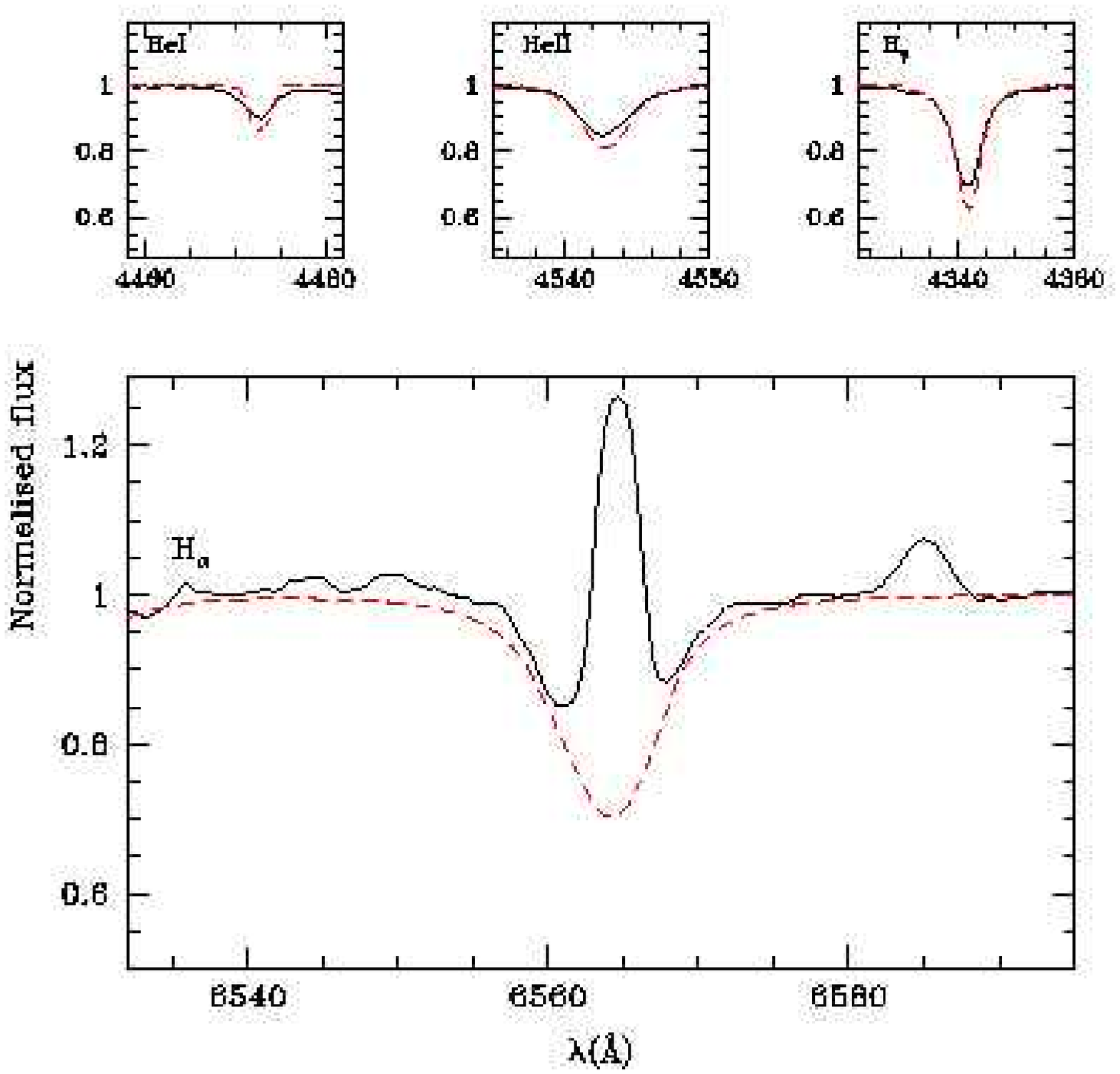,width=9cm}}
\caption{Best fit (red dashed line) of the observed He 
  and \ha\ lines 
  (black solid line) of HD 93204. The effective temperature is 40000 K, 
  $\log g$ = 4.0, \vsini\ = 130 \kms \ and \mdot\ = $10^{-6.75}$
  \myr.  
}
\label{hd93204_haopt}
\end{figure}

\begin{figure}
\centerline{\psfig{file=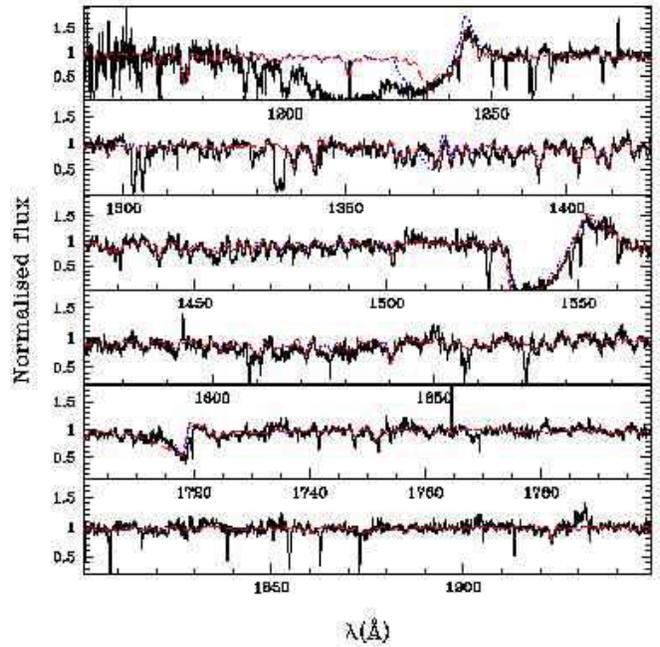,width=9cm}}
\caption{Best fit of the UV spectrum (black solid
  line) of HD 93204. For this model, \mdot\ = $10^{-6.75}$ \myr and
  \vinf\ = 2900 \kms. The red dashed line is a clumped model with
  $f_{\infty} = 0.1$ and the blue dotted line is a homogeneous model. 
}
\label{hd93204_UV}
\end{figure}

HD 93204 (O5V((f))) is a member of the young cluster Trumpler 16 in the Carina
complex.  
We adopt the value 130 km
s$^{-1}$ for \vsini\ in our fits, which helps to derive an effective
temperature of 40000 K (see Fig.\ \ref{hd93204_haopt})
A gravity of $\log g = 4.0$ is compatible with the observed Balmer
lines. 

Fig.\ \ref{hd93204_UV} shows the fit of the UV spectrum.
To fit reasonably all the UV lines, we had to use clumped 
models. This is especially true for \ov\ since as previously shown by
Bouret et al.\ (\cite{jc03}. \cite{jc05}) this line is predicted too strong in
homogeneous models. In our case, the use of clumping with the law
given in Sect.\ \ref{s_wind_param} and $f_{\infty} = 0.1$ improves the 
fit of \ov\ as well as \niv, as shown in Fig.\ \ref{hd93204_UV}.
Reducing \teff\
does not solve the problem since in that case \ov\ is weaker but \niv\ gets
stronger. We derive a mass loss rate of 10$^{-6.75}$ \myr\ and a
terminal velocity of 2900 \kms. Due to the high level of
nebular contamination of \ha, we can not use this line to
constrain \mdot\ (see Fig.\ \ref{hd93204_haopt}). 
Our value of \mdot\ is slightly smaller (factor 4) than that of 
Howarth \& Prinja (\cite{hp89}) (Table \ref{tab_comp}) mainly due to 
the inclusion of clumping in our models.
As for \vinf, our estimate is well within the range of values
previously derived (see Table \ref{tab_comp}).

\subsection{HD15629}
\label{s_hd15629}

\begin{figure}
\centerline{\psfig{file=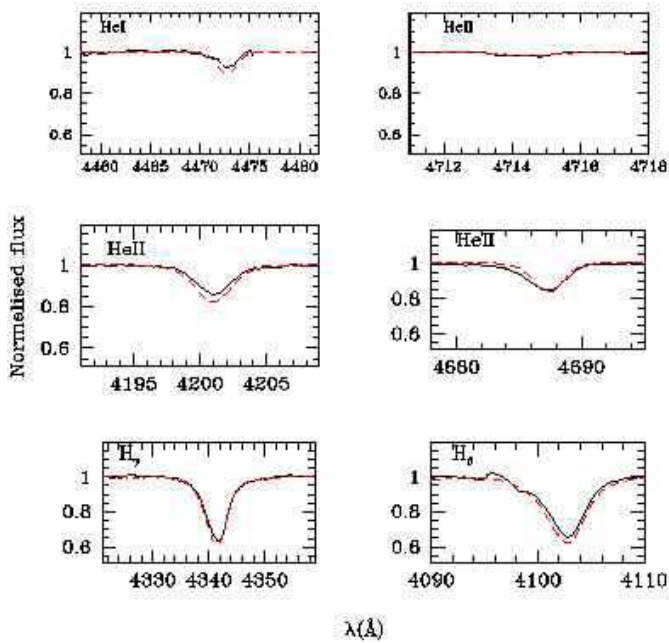,width=9cm}}
\caption{Best fit (red dashed line) of the observed He 
  and H lines 
  (black solid line) of HD 15629. The effective temperature is 41000 K, 
  $\log g$ = 3.75 and \vsini\ = 90 \kms.
}
\label{hd15629_opt}
\end{figure}

\begin{figure}
\centerline{\psfig{file=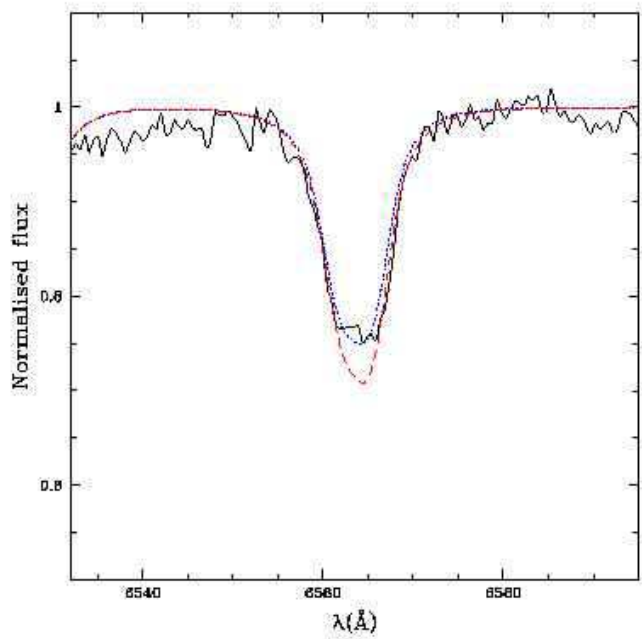,width=9cm}}
\caption{Best fit (red dashed line) of the observed \ha\ line
  (black solid line) of HD 15629. For this model, \mdot\ = $10^{-6.5}$ \myr, 
  \vinf\ = 2800 \kms\ and $f_{\infty} = 0.1$. We also show a model
  with \mdot\ = $10^{-5.89}$ \myr\ and no clumping (blue dotted line) as derived by
  Repolust et al.\ (\cite{repolust}). 
}
\label{hd15629_Ha}
\end{figure}

\begin{figure}
\centerline{\psfig{file=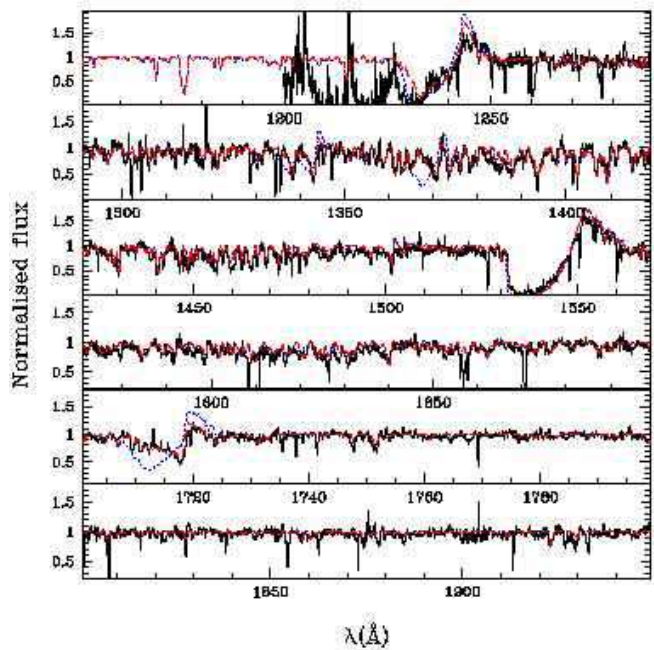,width=9cm}}
\caption{Best fit (red dashed line) of the UV spectrum (black solid
  line) of HD 15629. The wind parameters are the same as in Fig.\
  \ref{hd15629_Ha}. The IUE spectrum below
  1200 \AA\ is not shown since the low S/N ratio does not allow any
  reliable comparison. The blue dotted line is the model with the
  Repolust et al.\ (\cite{repolust}) parameters: note the too strong 
  \ov\ and \niv\ lines.   
}
\label{hd15629_UV}
\end{figure}

HD 15629 is classified as O5V((f)) and belongs to the star cluster IC 1805.
The projected rotational velocity is found 
to be 90 \kms\ by several authors and we adopted this value which give a 
good fit of optical and UV photospheric lines.
The optical spectrum presented in Fig.\ \ref{hd15629_opt} indicates an 
effective temperature of 41000 K.
This is in good agreement with the recent determination of Repolust 
et al.\ (\cite{repolust}) who found 40500 K. 
We adopted $\log g =3.75$ since it gives a reasonable fit of Balmer lines 
(Fig.\ \ref{hd15629_opt}) 
and it is close to the value derived by Repolust et al. (\cite{repolust}) 
who derived $\log g =3.70$.

The best fit model of the UV spectrum is shown in Fig.\
\ref{hd15629_UV}, and \ha\ is displayed in Fig.\ \ref{hd15629_Ha}. The
main parameters for this model are \mdot\ = $10^{-6.5}$ \myr, \vinf\ =
2800 \kms\ and $f_{\infty} = 0.1$. We also show on these figures a
model without clumping and with the mass loss rate of Repolust et al.\
(\cite{repolust}) which is higher - 10$^{-5.89}$ \myr- than our
derived value. Once again the inclusion of clumping is necessary to
correctly reproduce both \ov\ and \niv. With the Repolust et al.\
(\cite{repolust}) \mdot\ and no clumping, CNO abundances have to be
reduced by a factor of 3 to give reasonable fits, and even in that
case the \ov\ line is too strong. Such a reduction of the abundances
is unlikely for a Galactic star. For our best fit, we have adopted the
CNO solar abundances recently claimed by Asplund (\cite{asplund04})
since they are slightly lower than those of Grevesse \& Sauval
(\cite{gs98}) and allow a fit of the UV lines with a slightly higher
(0.25 dex) mass loss rate compared to the later values. Note that in
our final best fit, the core of \ha\ is not perfectly fitted. However,
we suspect that the strange squared shape of the observed line core is
probably contaminated by weak nebular emission. In support of the
nebular contamination we note the following: if we adopt the mass loss
rate of Repolust et al.\ (\cite{repolust}), the flux level in the line
core is correct, but the line is slightly narrower in the remainder of
the profile compared to the observed profile, while with our \mdot,
the line is well fitted except in the very core. Increasing the flux
level in the core in models with our \mdot\ requires the adoption of
$\beta$ = 1.7 which is high for a dwarf. In that case again, although
the flux level in the core is correct, the synthetic line profile is
too narrow. We are then rather confident that the observed line core
is somewhat contaminated and that our mass loss rate is correct. The
use of clumping explains partly the discrepancy with the result of
Repolust et al.\ (\cite{repolust}).  Concerning the terminal
velocities, previous extimates range from 2810 to 3220 \kms\ in
reasonable agreement with our value.

\subsection{HD46223}
\label{s_hd46223}

\begin{figure}
\centerline{\psfig{file=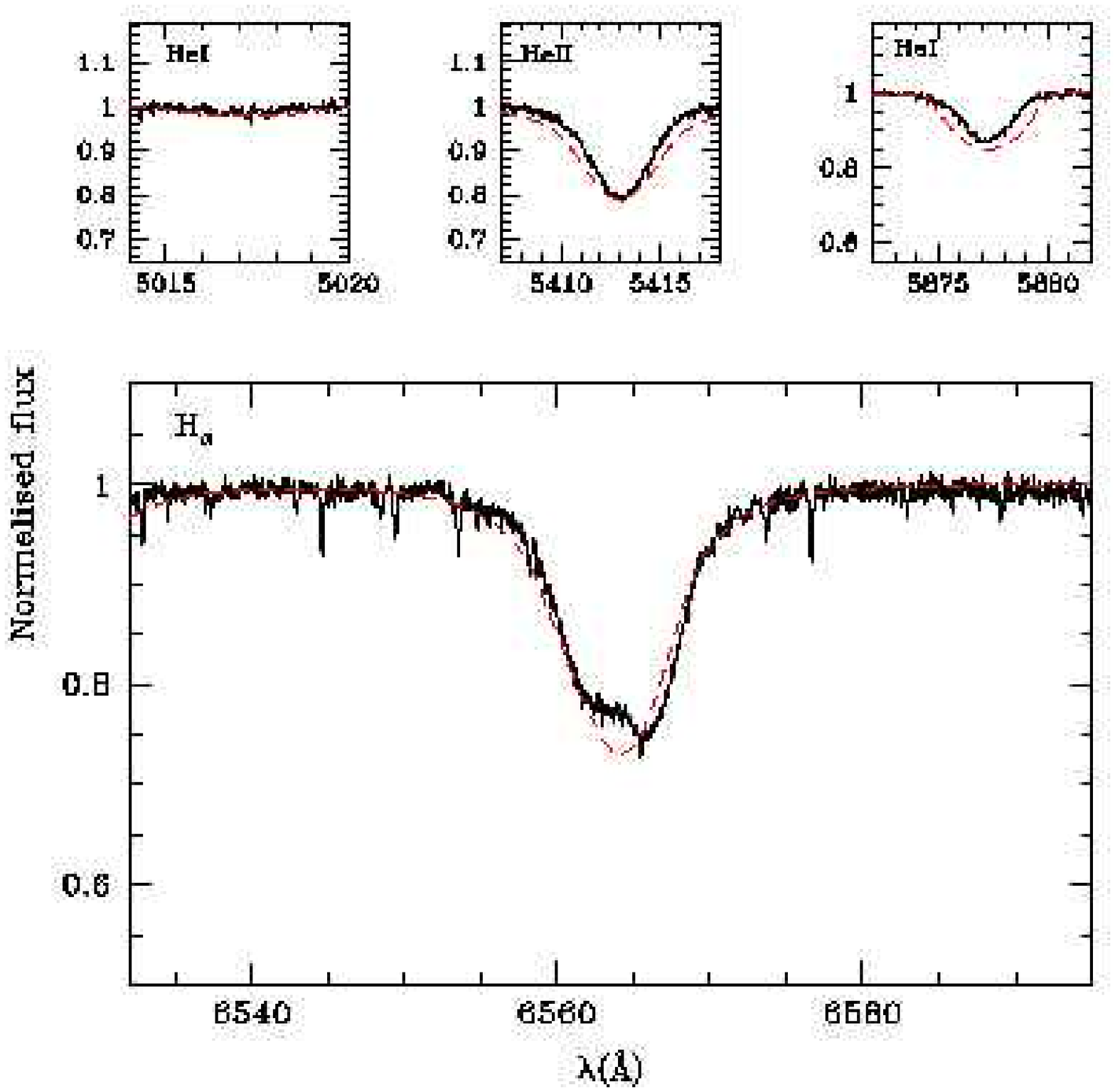,width=9cm}}
\caption{Best fit (red dashed line) of the observed He 
  and \ha\ lines 
  (black solid line) of HD 46223. The effective temperature is 41500 K, 
  $\log g$ = 4.0 and \vsini\ = 130 \kms. 
}
\label{hd46223_haopt}
\end{figure}

\begin{figure}
\centering{\psfig{file=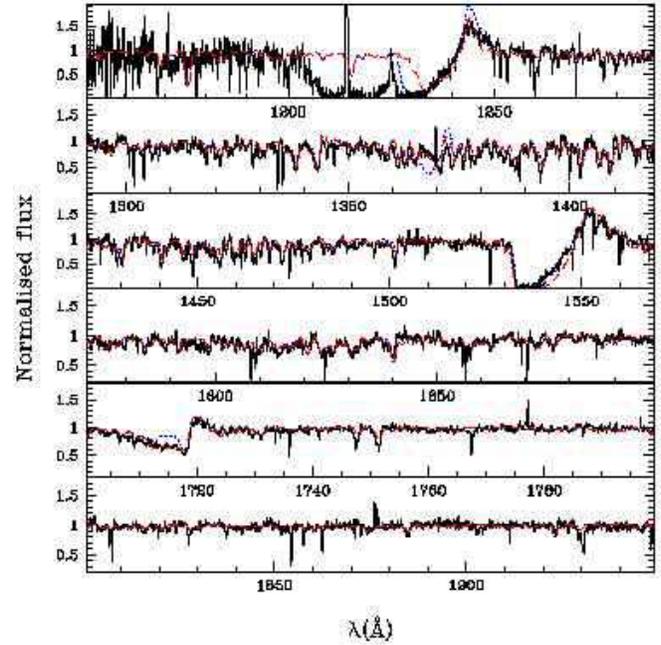,width=9cm}}
\caption{Best fit of the UV spectrum (black solid
  line) of HD 46223. For this model, \mdot\ = $10^{-6.5}$ \myr and
  \vinf\ = 2800 \kms. The red dashed line is a clumped model with
  $f_{\infty} = 0.1$ and the blue dotted line is a homogeneous model. 
}
\label{hd46223_UV}
\end{figure}

HD 46223 belongs to the Rosette cluster (NGC 2244) and has 
a spectral type O4V((f+)). 
A projected rotational velocity of 130 \kms\ was adopted from the fit 
of optical lines.
The upper panels of Fig.\
\ref{hd46223_haopt} show the fit of He optical lines with a model for
which \teff\ = 41500 K.  
Note that this effective temperature also gives a reasonable fit of 
the UV spectrum (see Fig.\
\ref{hd46223_UV}). 
The subsequently derived stellar parameters are gathered in Table \ref{tab_prop}.
As we do not
have reliable gravity indicators, we adopt $\log g = 4.0$

As regards the terminal velocity, we find \vinf\ = 2800 \kms\
from the UV resonance lines. This is in fairly good agreement with
previous estimates (see Table \ref{tab_comp}). 
The mass loss rate
is derived from \ha\ and the UV resonance lines. The adopted
value for \mdot\ is 10$^{-6.5}$ \myr. As for HD 93204, clumping was 
necessary to fit \ov\ and \niv. Since the inclusion of
clumping leads to mass loss rates lower than in homogeneous winds,
this explains partly why our estimate is nearly a factor 5 lower than
most previous estimates for this star which did not use clumping 
(see Table \ref{tab_comp}).
Note that in our models, the inclusion of clumping reduces the strength of
\nv\ which is then less well fitted than in the case of the
homogeneous model. However, the very blue part of the absorption 
profile is contaminated by interstellar Lyman absorption rendering 
the exact line profile uncertain.

\subsection{HD93250}
\label{s_hd93250}

\begin{figure}
\centerline{\psfig{file=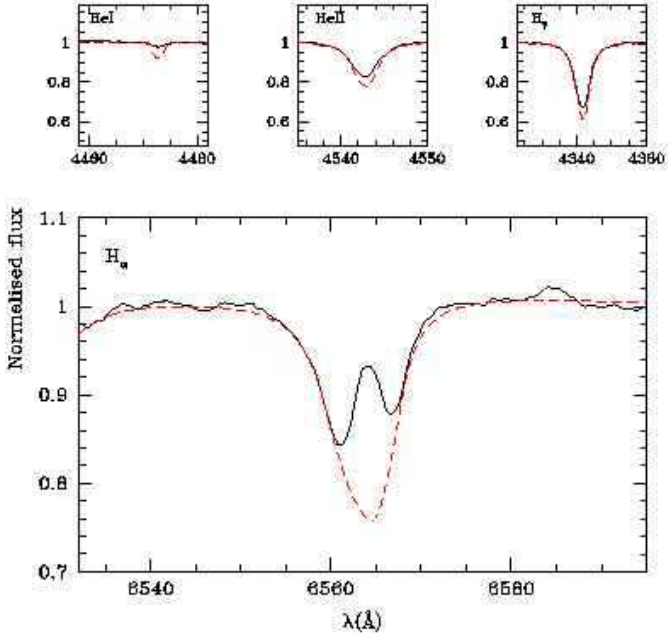,width=9cm}}
\caption{Best fit (red dashed line) of the observed He 
  and \ha\ lines 
  (black solid line) of HD 93250. The effective temperature is 46000 K, 
  $\log g$ = 4.0, \vsini\ = 110 \kms \mdot\ = $10^{-6.25}$
  \myr, \vinf\ = 3000 \kms\ and $f_{\infty} = 0.01$.  
}
\label{hd93250_Haopt}
\end{figure}

\begin{figure}
\centerline{\psfig{file=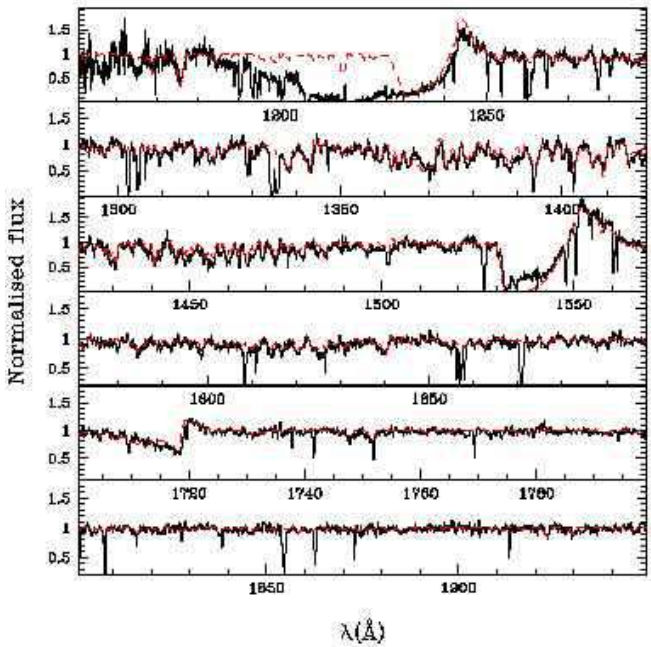,width=9cm}}
\caption{Best fit of the UV spectrum (black solid
  line) of HD 93250. For this model (red dashed line) \mdot\ = $10^{-6.25}$ \myr,
  \vinf\ = 3000 \kms and $f_{\infty} = 0.01$. 
}
\label{hd93250_UV}
\end{figure}

HD 93250 is a well studied O dwarf of the Trumpler 16 cluster in the 
Carina region. It is a prototype of the recently introduced O3.5 
subclass (ST O3.5((f+)) Walborn et al.\ \cite{walborn02}). 

\begin{figure}[!h]
\centerline{\psfig{file=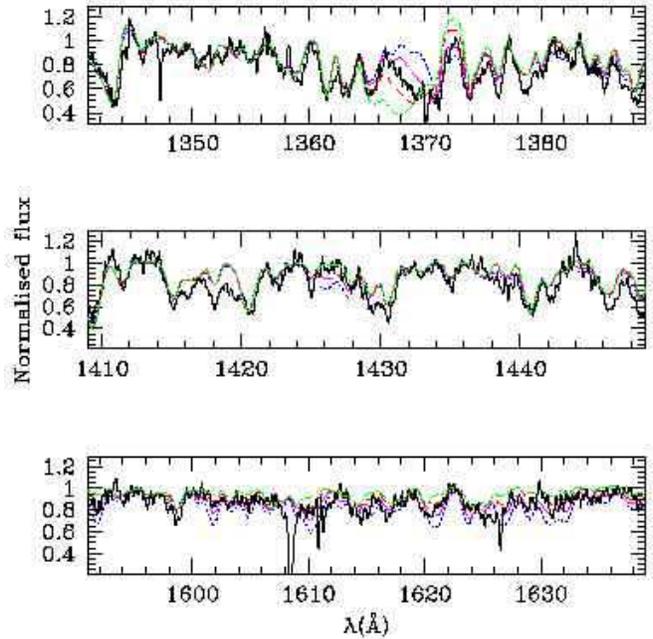,width=9cm}}
\caption{Determination of effective temperature from UV Fe line
  forests. Solid line is the observed spectrum, dotted line a model
  with \teff\ = 40000 K, long dashed line a model with \teff\ = 42000
  K, short dashed line a model with \teff\ = 44000 K and dot-dashed
  line a model with \teff\ = 46000 K. See text for discussion 
}
\label{hd93250_Fe_UV}
\end{figure}

Optical lines indicate a projected rotational velocity of 100 \kms\ 
and an effective temperature 
of $\sim$ 46000 K (mainly from the strength of \hei, see Fig.\
\ref{hd93250_Haopt}). However, Fe line forests in the UV are more
consistent with a value of 42000-44000 K as displayed in Fig.\
\ref{hd93250_Fe_UV}. For such a \teff\, \hei\ is a little too strong
in the model. However, this seems to be the case of all H and He
optical lines, possibly due to the fact that HD 93250 may be a binary
(see Repolust et al.\
\cite{repolust}) which may also be advocated from the fact that the 
absorption of \civ\ is not black despite the strength of the line 
(allowing the study of discrete absorption components). 
Hence, we rely mainly on the UV and we adopt a
value of 44000 K for the effective temperature of HD 93250. 
This value is in reasonable agreement with the 
determination of Repolust et al.\ (\cite{repolust}) who found 46000
K. We adopted $\log g = 4.0$ from Repolust et al.\ (\cite{repolust}) 
and our fit of H$\gamma$. Note that the estimated mass for this star 
is especially high (144 M$_{\odot}$), which may make HD 93250 one of 
the highest mass stars known. However, the uncertainty on the mass 
determination is huge, and HD 93250 is also suspected to be a binary. 
Hence, we caution that the mass given in Table \ref{tab_prop} is only 
indicative.

The determination of the wind parameters relies on \ha\ and on several
strong UV lines: \nv, \oiv, \ov, \civ, \heiiuv\ and \niv. The terminal
velocity deduced mainly from \civ\ is 3000 \kms, slightly lower than
the previously derived values which are between 3250 \kms\ (Repolust
et al.\ \cite{repolust}) and 3470 \kms\ (Bernabeu
\cite{bernabeu}). However, we use a microturbulent velocity of 200
\kms\ in the outer part of our model atmosphere for this star, so that
in practice, the absorption extends up to 3200 \kms. As regards the
mass loss rate, we actually found that it was impossible to find a
value for \mdot\ which would produce reasonable fits of all UV lines
in homogeneous winds. Indeed, \ov\ was always too strong and \niv\ too
weak. Reducing the effective temperature does not improve the
situation, since values as low as 40000 K are required to fit \ov, and
in that case the other UV lines are not correctly fitted so that
again, we had to include clumping.  In the end, we find that a mass
loss rate of 10$^{-6.25}$ \myr with a clumping factor $f_{\infty} =
0.01$ gives a reasonable fit, as displayed in Fig.\
\ref{hd93250_UV}. This value of \mdot\ is lower than the determination
of Repolust et al.\ (\cite{repolust}) -- 10$^{-5.46}$ \myr -- relying
only on \ha. We will return to this in Sect.\ \ref{disc_mdot}.  Note
that the value of $f_{\infty}$ we derive is quite small, but not
completely unrealistic in view of recent results presented by Bouret
et al.\ (\cite{jc05}) indicating $f_{\infty} = 0.02$ and $0.04$ for
two O4 stars.

\section{Role of X-rays and magnetic field in weak-wind stars}
\label{X_rays}

Several of our sample stars have published X-ray fluxes. Chlebowski \&
Garmany (\cite{chleb}) report X-ray measurements for HD 38666, HD
46202, HD 152590, HD 42088 and HD 46223, while Evans et al.\
(\cite{evans03}) give X-ray luminosities for HD 93204 and HD
93250. These high energy fluxes may have important consequences on the
atmosphere structure since, as shown by MacFarlane et al.\
(\cite{macfarlane}), the ionisation fractions may be significantly
altered. These authors also demonstrated that the effect of X-rays was
higher in low-density winds: ionisation in early O stars is almost
unchanged by X-rays, while in early B-stars changes as large a factor
10 can be observed between models with and without X-rays. The reason
for such a behaviour is that 1) X-rays produce higher ionisation state
through single ionisation by high energy photons and the Auger process
and 2) the ratio of photospheric to X-ray flux decreases when
effective temperature decreases, implying an increasing role of X-rays
towards late type O and early B stars (see MacFarlane et al.\
\cite{macfarlane}).  Moreover, the lower the density, the lower the
recombinations to compensate for ionisations so that we expect
qualitatively an even stronger influence of X-rays in stars with low
mass loss rate.  Since some of our sample stars are late type O stars
with low density winds, X-rays can not be discarded in their
analysis. Indeed, the Carbon ionisation fraction -- and thus the
strength of the \civ\ line and the derived mass loss rates -- can be
altered.

\begin{figure}[!h]
\centerline{\psfig{file=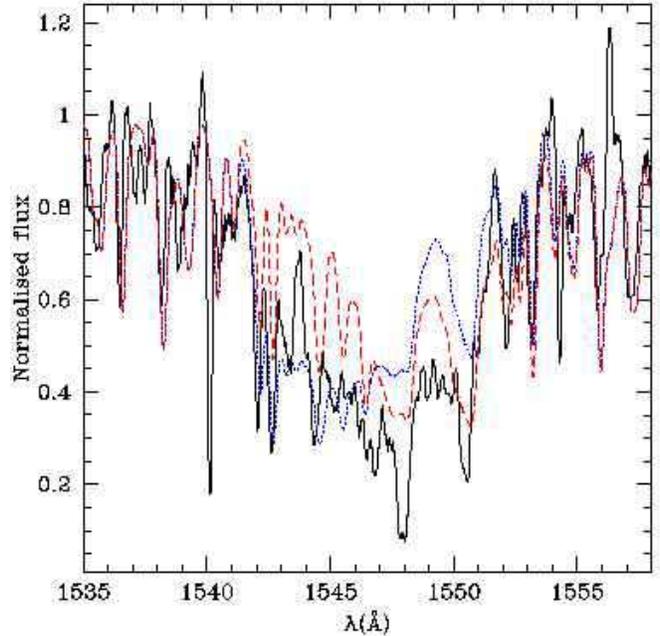,width=9cm}}
\caption{Effect on X-rays on the \civ\ line. The observed 
profile is the solid line, the initial model is the dotted 
line and the model with X-rays and the same \mdot\ is the 
dashed line. See text for discussion.
}
\label{CIV_HD46202_Xrays}
\end{figure}

\begin{figure}
\centerline{\psfig{file=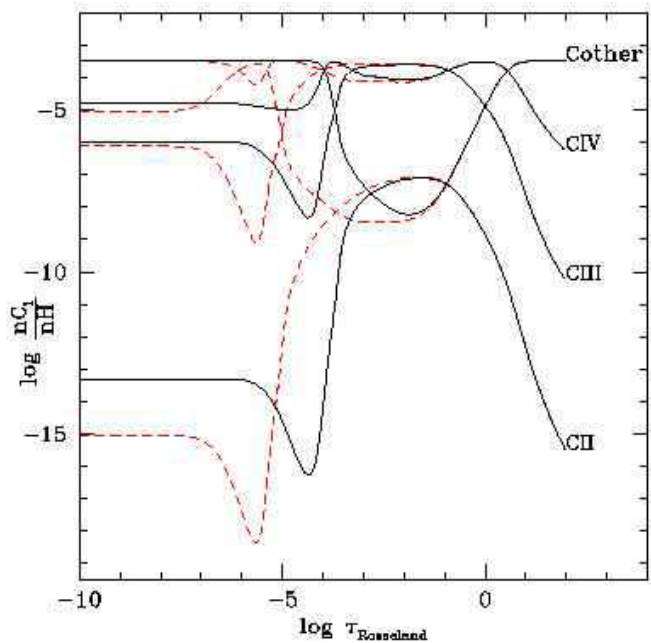,width=9cm}}
\caption{Effect of the inclusion of X-rays on the Carbon ionisation structure. Shown is the ratio of the population of C ionisation state $i$ to the total H population as a function of Rosseland optical depth. The red dashed line corresponds to the model giving the best fit to the UV spectrum when X-rays are not included, while the black solid line is the best fitting model with X-rays. In the former case, C~{\sc iv} is the dominant ionisation state in the wind, while it is no longer the case in the latter model (C~{\sc V} being the dominant). Note that C$_{\rm other}$ refers to the difference between the total C population and the sum of C~{\sc ii},  C~{\sc iii} and  C~{\sc iv}, and is mainly C~{\sc v} here. 
}
\label{ion_struct_C_X}
\end{figure}

In this context, we have first run test models for HD 46202 and HD
93250.  For HD 93250, the inclusion of X-rays did not lead to any
significant change of the ionisation structure as expected from the
above discussion. Indeed, the main wind line profiles were not
modified (see also MacFarlane et al.\ \cite{macfarlane}, Pauldrach et
al.\ \cite{pauldrach94}, \cite{pauldrach01}), indicating that X-rays
are not crucial for the modelling of \textit{these} lines in such high
density winds.  Of course, other lines are well known to be influenced
by X-rays (e.g. \ovi) but are not used in this study to derive the
stellar and wind parameters.  However, in the case of HD 46202 the
ionisation structure in the wind is strongly modified which leads to a
weaker \civ\ line (for a given \mdot) as displayed in Fig.\
\ref{CIV_HD46202_Xrays}. Indeed, the ionisation fraction of C~{\sc IV}
is reduced: this is displayed in Fig.\ \ref{ion_struct_C_X} where we
see that in the model giving the best fit without X-rays, C~{\sc iv}
is the dominant ionisation state, while when X-rays are included, it
is no longer the case. Fitting \civ\ thus requires a higher mass loss
rate. In practice, the change in the \civ\ profile when X-rays are
included is equivalent to a reduction of the mass loss rate by a
factor of $\sim$ 10 in models without X-rays. Given this result, we
have included X-rays in our modelling of the atmosphere of HD 38666,
HD 46202, HD 34078 and HD93028. For the two former stars, X fluxes
from the literature have been used while for the two latter ones, we
simply adopted $\log \frac{L_{\rm X}}{L_{\rm bol}} = -7.0$.

A question which remains to be answered concerning the X properties of
such weak wind stars is the origin of the X-ray emission. Indeed, it
is usually believed that shocks in the wind due to instabilities in
the line driving mechanism are responsible for the production of such
high energy photons (Lucy \& White \cite{lw80}, Owocki, Castor \&
Rybicki \cite{ocr88}). This scenario seems to apply to the strong wind
star Zeta Pup (Kramer et al.\ \cite{kramer03}).  However, recent
observations by Chandra have revealed that for the B0V star $\tau$ Sco
and for the Trapezium stars, most of the lines emitted in the X-ray
range were too narrow to have been produced in the wind up to
velocities of the order \vinf\ as expected in the wind-shock scenario
(see Cohen et al.\ \cite{cohen03}, Schulz et al.\
\cite{schulz03}). And these lines are also not formed very close to
the photosphere as predicted by a model in which the X-ray emission is
due to a hot corona (e.g. Cassinelli \& Olson \cite{co79}).  Actually,
such lines are more likely to be formed in an intermediate
region. This may be explained in the context of magnetically confined
winds: in this scenario, the presence of a magnetic field confines the
outflow and channels it into the equatorial plane where shocks produce
X-ray emission above the photosphere but not in the upper atmosphere
(See Babel \& Montmerle \cite{bm97}). This model has been recently
refined by Ud'Doula \& Owocki (\cite{uddoula}) who have investigated
the structure of both the wind outflow and the magnetic field through
time dependent hydrodynamic simulations. In particular, they estimated
from simple arguments the strength of the magnetic field required to
confine the wind (hereafter $B'$) and thus to lead to shocks in the
equatorial plane.

In Table \ref{tab_X}, we have gathered different properties of the
stars of our sample showing X-ray emission: the X-ray luminosity
($L_{\rm X}$), the mechanical wind luminosity ($L_{\rm wind} =
\frac{1}{2} \dot{M} v_{\infty}^{2}$) for our \mdot\ and \mdot\ from
Vink et al.\ (\cite{vink01}), and $B'$ again considering our derived
\mdot\ and Vink's \mdot. We see that for weak winds, $\frac{L_{\rm
X}}{ L_{\rm wind}}$ becomes of the order unity which shows the
increasing importance of X-rays as the wind becomes less and less
dense.  In addition, Table \ref{tab_X} shows that the magnetic field
strength required to confine the wind is low for weak-wind stars,
showing the increasing role of magnetic field when \mdot\
decreases. Given these results and the above discussion, we may
speculate that our weak wind stars may have magnetically confined
winds (although no detections of magnetic field exist for them). In
that case, one may wonder how our results would be modified. Fig.\ 8
of Ud'Doula \& Owocki (\cite{uddoula}) shows that the mass flux ($\rho
v$) is reduced close to the pole and enhanced near the equator, but
their Table 1 reveals that the total mass loss is only reduced by a
factor $<$ 2 even in the case of strong confinement. Hence, using
classical 1D atmosphere models should lead to correct values for the
mass loss rates within a factor of two, even if magnetic confinement
exists.

As regards this last point, a comment on the shape of the line
profiles is necessary. Indeed, the only O stars with a detected
magnetic field - Theta1 Ori C (Donati et al.\ \cite{donati}) - shows 
unusual features which may be
related to the geometry of magnetically confined wind (Walborn et al.\ 
\cite{walbornIUE}, Gagn\'e et al., \cite{gagne}). The absence of
such unusual features in the spectra of our stars with weak winds may
argue against such a confinement. However, as Theta1 Ori C is the only
example and as there is no theoretical prediction of the change in the
shapes of wind lines in the presence of magnetic confinement - which
also probably depends on the tilt angle between the nagnetic and
rotation axis -, we can not completely rule out the existence of
magnetic confinement in our weak wind stars.

\begin{table*}
\caption{X-ray properties of our sample stars with known X-rays fluxes. 
$L_{\rm X}$ is from Chlebowski \& Garmany (\cite{chleb}) for HD 38666, 
HD 46202, HD 152590, HD 42088 and HD 46223, and from Evans et al.\ 
(\cite{evans03}) for HD 93204 and HD 93250. $L_{\rm wind}$ is the 
mechanical wind luminosity. Values with ``Vink'' are 
those for which \mdot\ is taken from Vink et al.\ (\cite{vink01}) 
mass loss recipe. $B'$ is the value of the magnetic field for which 
confinement begins (corresponding to $\eta_{\star} = 1$ in the formalism of 
Ud'Doula \& Owocki \cite{uddoula}).}
\label{tab_X}
\small
\center
\begin{tabular}{ll|llllllll}
HD & & $\log L_{\rm X} $ & $\log \frac{L_{\rm X}}{L_{\rm bol}} $ & $\log L_{\rm wind}$ & $\log L_{\rm wind}^{Vink}$ & $\log \frac{L_{\rm X}}{ L_{\rm wind}}$ & $\log \frac{L_{\rm X}}{L_{\rm wind}^{Vink}}$ & $B'$ & $B_{Vink}'$ \\
  & & [erg s$^{-1}$] & & [erg s$^{-1}$] & [erg s$^{-1}$] & & & [G] & [G] \\
\hline
38666  & & 31.37 & -6.87 & 32.16 & 34.25 & -0.79 & -2.88 & 7 & 75 \\
46202  & & 32.40 & -6.05 & 32.76 & 34.43 & -0.36 & -2.03 & 11 & 72 \\
152590 & & 32.51 & -5.86 & 34.20 & 34.83 & -1.69 & -2.32 & 60 & 125 \\
42088  & & 32.38 & -6.43 & 34.06 & 35.89 & -1.68 & -3.51 & 33 & 269 \\
93204  & & 32.06 & -7.03 & 35.67 & 36.31 & -3.61 & -4.25 & 137 & 286 \\
46223  & & 32.62 & -6.53 & 35.89 & 36.42 & -3.27 & -3.80 & 180 & 332 \\
93250  & & 33.22 & -6.53 & 36.20 & 37.20 & -2.98 & -3.98 & 153 & 485 \\
\normalsize
\end{tabular}
\end{table*}

\section{Sources of uncertainty for the \mdot\ determination}
\label{s_uncertainty}

In this section, we investigate the various sources of uncertainty of
our determinations of mass loss rates both on the observational side
and on the modelling side.

\subsection{Observational uncertainties}
\label{uncert_obs}

Under the term ``observational uncertainty'', we gather all the
effects which can influence the shape of the observed line profiles,
especially \ha. 
The first source of uncertainty is the S/N ratio. However, in most of
the stars studied here, this ratio is good ($\sim$ 100) and does not affect the
analysis. The second source of uncertainty comes from the
normalisation of the spectra. This is a general and well known problem 
which can affect the strength of lines, especially in the case of weak 
lines. In our spectra the main difficulties arise in the \nv\ and \ha\ 
regions. For the former, this is due to the presence of the broad Lyman $\alpha$
absorption around 1216 \AA\ which renders uncertain the exact position
of the continuum. 
We simply check that the 
strength of the emission part of the profile in the models is on average 
consistent with the
observed line, leaving aside the bluest part of the absorption.
The case of \ha\ is more critical. The normalisation can be hampered 
by the S/N ratio: a low
ratio will not allow a good identification of the continuum
position. The use of echelle spectra renders also difficult the identification
of the continuum since the wavelength range around the line
of interest in a given order is limited to $\sim$ 60 \AA. We estimate
that taken together, these effects induce an uncertainty $\lesssim$
0.02 on the absolute position of the \ha\ core. Of course, \ha\ is
also contaminated by nebular emission. When present, such an emission
precludes any fit of the very core of the line. But the high
resolution of our spectra allows a fit of $\sim 80-90 \%$ of the
stellar profile, excluding the very core.

\subsection{Photospheric \ha\ profile}
\label{photosph_Ha}

Our estimates of \mdot\ rely on the fit of both the UV wind sensitive
lines \textit{and} \ha. In low density winds, \ha\ is
essentially an absorption profile for which only the central core is 
sensitive to mass loss rate. In order to derive reliable values of
\mdot\, it is thus important to know how robust the prediction of 
the photospheric profile is, since it will dominate over the wind
emission. This is of much less importance in high density winds 
where the lines are dominated by wind emission.

\begin{figure}
\centerline{\psfig{file=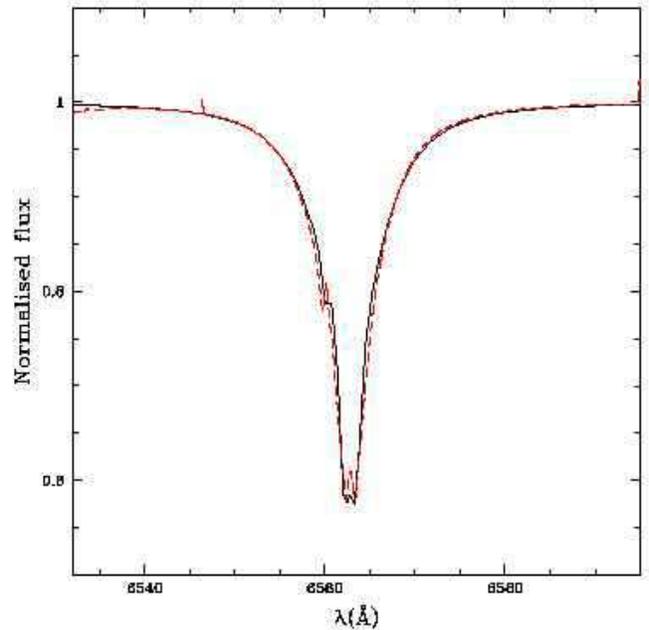,width=9cm}}
\caption{Comparison between CMFGEN (red dashed line) and FASTWIND
  (black solid line) \ha\ profile. The model is for \teff\ = 35000 K,
  $\log g =4.0$, \mdot\ $= 10^{-9}$ \myr\ and $\beta = 0.8$. The
  agreement between both codes is good.
}
\label{fast_cmf_ha}
\end{figure}

To check the CMFGEN prediction in a low density wind, we
have compared the \ha\ line with that predicted by FASTWIND, the other 
non-LTE atmosphere code including wind and line-blanketing widely used 
for optical spectroscopic analysis of massive stars (see Santolaya-Rey 
et al.\ \cite{sr97}, Puls et al.\ \cite{puls05}).
The test model
was chosen with the following parameters: \teff\ = 35000 K, $\log g =
4.0$, \mdot\ $= 10^{-9}$ \myr\ and $\beta = 0.8$. This set of
parameters is typical of the stars with weak winds analysed in the
present study, and \ha\ should not be too much contaminated by wind
emission. 
The result of the comparison between CMFGEN and FASTWIND
is given in Fig.\ \ref{fast_cmf_ha}. We see that the
agreement between both codes is very good. This is not a proof that
the predicted profile is the correct one, but it is at least a kind of 
consistency check.

\begin{figure}
\centerline{\psfig{file=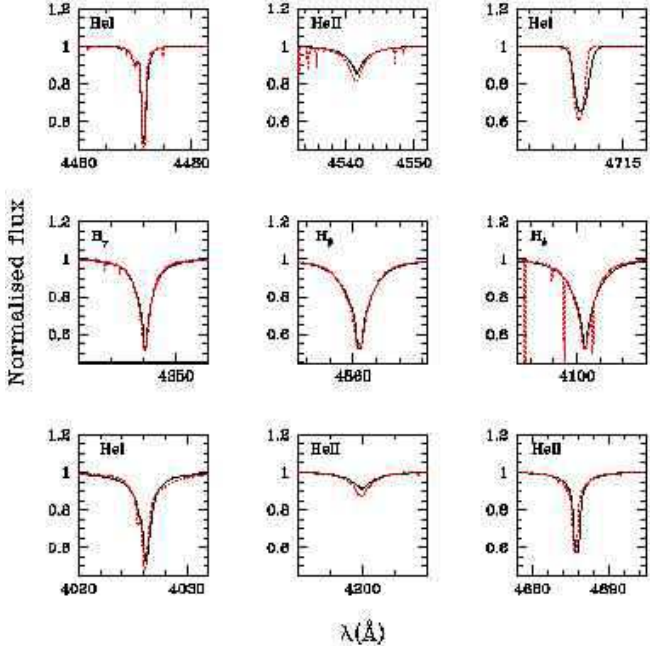,width=9cm}}
\caption{Same as Fig.\ \ref{fast_cmf_ha} but for H and He optical
  lines. The agreement between both codes is good. 
}
\label{fast_cmf_opt}
\end{figure}

We have also investigated another effect which can alter the shape of
the \ha\ line core: the number of depth points included in the
models. Indeed, the thinner the spatial sampling, the better the line
profile. This means that a too coarse spatial grid should introduce
errors in the determination of \mdot\ from \ha. We have run a
test model taking the best fit model for HD 34078 and increasing the
number of depth points from 72 to 90: 
a thinner spatial grid leads to a slightly less 
deep line core, but the difference is only of 0.01 in terms of
normalised flux. This is lower than any other observational
uncertainty (see Sect.\ \ref{uncert_obs}) so that we have adopted
$\sim$ 70 depth points in all our computations \footnote{choosing 90
  depth points significantly increases the resources
  required for the computation.}.

In conclusion, there is no evidence that the photospheric \ha\ profile
is not correctly predicted by our models.

\subsection{Ionisation fraction}
\label{ion_frac}

\begin{figure*}
\begin{minipage}[c]{1.0\textwidth}
\centerline{\hbox{
                  \psfig{file=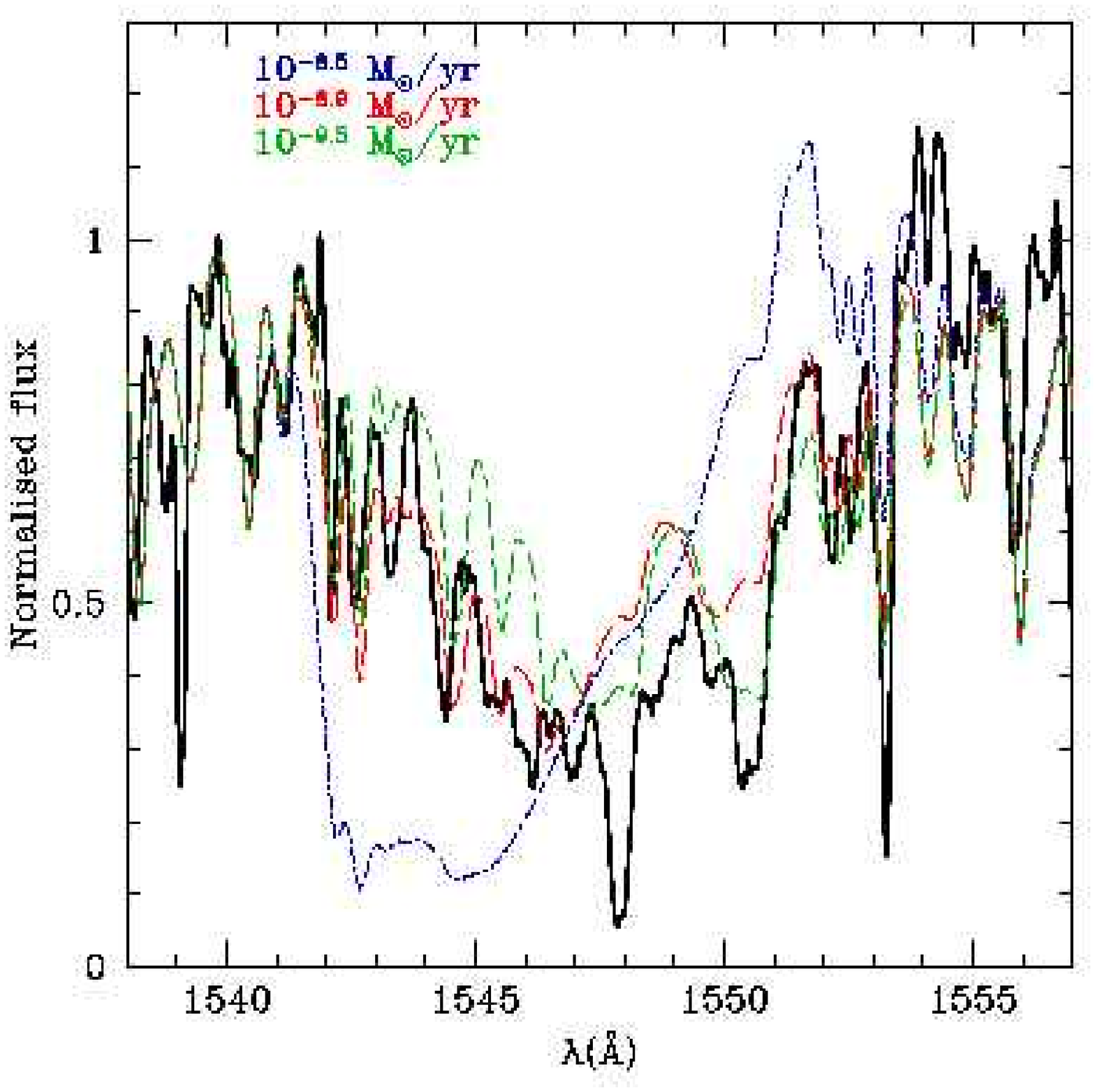,width=9.0cm}
                  \hspace{0.0cm}
                  \psfig{file=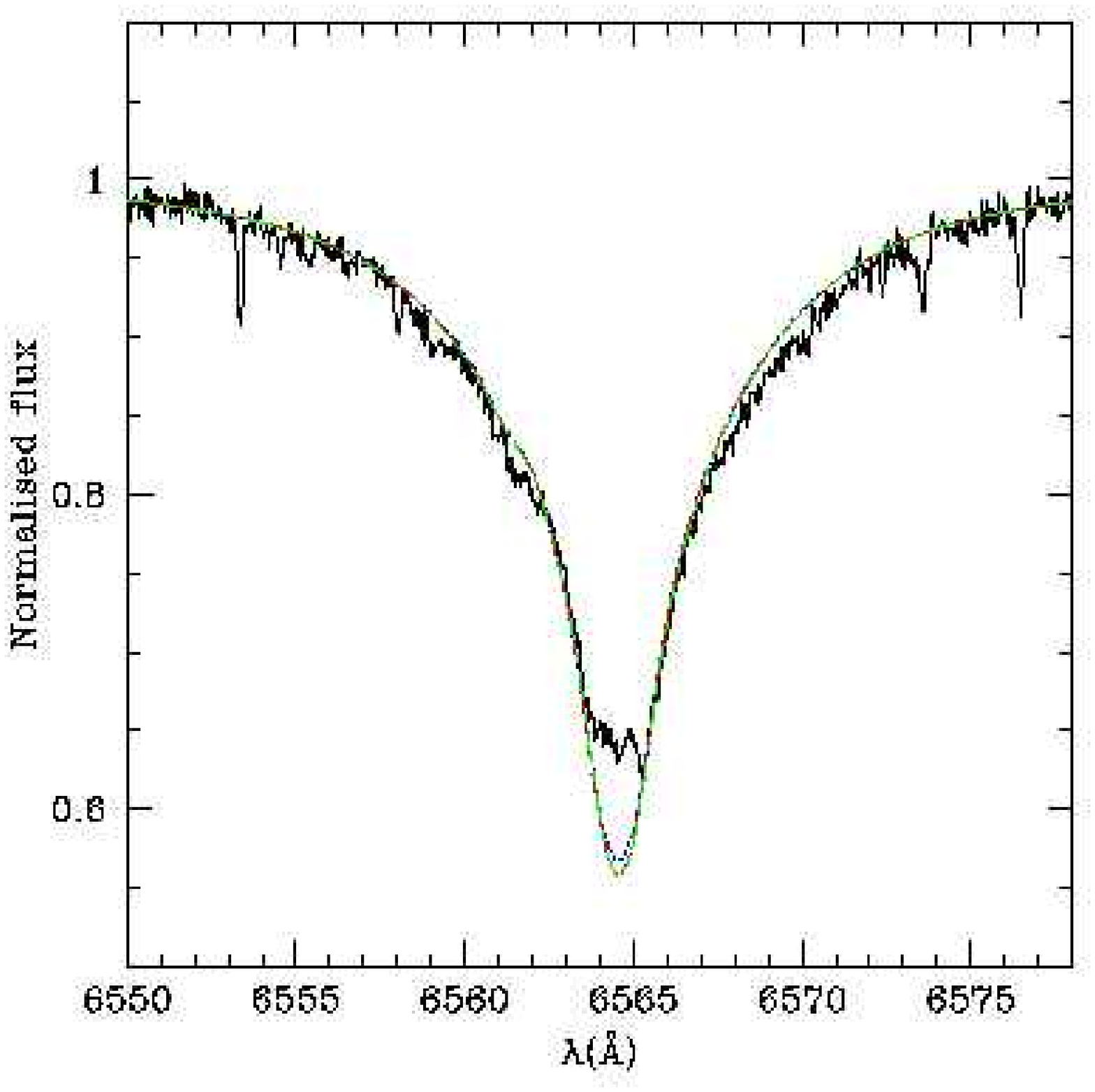,width=9.0cm}
                 }}
\vspace{0.0cm}
\hbox{\hspace{3.5cm} (a) \hspace{8.0cm} (b)} 
\vspace{0.1cm}
\caption{Determination of \mdot\ in low density winds (HD 46202). (a) shows the
  variation of the \civ\ line profile when \mdot\ is reduced (solid line:
  observed line; dash-dotted line: \mdot\ = $10^{-8.5}$ \myr; long dashed
  line: \mdot\ = $10^{-8.9}$ \myr; short-dashed long-dashed line: \mdot\ =
  $10^{-9.5}$ \myr). Notice that the weaker the line, the more prominent 
the presence of photospheric lines (mostly from iron) superimposed to the 
C~{\rm iv} absorption. (b) shows the behaviour of \ha\ under the same
  changes. See text for discussion. 
}
\label{Mdot_HD46202}
\end{minipage}
\end{figure*}

In the low density winds, which correspond to late O type dwarfs in the 
present study, the final word concerning the mass loss rate 
is often given by \civ. Indeed, \ha\ becomes almost insensitive to
\mdot\ in these cases, and the other main wind
sensitive UV line, \nv, is almost absent from 
the spectra due to the reduced effective temperature. Other indicators 
such as \siiv\ or \niv\ are still present, but they are
weaker than \civ\ and become rapidly insensitive to any change of the
mass loss rate. For more standard winds, almost all indicators can be used
together to derive \mdot. We show in Fig.\ \ref{Mdot_HD46202} the
variation of the \civ\
line profile when the mass loss rate is decreased from 10$^{-8.5}$
down to 10$^{-9.5}$ \myr\ for the case of star HD 46202. We clearly
see that \civ\ is still sensitive to changes in \mdot\ even for such
low values. In parallel, we see that \ha\ is essentially unchanged in
this regime of \mdot.

\begin{figure}
\centerline{\psfig{file=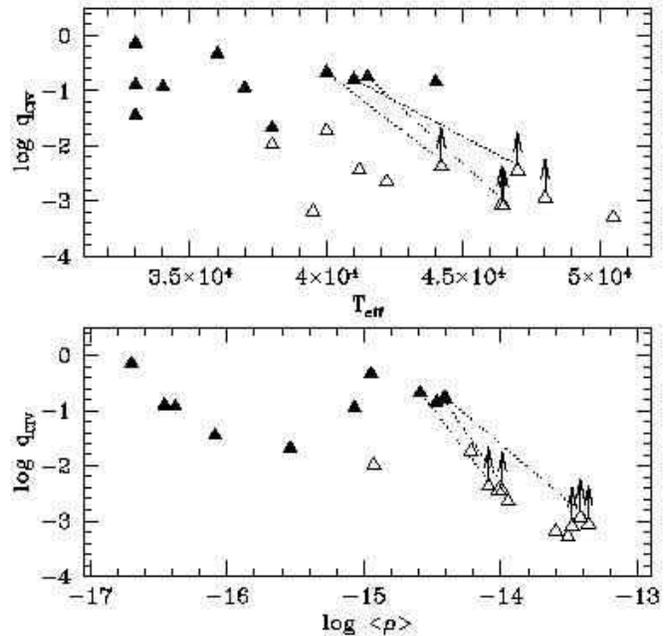,width=9cm}}
\caption{C~{\sc IV} ionisation fractions in CMFGEN models (filled symbols) and
  from Lamers et al.\ \cite{lamers99} (open symbols) for dwarfs as a function of
  effective temperature (upper panel) and mean wind density (lower
  panel). The dotted lines link objects in common between this work
  and the study of Lamers et al.\ (\cite{lamers99}). See text for discussion.
}
\label{ion_frac_CIV}
\end{figure}

However, relying on only one line to assign final mass loss rates may
be risky. We have highlighted in paper I that erroneous
mass loss rates may be derived in the case where the C~{\sc IV} ionisation fraction
is incorrectly predicted. This is still true here, since fitting the
observed profile gives the right value of $\dot{M} \times q_{C~{\sc
IV}}$ ($q_{C~{\sc IV}}$ being the ionisation fraction of C~{\sc
IV}) but not necessarily the right \mdot. In Fig.\ \ref{ion_frac_CIV}, 
we compare the ionisation fractions predicted by the CMFGEN best fit
models to the values derived by Lamers et al.\ \cite{lamers99} (hereafter L99) for
dwarfs. The ionisations fractions are defined by

\begin{equation}
q_{C~{\sc IV}} = \frac{ \int^{1}_{0.2} n_{C~{\sc IV}}(x) dx}
  {\int^{1}_{0.2} n_{C}(x) dx}
\end{equation}

\noindent where $x=\frac{v}{v_{\infty}}$ and $n_{C~{\sc IV}}$ and $n_{C}$ are the
number densities of C~{\sc IV} and C respectively. At first glance, the CMFGEN
ionisation fractions seem to be $\sim$ 2 orders of magnitude higher
than the L99 results, and that in spite of the few lower
limits in the latter data. However, several comments can be
made:

- First, the work of L99 is based on previous mass loss rate
determinations, mainly from \ha\ (Puls et al.\ \cite{puls96}, Lamers
\& Leitherer \cite{ll93}) or from predictions for their dwarf
subsample (Lamers \& Cassinelli \cite{lc96}). In the latter case,
\mdot\ is derived from the modified wind momentum - luminosity
relation, so that any error in the calibration can lead to incorrect
mass loss rate. Moreover, the uncertainty of such a method due to the
fact that a given star can deviate from a mean relation may introduce
a bias in the derived ionisation fraction. Concerning the mass loss
rates derived from \ha, Lamers \& Leitherer (\cite{ll93}) use the line
emission strength to determine \mdot. However, in most O dwarfs \ha\
is in absorption so that the determination of the emission part of the
line filling the photospheric profile may be uncertain. Puls et al.\
(\cite{puls96}) also use \ha\ to derive \mdot\ but give only upper
limits in the cases of thin winds. As L99 adopt these upper limits as
the real values, we should expect the derived ionisation fractions to
be lower limits.

- Second, there is a significant shift in terms of parameter space
sampled by our results and that of L99: we have stars with $ 33000 <
\teff < 44000$ K and $ -17.3 < \log <\rho> < -14.4$ (where $<\rho> =
\frac{\dot{M}} {4 \pi r^{2}_{0.5}} \frac{2}{v_{\infty}}$ and $r_{0.5}$
is the radius at which the velocity reaches half the terminal
velocity) while L99 have $38000 < \teff < 50500$ K and $ -15 < -log
<\rho> < -13.4$, although both studies have stars of late and early O
spectral types. Concerning effective temperatures, part of the
discrepancy comes from the use of line-blanketing in our models, which
is known to reduce \teff\ compared to unblanketed studies.  But for
densities, the explanation may again come from the fact that the
adopted mass loss rates (and consequently the densities) in one or the
other study are not correct.  Can we discriminate between them? An
interesting point is that 3 stars are common to our study and that of
L99: they are shown linked by dotted lines in Fig.\
\ref{ion_frac_CIV}. If we consider the fact that line-blanketing may
explain the lower \teff\ in our study, and the fact that for these
stars the ionisation fractions derived by L99 are only lower limits,
then the ionisation fractions predicted by CMFGEN are not necessarily
too high.  And if in addition we argue that our study investigates a
density range not explored by L99, then we can not conclude that the
ionisation fractions predicted by CMFGEN are wrong since no comparison
can be made for very low mean densities.

How could we test more strongly the wind ionisation fractions of our
atmosphere models? One possibility is offered by the analysis of far
UV spectra. Indeed, this wavelength range contains a number of lines
\textit{formed in the wind} from different ions of the same
elements. Such a test will be done in a subsequent paper, based on
FUSE observations of Vz stars in the LMC. But we can already mention
that several studies of supergiants in the Magellanic Clouds using
FUSE + optical data do not reveal any problem with the CMFGEN wind
ionisation fractions, except that clumping must be used to reproduce a
couple of (but not all) lines (see Crowther et al.\ \cite{paul02},
Hillier et al.\ \cite{hil03}, Evans et al.\ \cite{evans04}). Hence in
the following, we assume that the ionisation fractions given by CMFGEN
are correct. \\

\subsection{Abundances}
\label{abundances}

Although our mass loss determination relies on both \ha\ and UV lines, 
we usually give more weight to the UV diagnostics since the absorption
profile of \ha\ can be shaped by other parameters than \mdot\ ($\beta$,
clumping). But the UV lines depend more strongly on abundances than
\ha. Hence, we have to estimate the error we make on the \mdot\
determination from UV lines due to uncertain abundances. We have
already seen in Sect.\ \ref{s_hd15629} that adopting the CNO
abundances of Asplund (\cite{asplund04}) instead of those of Grevesse \& 
Sauval (\cite{gs98}) -- which corresponds on average to a reduction by a 
factor of $\sim$ 3/4 -- leads to an increase of \mdot\ by 0.25 dex. 
We have also run test models for a low luminosity star (HD 46202). It
turns out that reducing the CNO abundances by a factor 2 implies
an increase of the mass loss rate of the order of 2-2.5 in order 
to fit \civ\ since this line is not saturated in low density winds 
and thus its strength is directly proportional to the number of absorbers.. 

How different from solar could the CNO abundances of our sample stars be ?
Given the estimated distances, it turns out that all stars are within 
3 kpc from the the sun. Determinations of abundances through spectroscopic 
analysis of B stars (Smartt et al.\ \cite{smartt_ab}, Rolleston et al.\ 
\cite{rolleston}) reveal the following gradients: $-0.07$ dex/kpc for C, 
O and Si, and $-0.06$ to $-0.09$ dex/kpc for N. Similarly, Pilyugin et al.\ 
(\cite{pilyugin}) derive an Oxygen abundance gradient of $-0.05$ dex/kpc 
from studies of HII regions. Taken together, these results indicate that 
on average we do not expect variation of CNO abundances by more than $\pm$ 
0.25 dex for our sample stars. This means that adopting a solar metallicity 
leads to an error of at most 0.3 dex on the mass loss rate determination.
 
Given the above discussion, we estimate the error on \mdot\ due to 
uncertainties in the CNO abundances to be of $\sim$ 0.3 dex.

\subsection{Advection / adiabatic cooling}
\label{adv_adcool}

In low density winds, two processes may affect the ionisation
structure: advection and adiabatic cooling. The former is rooted in
the fact that for low densities, the timescale for recombinations
becomes longer than the timescale for transport by advection. Thus
the ionisation structure can be significantly changed. The latter
process (adiabatic cooling) lowers the temperature in the outer part of 
the atmosphere where the heating processes (mainly photoionisations)
are less and less efficient due to the low density, implying also a
modification of the ionisation structure (see also Martins et al.\
\cite{n81}). 
We have tested the influence of those two effects in one of our low
\mdot\ models for HD 46202. Their combined effects lead to an
increased ionisation in the outer atmosphere, the mean ionisation fraction
of C~{\sc IV} being lowered by $\sim$ 0.1 dex (which does not modify
the conclusions of Sect.\ \ref{ion_frac}). This slightly changes the
UV line profiles, especially \civ\, which for a given \mdot\ shows a
smaller absorption in the bluest part of the profile. Quantitatively,
the inclusion of advection and adiabatic cooling is equivalent to an
increase of \mdot\ by $\sim$ 0.15 dex. We have thus included these two 
processes in our models for low density winds (HD 38666, HD 34078, 
HD 46202, HD 93028).\\

Given the above discussions, we think our \mdot\ determinations have a
very conservative error bar of $\pm 0.7$ dex (or a factor $\pm$
5). This is a quite large uncertainty which however does not modify
qualitatively our results, namely the weakness of O dwarfs with low
luminosity (see Sect.\ \ref{disc_mdot}).

\section{Discussion}
\label{s_disc}

\subsection{Evolutionary status}
\label{disc_evol}

\begin{figure}
\centerline{\psfig{file=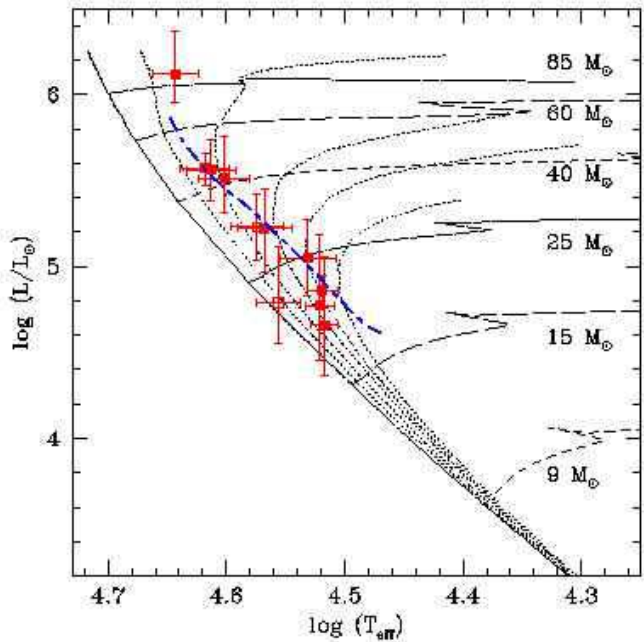,width=9cm}}
\caption{HR diagram of the Galactic stars. Evolutionary tracks are
  from Lejeune \& Schaerer (\cite{lejeune}) and Z = Z$_{\odot}$. 
  Isochrones for 0, 1.. 5 Myrs
  are indicated together with evolutionary paths of stars of different 
  masses. The blue dashed line is our new relation \teff\ - luminosity for
  dwarfs (Martins et al.\ \cite{calib05}). Squares are the stars studied
  here (open symbols are for Vz stars). 
}
\label{HR_diag}
\end{figure}

Fig.\ \ref{HR_diag} shows the HR diagram of the our sample stars. 
Overplotted on Fig.\ \ref{HR_diag} is our new
calibration \teff\ - luminosity (Martins et al.\ \cite{calib05}, solid
line) for dwarfs: most stars of our sample agree more or less with
this relation (within the error bars). 
The latest type stars of our sample, which are also the stars showing the 
weakest winds, may be slightly younger than ``standard'' dwarfs of the same 
spectral type (or \teff). Notice that this does not mean that these stars are 
the youngest in terms of absolute age, but that they are less evolved than classical 
dwarfs. Indeed, the youngest stars of our sample are those of Trumpler 16 
(HD 93250 and HD 93204) for which we derive an age of 1 to 2 Myrs, compatible 
with the $L -$ \teff\ relation. In comparison, HD 38666 and HD 34078 may be 2 to 
4 Myrs old according to our HR diagram (although given the error bars, we can 
not exclude younger ages), slightly less than for standard late type dwarfs. In 
the scenario where these two stars originated from a binary and were ejected 
in a dynamical interaction, Hoogerwerf et al.\ 
(\cite{hbz01}) estimate a travel time of 2.5 Myrs, while van Rensbergen et al.\ 
(\cite{vr96}) found travel times of $\sim$ 3.5 Myrs for HD 38666 and $\sim$ 2.5 
Myrs for HD 34078. These estimates are in good agreement with our results. We also 
derive an age of $\sim$ 3-5 Myrs HD 46202, one of our weak wind stars. Note that 
this star is in the same cluster as HD 46223 which is likely 1-2 Myrs old 
according to Fig.\ \ref{HR_diag}. A similar age should be expected for these two 
stars in case of a burst of star formation, but an age spread of 1-2 Myrs (common 
in star clusters) can explain the difference. The same is true for the stars of 
Cr 228: HD 93146, the brightest star, may be slightly younger than HD 93028. 

HD 152590 behaves
differently, being less luminous than other dwarfs of same \teff. It
is interesting to note that this
star is classified as Vz. Taking literally the result of Fig.\
\ref{HR_diag}, it seems indeed that it is younger than other dwarfs (but again, 
the error bars are large),
confirming the fact that Vz stars are supposed to lie closer to the
ZAMS than typical dwarfs. However, HD 42088 is another Vz star of our
sample, and it has a more standard position on the HR diagram. This
poses the question of the exact evolutionary status of Vz
stars. Indeed, they are defined by stars having \heiivz\ stronger than 
any other He~{\sc II} lines which is thought to be a characteristic
of youth since this line is filled with wind emission when the star
evolves. In fact the Vz characteristics may be more related to the wind 
properties than to the youth of the star. Indeed, HD 42088 seems to have the same 
stellar properties as HD 93146, but the former is classified Vz (not the latter) 
and has a weaker wind (\mdot\ = $10^{-8}$ \myr\ compared to $10^{-7.25}$ \myr\ for 
HD 93146. Note however that the distance (and thus luminosity) of HD 42088 
is highly uncertain. Obviously, more studies are required to better understand
the physics of Vz stars.

To summarise, there may be a hint of a link between a relative youth and the 
weakness of the wind \textit{if by youth we mean an evolutionary state earlier 
than for standard stars and not an absolute age}, standard stars meaning stars 
with the average properties of dwarfs studied so far. But the present results are 
far from being conclusive. A 
forthcoming study of Vz stars in the LMC will probably shed more light on this 
issue.

\subsection{Wind properties}
\label{disc_ww}

\subsubsection{Terminal velocities}
\label{disc_vinf}

\begin{figure}
\centerline{\psfig{file=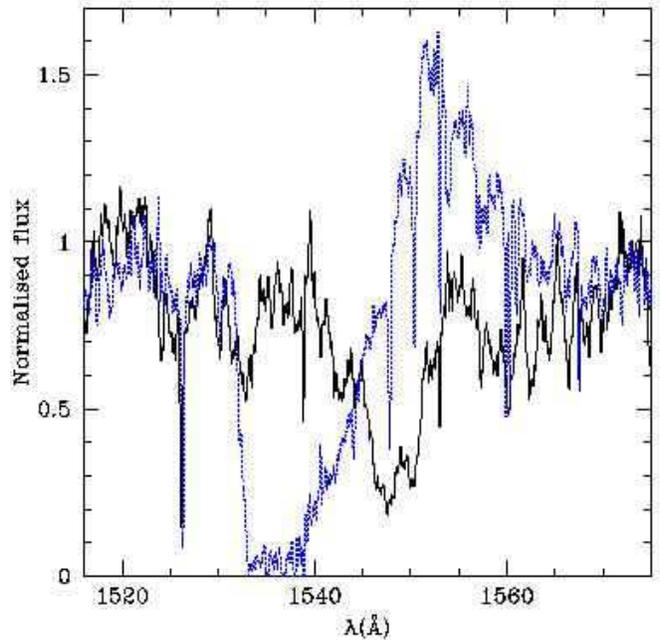,width=9cm}}
\caption{Comparison between \civ\ line profiles in a star 
with weak (HD 38666, solid line) and strong (HD 46223, 
dotted line) wind. The rise of the flux level in the very 
blue part of the absorption features is slightly shallower 
in HD 38666. See text for discussion. 
}
\label{comp_CIV}
\end{figure}

\begin{figure}
\centerline{\psfig{file=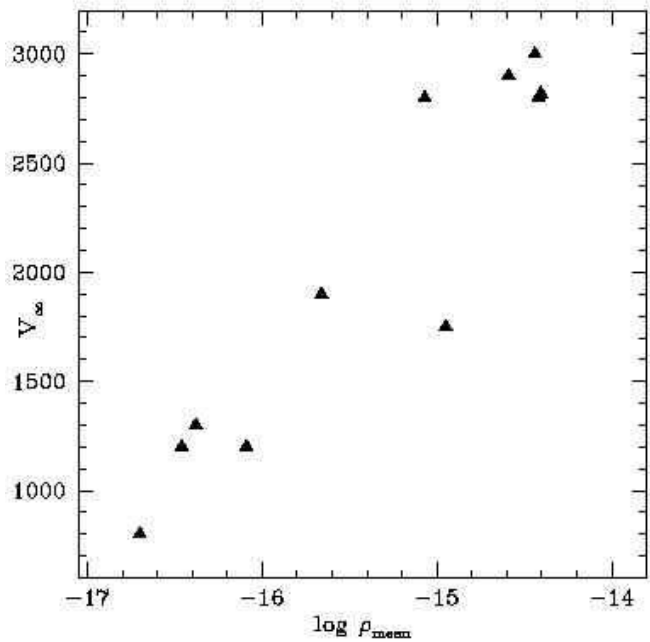,width=9cm}}
\caption{Terminal velocity as a function of mean density 
in the wind for the stars studied here. There is a clear 
trend of lower \vinf\ in lower density winds. 
}
\label{CIV_rho}
\end{figure}

Some of the terminal velocities we derive are surprisingly low (see
Table \ref{tab_prop}) reaching values lower than the escape velocity
in a one case (HD 34078). What could explain this behaviour? First, the most
obvious reason could be an underestimation of \vinf. We have argued
in paper I that in stars with weak winds the density in the upper
parts of the atmosphere may be so low that almost no absorption takes
place in strong lines usually formed up to the top of the
atmosphere. This explanation was also given by Howarth \& Prinja
(\cite{hp89}) to justify the low \vinf\ they obtained in some
stars. If this is indeed the case, one would expect a smooth decrease
of the absorption strength in the blue part of P-Cygni profiles due to 
the reduction of the density as we move outwards, and
not a steep break as seen in dense winds. Is there such a transition?
Fig.\ \ref{comp_CIV} shows the \civ\ line profiles of HD 34078 and 
HD 46223 and reveals that although the increase of the flux level 
from the deepest absorption to the continuum level in the bluest 
part of the profile extends over a slightly larger range in the 
case of the weak wind star (3 \AA\ for HD 38666 instead of 2 \AA\ 
for HD 46223), it is difficult to draw any final conclusion as 
regards the reduction of the \civ\ absorption in the outer wind 
of low density wind stars from this simple eye estimation given 
also that blending is clearly apparent in the line of HD 38666. 
More information is given by Fig.\ \ref{CIV_rho} which shows 
the derived terminal velocities as a function of mean density 
in the wind (see Sect.\ \ref{ion_frac} for definition). There 
is an obvious trend of lower terminal velocities with lower 
wind densities. This is not a proof of the fact that absorption 
in strong UV lines extends to larger velocities since low 
densities also mean low mass loss rate and correspond to 
stars with lower radiative acceleration. However, it is an 
indication that underestimations of \vinf\ are certainly more 
likely to happen in such low density stars.

In view of the above discussion, it is not clear whether the lower
density in the outer atmosphere of weak wind stars is responsible for
an underestimation of the terminal velocities.  But we can not exclude
that our estimates of \vinf\ are lower than the true values in low
density winds. Now, with this in mind, let us now assume for a moment
that the derived values are real terminal velocities: what are the
implications? The radiation driven wind theory predicts that \vinf\ is
tightly correlated to the escape velocity (\vesc) according to

\begin{equation}
v_{\infty}^{2} = v_{esc}^{2} I(\alpha) \frac{\alpha}{1-\alpha}
\label{eq_vinf}
\end{equation}

\noindent where $\alpha$ is the usual parameter of the Castor, Abbott \& 
Klein (\cite{cak}) formalism and $I(\alpha)$ is a correction 
factor to take into account effects of the finite cone angle 
of the star disk (see Kudritzki et al.\ \cite{kud89}). In practice, 
it is possible to derive values of $\alpha$ from this equation 
once the stellar parameters and \vinf\ are known.
 The only problem comes from $I(\alpha)$ which is a complex 
function of (among other parameters) $\alpha$. However, it is 
possible to solve this problem with the following procedure: 
we first assume a given value of $\alpha$, then estimate $I(\alpha)$ 
which is subsequently used to find a new $\alpha$ using 

\begin{equation}
\alpha = \frac{\frac{v_{\infty}^{2}}{v_{esc}^{2}}}{I+\frac{v_{\infty}^{2}}{v_{esc}^{2}}}
\label{eq_alpha}
\end{equation}

\begin{figure}
\centerline{\psfig{file=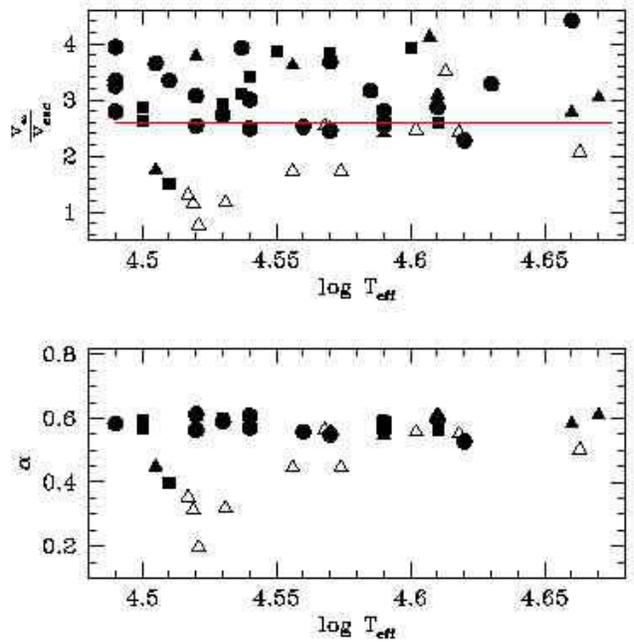,width=9cm}}
\caption{Upper panel: ratio of terminal to escape velocity in our 
sample stars (open symbols) and stars studied by Herrero et al.\ 
\cite{her00}, \cite{hpn02}, Repolust et al.\ \cite{repolust} and 
Markova et al.\ \cite{mark04} (filled symbols). The solid line 
indicates the classical value 2.6 derived for stars hotter than 
21000 K (Lamers et al.\ \cite{lamers95}). Dwarfs (giants, supergiants) 
are shown by triangles (squares, circles). Lower panel: derived 
$\alpha$ parameter from the estimated terminal and escape velocities. 
See text for discussion.
}
\label{vinf_alpha}
\end{figure}

\noindent A few iterations should lead to the final value of the $\alpha$ 
parameter. We have used such a scheme to estimate $\alpha$ for 
our sample stars and for a number of stars studied elsewhere (Herrero et al.\ 
\cite{her00}, \cite{hpn02}, Repolust et al.\ \cite{repolust} and 
Markova et al.\ \cite{mark04}). 
Solutions are usually found with less than 10 iterations (and 
are essentially independent on the starting value of $\alpha$), except 
in the cases where \vinf/\vesc was larger than $\sim 3$: in that 
case, the iterative process did not converge but kept oscillating 
between two distinct values. The results for the cases where 
solutions could be found are shown in Fig.\ \ref{vinf_alpha} (lower panel). 
A majority of cases lead to $\alpha \sim 0.5-0.6$, in reasonable agreement with
(although slightly lower than) 
theoretical expectations (Puls, Springmann \& Lennon 
\cite{psl00}). However, for the stars of this work with weak 
winds (and low \vinf), lower values are deduced ($\alpha \sim 0.3$). Hence, 
\textit{if the derived low terminal velocities 
correspond to real \vinf}, they may be due to low values 
of the $\alpha$ parameter (given Eq.\ \ref{eq_vinf}.) If true, this
may also have important 
implications for the scaling relations involving mass loss rates 
(see next section). Again, it may be possible that we underestimate 
the terminal velocities, but the above possibility is worth being 
discussed in view of the puzzle of the weak winds. Note that we have 
also plotted in Fig.\ \ref{vinf_alpha} the ratio of terminal to escape 
velocity which is usually of the order 2.6 for O stars with \teff\ $>$ 
21000 K (Lamers et al. \cite{lamers95}). We see that hottest stars of 
our sample follow this general trend whereas stars with weak winds have 
much lower ratios.

Another very interesting explanation for the low terminal velocities
we derive (again assuming that they are the real values) is the effect
of X-rays. Indeed, Drew et al.\ (\cite{drew94}) highlighted the fact
the cooling time in the outer atmosphere of massive stars with
relatively weak winds (late O / B stars) can become relatively high so
that in the case where X-rays possibly emitted by shocks heat the
outer atmosphere, this region remains hot. In that case, the
ionisation structure is strongly modified compared to the inner
atmosphere and in practice, the radiative force becomes negligible in
this hot region.  This means that the wind keeps expanding at the
velocity reached at the top of the ``cool'' region which is lower than
the value predicted by the radiation driven wind theory. This effect
should be checked in future hydrodynamical simulations.

\subsubsection{Mass loss rates and modified wind momenta}
\label{disc_mdot}

\begin{figure}
\centerline{\psfig{file=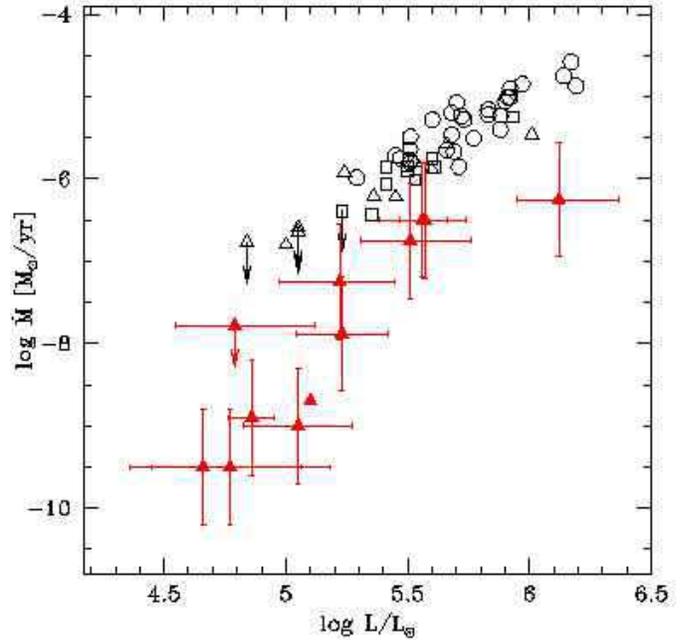,width=9cm}}
\caption{Mass loss rates as a function of Luminosity for Galactic O 
stars. The filled triangles are the dwarfs studied in the present paper 
(+ 10 Lac from paper I displayed by the filled triangle without error 
bars). Open symbols are data from Herrero et al.\ (\cite{her00}, 
\cite{hpn02}), Repolust et al.\ (\cite{repolust}) and Markova et al.\ 
(\cite{mark04}). Triangles (squares, circles) are for dwarfs (giants, 
supergiants). 
}
\label{Mdot_L}
\end{figure}

The mass loss of O stars has been known for a long time to depend on 
luminosity since due to the basic mechanism of radiatively driven 
winds, the more photons are available, the larger the acceleration 
and the larger the mass loss rate (e.g. Castor, Abbott \& Klein 
\cite{cak}, Kudritzki \& Puls \cite{kp00}). Fig.\ \ref{Mdot_L} shows 
mass loss rates for our sample stars (filled symbols) and stars from 
other studies (Herrero et al.\ 
\cite{her00}, \cite{hpn02}, Repolust et al.\ \cite{repolust} and 
Markova et al.\ \cite{mark04}, open symbols) as a function of 
luminosity. One sees that there is a good correlation between 
\mdot\ and $L$ for bright stars. Note however that our sample stars 
seem to show lower \mdot\ than what could be expected from the 
other studies (see also Table \ref{tab_comp}). For low luminosity
stars, the correlation still 
exists, but the scatter is much larger. Moreover, the slope of the 
relation seems to be steeper for these objects, the transition 
luminosity being $\log \frac{L}{L_{\odot}} \sim 5.2$. Although our 
work is the first to show such a behaviour based on quantitative 
modelling of atmosphere of O stars, this trend was previously 
mentioned by Chlebowski \& Garmany (\cite{chleb}) and Lamers \& 
Cassinelli (\cite{lc96}). This result confirms our finding of paper 
I in which we showed that the stellar components of the star 
forming region SMC-N81 displayed winds weaker than expected from 
the relation \mdot\ - $L$ at high luminosities. 
In paper I, we mentioned that a possible explanation of such a
weakness was the reduced metallicity of the SMC, 
but we also showed that the Galactic star 10 Lac had the same 
low mass loss rate. Here, we confirm that several Galactic stars 
with low luminosity indeed show low \mdot, rendering unlikely the 
effect of metallicity alone. 

\begin{figure}
\centerline{\psfig{file=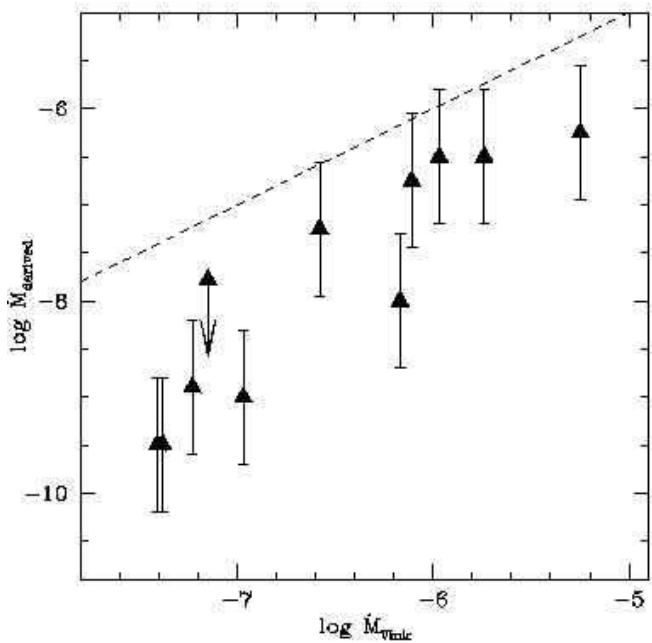,width=9cm}}
\caption{Comparison between derived mass loss rates and predictions of 
  hydrodynamical simulations (Vink et al.\ \cite{vink01}). 
}
\label{Mdot_vink}
\end{figure}

We also showed in paper I that the winds were weaker than predicted by
the current hydrodynamical simulations. Fig.\ \ref{Mdot_vink} extends
this trend for the Galactic stars studied here: in the ``worst''
cases, the difference between our derived \mdot\ and the mass loss
rates predicted by Vink et al.\ (\cite{vink01}) can reach 2 orders of
magnitude! Note that even for bright stars our values are lower than
the predictions but only by a factor $\lesssim$ 5. This is mainly due
to the introduction of clumping in our models for these stars which
naturally leads to reduced mass loss rates (Hillier et al.\
\cite{hil03}, Bouret, Lanz \& Hillier \cite{jc05}). Using a different
approach, Massa et al. (\cite{massa03}) also hilighted the possibility
of lower mass loss rates due to the presence of clumping. We will come
back to this below.

One may also wonder why our values of \mdot\ are lower than other previous
studies. Indeed, as shown by Table \ref{tab_comp} our mass  loss rates are 
systematically lower than derived so far for all stars of the sample. How 
can we explain this behaviour? First, let us recall that the mass loss 
rates gathered in Table \ref{tab_comp} are estimated from either pure 
\ha\ analysis (Leitherer \cite{leith88}, Lamers \& Leitherer \cite{ll93}, 
Puls et al.\ \cite{puls96}, Repolust et al.\ \cite{repolust}, Markova 
et al.\ \cite{mark04}) or from pure UV analysis (Howarth \& Prinja 
\cite{hp89}, Chlebowski \& Garmany \cite{chleb}). 

The \ha\ study of 
Leitherer (\cite{leith88}) and Lamers \& Leitherer (\cite{ll93}) 
relied on measurement of \ha\ emission equivalent widths. They are 
linked with \ha\ luminosities which are themselves related to mass 
loss rates. The relation L(\ha) - \mdot\ is based on estimates of 
the population of the third level of Hydrogen for which departure 
coefficients from LTE are taken from the pure H He computations of 
Klein \& Castor (\cite{kc78}). The \ha\ emission equivalent width is 
calculated from the total equivalent width to which a photospheric 
profile from the plane-parallel pure H He models of Auer \& Mihalas 
(\cite{am72}) is subtracted. This procedure may suffer from various 
approximations: the use of pure H He models may introduce errors in 
the prediction of departure coefficients since line-blanketing is 
known to affect the ionisation (and excitation) structure; moreover, 
photospheric profiles based on H He plane-parallel models may also 
be different from line-blanketed spherically extended models. In the 
case of stars with \ha\ in absorption, the estimate of the emission 
part may be risky since it may suffer from uncertainties in the 
photospheric component subtraction, from contamination by interstellar 
lines or from errors in normalisation process. Leitherer 
(\cite{leith88}) himself argues that the wind emission part of the
global (wind + photospheric) profile becomes 
almost undetectable in stars with \ha\ in absorption. Hence, the 
\ha\ mass loss rates of such objects based on this method is 
rather uncertain.
 
Another method relying on \ha\ is that of Puls et al.\ (\cite{puls96}) 
and Markova et al.\ (\cite{mark04}). It is again based on the emission 
part of \ha\ which is related to mass loss rate through an estimate of 
the H departure coefficient under the Sobolev approximation. Their 
method is accurate for values of their parameter A $> 10^{-4}$ (see 
Eq. (3) of Puls et al.\ for a definition of A) which, for typical values 
of the stellar parameters of O stars, corresponds roughly to \mdot\ 
$> 10^{-7}$ \myr. This is mainly the reason why the authors give only 
upper limits for stars with \ha\ in absorption. 
Finally, Repolust et al.\ (\cite{repolust}) used FASTWIND (see Sect.\ 
\ref{photosph_Ha}) to fit the \ha\ profile and estimate \mdot. In the 
case of weak winds, this method becomes less and less accurate since 
\ha\ is almost entirely photospheric and hardly depend on mass loss. 
Here again, the authors only give upper limits on \mdot\ in those cases. 

Concerning UV based determinations of the mass loss rate, Howarth \& 
Prinja (\cite{hp89}) use the column densities in several UV lines to 
estimate $\dot{M} \times qi$ ($qi$ being the ionisation fraction of 
the ion responsible for the line studied). Under the approximation 
that the ionisation fraction is independent of luminosity, they 
derive \mdot. The latter approximation may introduce errors in the 
mass loss rates estimate. Indeed, our modelling of massive stars 
atmospheres reveals that the ionisation fractions are not constant 
when \teff\ changes among dwarf stars (which in this case reduces 
to a change of luminosity, see Fig.\ \ref{ion_frac_CIV}). Since 
the largest ionisation fractions of C~{\sc IV} given by Howarth \& 
Prinja (\cite{hp89}) are of the order 10$^{-2.5}$ (see their Fig.\ 16) 
while we find values as high as $\sim$ 10$^{-0.5}$, a factor of 100 
between their \mdot\ and ours is possible.

The other UV analysis to which our results are compared is that of 
Chlebowski \& Garmany (\cite{chleb}). The authors use fits of the UV 
lines using the method of Olson (\cite{olson}) which is similar 
to the SEI method (Lamers et al.\ \cite{lamers87}). Basically, this 
method uses a parameterisation of the optical depth through the line 
profile to produce synthetic profiles which, once compared with observed 
spectra, allow a determination of \mdot. However, this latter step 
requires a few approximations. In particular, the ionisation fraction 
has to be estimated which involves the use of SED at high energies 
(i.e. close to ionisation thresholds of C~{\sc IV} and N~{\sc V}): a 
blackbody distribution is used in the computations of Chlebowski \& 
Garmany (\cite{chleb}). Moreover, only photoionisation and 
recombinations from/to the ground states are taken into account. 
Hence, once again the ionisation fractions may not be correctly 
predicted leading to errors on \mdot. However, it is interesting 
to note that the approach of Chlebowski \& Garmany (\cite{chleb}) 
is more accurate than that of Howarth \& Prinja (\cite{hp89}) as 
regards the ionisation fractions and leads to lower mass loss rates 
(see Table \ref{tab_comp}), so that if, as we can expect, the ionisation 
fractions are better predicted in the current atmosphere models and 
are in fact higher, lower mass loss rates are not too surprising.

Another important point highlighted in paper I was the behaviour 
of the so called modified wind momentum - luminosity relation (WLR) 
at low luminosities. Indeed, we showed that there seemed to be a 
breakdown of this relation below $\log \frac{L}{L_{\odot}} \simeq 
5.2$, at least for stars of the SMC (including the stars of paper I 
and 3 stars of NGC 346 studied by Bouret et al.\ \cite{jc03}). 
The Galactic star 10 Lac also showed a reduced wind momentum, 
indicating that this property could not be related to 
metallicity alone. 
In the present study, several characteristics of the WLR are 
highlighted.

\begin{figure}
\centerline{\psfig{file=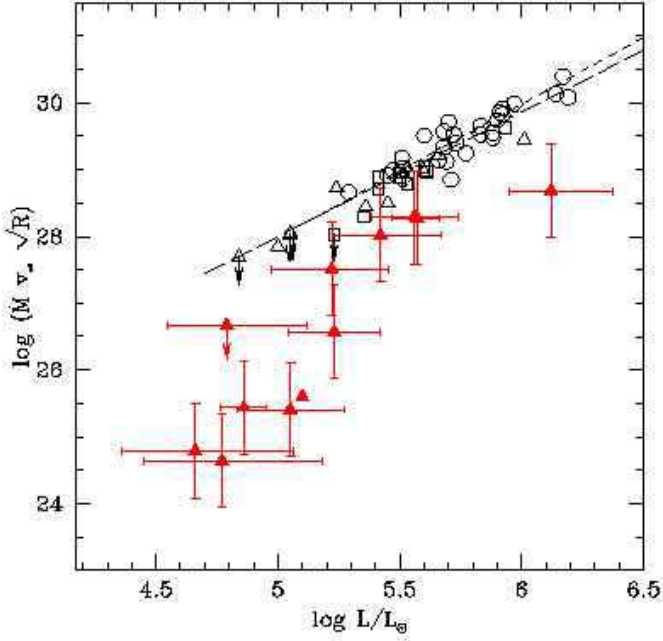,width=9cm}}
\caption{WLR for Galactic stars. Symbols and data are the same as 
in Fig.\ \ref{Mdot_L}. Note the breakdown of the relation below  
$\log \frac{L}{L_{\odot}} \sim 5.2$. See text for discussion. The 
short (long) dashed line is the regression curve for supergiants 
(giants + dwarfs) of Repolust et al.\ (\cite{repolust}). 
}
\label{WLR_gal}
\end{figure}

We first confirm that there is a breakdown - or at least a steepening- 
of the WLR 
below $\log \frac{L}{L_{\odot}} \simeq 5.2$. Indeed, Fig.\ 
\ref{WLR_gal} shows that most stars below this transition luminosity 
have wind momenta lower that what one could expect from a simple 
extrapolation of the WLR for bright stars. Indeed, for $\log 
\frac{L}{L_{\odot}} \sim 5.0$, the relation for dwarfs + giants 
found by Repolust et al.\ (\cite{repolust}) gives wind momenta of 
the order $10^{28}$ while we find values as low as $10^{25}$! There is in fact 
only one object which is marginally in agreement with the relation 
of Repolust et al.\ (\cite{repolust}), but we have only an upper 
limit for \mdot\ for this star (HD 152590).

\begin{figure}
\centerline{\psfig{file=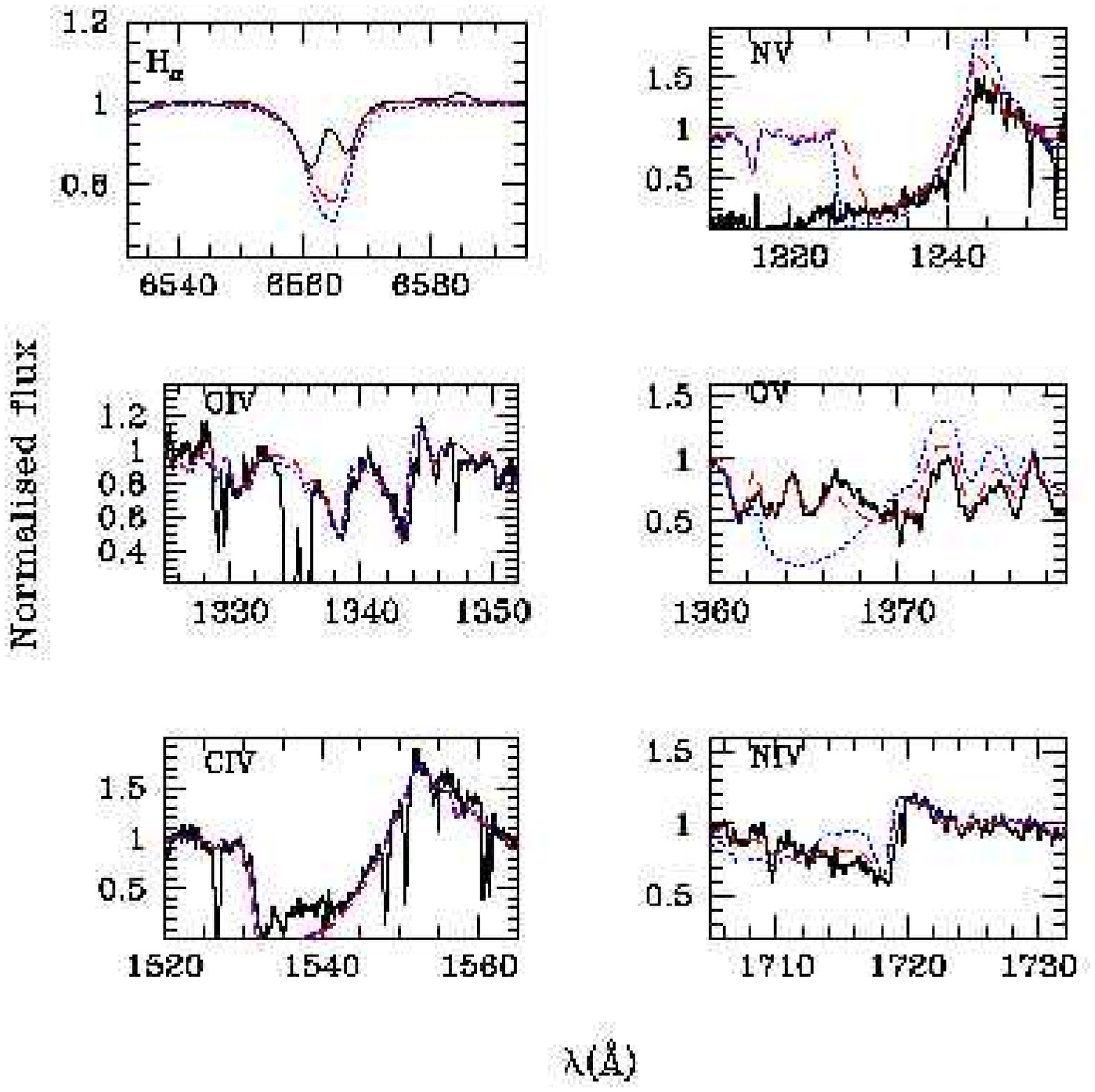,width=9cm}}
\caption{Influence of clumping on the determination of \mdot\ 
of HD 93250. The solid line is the observed spectrum, the 
dotted line is a model with \mdot\ = $10^{-5.46}$ \myr (mass 
loss rate derived by Repolust et al.\ \cite{repolust} from 
\ha) and no clumping, and the dashed line is a model with 
\mdot\ = $10^{-6.25}$ \myr and $f_{\infty} = 0.01$. The 
inclusion of clumping leads to a reduction of the mass loss
 rate derived from both UV lines and \ha. See text for discussion.
}
\label{mdot_hd93250}
\end{figure}

Second, we find that for the bright stars of this study, the modified
wind momenta are reduced compared to the pure \ha\ analysis on which
the WLR is established. The difference is on average a factor of
between 5 and 10, especially for the two objects we have in common (HD
15629 and HD 93250). The explanation of this discrepancy comes from
the use of clumping in our models for these stars. Indeed, it is
necessary to use inhomogeneous winds to correctly fit a number of UV
lines, especially \ov\ and \niv.  In Fig.\ \ref{mdot_hd93250} we show
the UV + \ha\ spectra of HD 93250 and two models: one with the mass
loss rates derived by Repolust et al.\ (\cite{repolust}) from \ha\
only ($\dot{M} = 10^{-5.46}$ \myr), and the other with our estimate
of \mdot\ relying on both \ha\ and UV lines and taking clumping into
account ($\dot{M} = 10^{-6.25}$ \myr).  One sees clearly that although
both models are acceptable for \ha\ (in view of the nebular
contamination one can not exclude one or the other possibility), UV
lines are overpredicted with $\dot{M} = 10^{-5.46}$ \myr. Note that
Repolust et al.\ (\cite{repolust}) put forward the fact that the
presence of clumping may lead to an overestimation of the derived mass
loss rates if unclumped models are used (see also Massa et al.\
\cite{massa03}). Their argument is mainly based on the larger modified
wind momenta of stars with \ha\ in emission compared to stars with
\ha\ in absorption which can be explained by the neglect of clumping
in the former. However, they do not exclude the existence of clumping
in the latter, but claim that its effects can not be seen due to low
optical depth. In our case, all stars have \ha\ in absorption, and we
deduce the presence of clumping from UV lines.

Note that in our study, we had to include clumping only in the
more luminous stars to correctly fit the UV spectra. Does it mean that the 
winds of fainter stars are homogeneous? Not necessarily. Indeed,
clumping is required to reproduce \ov\ and \niv. But it turns out that 
in late type O stars (i.e. low $L$ stars) \ov\ is absent and \niv\ is
mainly photospheric so that homogeneous winds give reasonable
fits. Clumping may be present, but it can not be seen from the UV and
optical spectra. In any case, if clumping were to be included in the
models, the mass loss rates would have to be reduced to fit the
observed spectra (see above). Thus, \mdot\ would have to be further
reduced compared to the already low values we obtain, making the winds 
of low luminosity O dwarfs even weaker!

In spite of the global shift of our WLR for bright O dwarfs compared
to pure \ha\ studies
when clumping is included, the slope of the relation is roughly 
the same as that found by Repolust et al.\ (\cite{repolust}). This 
is important since it shows that the breakdown of the WLR we 
find at low luminosities is not an artifact of our method. 
Equivalently, this means that even if we underestimate the 
mass loss rates (due to ionisation fractions, see Sect.\ 
\ref{ion_frac}), there is however a qualitative change of 
the slope of the WLR near $\log \frac{L}{L_{\odot}} \sim 5.2$.

Can we estimate the value of the slope of the WLR in this low $L$ 
range? The number of stars studied is still too low to give a 
reliable value, but if we do a simple by eye estimate, excluding 
star HD 152590 (due again to the fact that it is a possible 
member of a binary), we find a slope of the order of 4.3. As this 
slope is in fact equal to 1/$\alpha'$ (e.g. Kudritzki \& Puls 
\cite{kp00}), where $\alpha' = \alpha - \delta$ and $\delta = 0.005..
0.1$,  we deduce $\alpha' \sim 0.25$ and $\alpha \sim 0.30$ which is
very low 
compared to the classical value of $\sim$ 0.6, but which is 
consistent with our finding based on the ratio of terminal to 
escape velocities (Sect.\ \ref{disc_vinf}). This does not mean 
that $\alpha$ is indeed this low for these stars since both 
\vinf\ and \mdot\ may be underestimated, but it is at least a 
kind of consistency check. 

\subsubsection{Origin of weak winds}
\label{disc_origin_ww}

In view of the above results, what can we conclude as regards the 
origin of the weakness of the winds observed in some O dwarfs? 
The main possibilities have been detailed in paper I. Among them, 
metallicity was the first to come to mind since at that time, as most 
stars with weak winds were found in the SMC (paper I, Bouret et al.\ 
\cite{jc03}). The present study clearly 
shows that metallicity cannot explain the reduced wind strength observed 
since several Galactic stars show mass loss rates and terminal velocities 
as low as SMC objects. On the contrary, it becomes more and more evident 
that there is a transition in the wind properties near $\log 
\frac{L}{L_{\odot}} = 5.2$, although the reasons for such a change of 
behaviour remain unclear at present. 
To say things more clearly, we do not state that metallicity has no effect 
on the wind strength (this is now well established): we simply show that 
low luminosity dwarf O stars have winds much weaker than expected from 
hydrodynamical simulations and than so far observed for O stars, 
\textit{independent of metallicity}.

A possible explanation is the reduction of the $\alpha$ 
parameter which, if it were as low as $\sim$ 0.25, could explain both the 
reduced mass loss rates and terminal velocities. However, what could
be the reason for such a low $\alpha$ ? A nice possibility was highlighted by
Puls et al.\ (\cite{psl00}) in their very detailed analysis of the line
statistics. They first show that under fairly general conditions, the
classical $\alpha$ parameter, i.e. the one entering the slope of the
line strength distribution function, and the $\hat{\alpha}$ parameter
used to express the radiative acceleration according to

\begin{equation}
g^{rad} \propto t^{-\hat{\alpha}}
\label{eq_grad}
\end{equation}

\noindent in the CAK formalism (see
Castor, Abbott \& Klein \cite{cak}) are the same (this is the basics of 
the radiation driven wind formalism). However, the line strength
distribution function may not have a constant slope, and in that case
the value of $\alpha$ such that $\hat{\alpha} = \alpha$
must be derived at the point where the line strength is equal
to $t^{-1}$. This has two important consequences: first, since $t =
s_{e} v_{th} \frac{\rho}{dv/dr}$ (with $s_{e}$ the Thomson scattering
opacity and $v_{th}$ the typical thermal velocity), it will be
different from star to star; and second, $t$ is not constant in a
given star's atmosphere. In practise, this means that $\alpha$ should
be different not only from star to star, but also from point to point
in the atmosphere of a given star! However, Puls et al.\ (\cite{psl00}) 
have shown that the slope of the line
strength distribution function is constant over a large range of $t$
values, implying that $\alpha$ keeps a constant value close to 0.6.
But in extreme cases, we may reach a range where this slope varies: 
in that case, $\alpha$ is reduced. This is shown by Fig.\ 27 
of Puls et al.\ (\cite{psl00}) where we see that for $t$ lower than
10$^{-6}$ $\alpha$ deviates significantly from 0.6. The interesting
thing is that this situation corresponds to low densities (see the
definition of $t$ above). In conclusion, $\alpha < 0.6$
is expected in low density winds, which is consistent with our
findings.

Given this fact, the main problem would come from the strong
disagreement between the results of spectroscopic analysis and the
predictions of hydrodynamical simulations. Could such simulations
overestimate mass loss rates? It is indeed possible since they 
neglect velocity curvature terms in the computation 
of radiative accelerations (Owocki \& Puls \cite{op99}), which can
affect the final results. As discussed in 
paper I, O dwarfs with low luminosity are the most sensitive to such effects 
but test models for a 40 \msun\ star performed by Owocki \& Puls (\cite{op99}) 
lead to downward revision of the mass loss rate by only a factor 1.5 while we 
would need a factor $\sim$ 100! 

One can also wonder what is the effect of X-rays on the driving of winds. 
We have already seen in Sect.\ \ref{disc_vinf} that low terminal velocities 
can be expected when X-rays are present due to changes in the outer 
atmosphere ionisation structure. Further insights can be found in Drew 
et al.\ (\cite{drew94}). These authors have 
studied two B giants and have found mass loss 
rates 5 times lower than values expected for a simple extrapolation of 
the \mdot\ - $L$ relation of Garmany \& Conti (\cite{gc84}) established 
for O stars. They also detected X-ray emission in both stars, and argued 
that such X-rays, likely formed in the outer atmosphere, can propagate 
towards the inner atmosphere and change the ionisation structure here too, 
reducing the total radiative acceleration and thus the mass loss rate. 
This may partly explain why the mass loss predictions of Vink et al.\ 
(\cite{vink01}), which does not take X-rays into account, are higher than 
our derived values.
Obviously, hydrodynamical models including X-rays are needed to test 
this attractive hypothesis.

We highlighted in paper I that decoupling may be an alternative explanation 
although no conditions for it to take place were fulfilled in the N81 atmospheres. 
Here, the stellar and wind parameters of the weak winds stars being similar to 
those of the N81 stars, we have checked that such a process is not likely to 
be at work either.

\section{Conclusion}
\label{s_conc}

We have derived the stellar and wind properties of Galactic O dwarfs 
with the aim of tracking the conditions under which weak winds such as 
observed in SMC-N81 (Martins et al.\ \cite{n81}) develop. Atmosphere 
models including non-LTE treatment, spherical expansion and line-blanketing 
were computed with the code CMFGEN (Hillier \& Miller \cite{hm98}). 
Optical H and He lines provided the stellar parameters while both UV 
lines and \ha\ were used to determine the wind properties. The main results 
can be summarised as follows:

\begin{itemize}

\item[$\diamond$] The O dwarfs studied here are 1 to 2 Myrs old for the 
hottest and 2 to 4 Myrs old for the coolest. Except for the faintest, 
they have luminosities in reasonable agreement with the new calibration \teff\ 
- Luminosity of Martins et al.\ (\cite{calib05}). 

\item[$\diamond$] Stars with luminosities below a certain transition 
luminosity ($\log \frac{L}{L_{\odot}} \la 5.2$) have mass 
loss rates of the order of 10$^{-8..-9.5}$ \myr\ and low terminal velocities 
(down to 800 \kms). The mass loss rates are lower by nearly a factor of 100 
compared to the hydrodynamical predictions of Vink et al.\ (\cite{vink01}). 
Uncertainties in the determination of \mdot, discussed here in detail, are 
not expected to qualitatively alter the results.

\item[$\diamond$] Stars with $\log \frac{L}{L_{\odot}} \ga 5.2$ are found to 
have reduced mass loss rates compared to both hydrodynamical predictions and 
previous analysis based only on \ha. The main reason is the inclusion of 
clumping in our models in order to fit \ov\ and \niv\ in the IUE range. The 
adoption of pure \ha\ based mass loss rates does not allow fits of most
of the UV lines. 

\item[$\diamond$] The modified wind momentum - luminosity relation shows a break 
down around  $\log \frac{L}{L_{\odot}} = 5.2$. Below this transition value, 
the slope corresponds to a value of the $\alpha$ parameter of the order of 0.3, 
which is consistent with the low terminal velocities observed. Such a
low $\alpha$ is expected in low density winds (Puls et al.\
\cite{psl00}). 

\item[$\diamond$] The origin of the weakness of the winds in low luminosity 
stars compared to hydrodynamical simulations is still unknown, but
metallicity effects can be excluded since all the 
stars of the present study are Galactic stars and show reduced winds similar 
to SMC stars (Bouret et al.\ \cite{jc03}, Martins et al.\ \cite{n81}). An earlier 
evolutionary state than in standard dwarfs may be responsible or not for the 
weakness: the present results can 
not resolve this issue given the error bars in the age estimates. 

\item[$\diamond$] Although their origin remains unclear, X-rays appear to play 
a very important role in the physics 
of weak winds. They may be due to magnetic mechanisms and affect the ionisation 
structure. This can possibly reduce the driving force and partly explain the low 
terminal velocities and low mass loss rates. Hydrodynamical simulations 
including X-rays should give more quantitative results. 

\end{itemize}

The low luminosity objects of our sample have not been studied
individually with quantitative spectroscopy before since the
atmosphere models allowing the analysis of weak wind stars have only
been available for a few years. Indeed, the detailed modelling of UV
wind sensitive lines requires a reliable treatment of
line-blanketing since most of these lines are from elements heavier
than He. This also explains why a number of previous quantitative analysis
relied essentially on \ha.

Now that the existence of weak winds has been established observationally both 
in the SMC and in the Galaxy, it would certainly be suitable to investigate the 
problem from a theoretical point of view with new hydrodynamical 
simulations. Apart from that, we still have to make sure that the ionisation 
fractions predicted by CMFGEN are correct since thay may alter the mass loss 
rate determinations. We will conduct such an investigation in a forthcoming paper 
using FUSE data.

\begin{acknowledgements}
We thank the referee, Alex Fullerton, for interesting and useful
suggestions and comments which helped to improve the quality of the
paper. We thank the ESO staff in La Silla for their help during the
observations. Some of the data presented in this paper were obtained
from the Multimission Archive at the Space Telescope Science Institute
(MAST). STScI is operated by the Association of Universities for
Research in Astronomy, Inc., under NASA contract NAS5-26555. Support
for MAST for non-HST data is provided by the NASA Office of Space
Science via grant NAG5-7584 and by other grants and contracts.  This
study is also based on INES data from the IUE satellite and made use
of the SIMBAD database, operated at CDS, Strasbourg, France. FM and DS
acknowledge financial support from the FNRS. We thank J. Puls for
providing FASTWIND models and A. Herrero and D.J. Lennon for the \ha\
spectra of HD 93204, HD 93250 and HD 15629. DJH thanks Janos Zsargo
for interesting discussions on the effects of X-rays and acknowledges
support from NASA LTSA grant NAG5-8211.
\end{acknowledgements}

{}

\end{document}